\documentclass[12pt,preprint]{aastex}
\usepackage{graphicx}
\usepackage{amssymb}
\usepackage{amsmath}
\usepackage{hyperref}
\usepackage{varwidth}
\usepackage{booktabs}
\usepackage{epsfig}
\usepackage{natbib}
\usepackage{graphicx}
\usepackage{slashbox}
\usepackage{multirow}
\usepackage{lscape}
\usepackage{mathrsfs,amssymb}
\usepackage{subfigure}
\usepackage{amssymb}
\usepackage{multirow}
\usepackage{tabularx}
\usepackage{rotating}
\usepackage{longtable}
\usepackage{ulem}
\usepackage{lineno}

\usepackage{lipsum}
\usepackage{changes}
\definechangesauthor[name={JHP}, color=orange]{JHP}
\definechangesauthor[name={XJY}, color=red]{XJY}
\hypersetup{
  colorlinks = true,
  urlcolor = blue,
  linkcolor = blue,
  citecolor = blue
}

\newcommand       \simgt        {\gtrsim}

\newcommand       \mum          {\,{\rm \mu m}}

\newcommand       \simali       {\sim\,}

\newcommand       \be           {\begin{equation}}
\newcommand       \ee           {\end{equation}}
\countdef\decade=200
\advance\decade by \year
\countdef\hours=201
\advance\hours by \time
\divide\hours by 60
\countdef\mins=202
\advance\mins by \hours
\multiply\mins by 60
\multiply\hours by 100
\countdef\miltime=203
\advance\miltime by \hours
\advance\miltime by \time
\advance\miltime by -\mins

\shorttitle{Aliphatically Deuterated Hydrocarbons in 30 Doradus}
\title{
\vspace*{-2.0em}
{\normalsize\rm {Accepted for publication in \it The Astrophysical Journal}}\\
\vspace*{1.0em}
Detection of Aliphatically Deuterated Aromatic Hydrocarbons
in the Large Magellanic Cloud 30 Doradus Star-Forming Complex
}
\author{Junhao Peng\altaffilmark{1},
             Xuejuan Yang\altaffilmark{1,2}
             and Aigen Li\altaffilmark{2}
             }
\altaffiltext{1}{Department of Physics,
                  Xiangtan University,
                  411105 Xiangtan, Hunan Province, China;
                       \sf{xjyang@xtu.edu.cn}}
\altaffiltext{2}{Department of Physics and Astronomy,
                        University of Missouri,
                        Columbia, MO 65211, USA;
                        {\sf lia@missouri.edu}}

\begin{document}

\begin{abstract}
The ``unidentified infrared (IR) emission'' (UIE)
bands at 3.3, 6.2, 7.7, 8.6, 11.3 and 12.7$\mum$
are ubiquitously seen in a wide variety of astrophysical
environments. While the exact assignment of these
UIE bands remains controversial, they are generally
ascribed to C--H and C--C stretching and bending
vibrations of aromatic hydrocarbon molecules.
Here, based on observations made with
the Near Infrared Spectrograph (NIRSpec)
and the Mid Infrared Instrument (MIRI)
aboard the James Webb Space Telescope (JWST),
we report that the UIE emitters in the 30 Doradus
star-forming complex in the Large Magellanic Cloud (LMC)
are deuterated and have an appreciable amount
of aliphatic content.
The spatially resolved NIRSpec and MIRI spectra of
30 Doradus reveal a widespread detection of
the 3.4 and 6.85$\mum$ emission features
attributed to aliphatic C--H stretch
and deformation, respectively,
as well as the 4.65$\mum$ feature
attributed to aliphatic C--D stretch.
Notably, the 6.85$\mum$ feature exhibits
three complex substructures at $\simali$6.83,
6.86 and 6.88$\mum$ that have never been reported before.
\end{abstract}



\section{Introduction} \label{sec:intro}
The ``unidentified infrared (IR) emission'' (UIE) bands
at 3.3, 6.2, 7.7, 8.6, 11.3 and 12.7$\mum$
are ubiquitously seen in a wide variety of
astrophysical regions \citep{Li2020}
in the Milky Way and nearby galaxies
as well as distant galaxies at redshifts $z\simgt4$
\citep{Riechers2014, Spilker2023}.
A large number of candidate materials have been
proposed as carriers of the UIE bands.
A large number of candidate materials have been
proposed as carriers of the UIE bands.
The prevailing interpretation attributes
the UIE bands to carbonaceous materials,
in which case the proposed carriers
can be divided into two broad categories:
gas-phase, free-flying polycyclic aromatic
hydrocarbon (PAH) molecules
\citep{Leger1984, Allamandola1985}
and amorphous solids with a mixed aromatic
and aliphatic composition
\citep[e.g., see][]{Papoular1989, Jones1990, Sakata1990, Kwok2011, Cataldo2013, Tokunaga2025}.\footnote{%
   {An alternative interpretation has recently been
proposed by \citet{Zagury2023, Zagury2025},
who argued that Raman scattering by atomic
hydrogen in photodissociation regions (PDRs)
might be responsible for the UIE bands
as well as some of the other unexplained
interstellar phenomena,
including the mysterious diffuse interstellar
bands and the ``extended red emission''
\citep[ERE; see][]{Witt2020}.
For the sake of discussion convenience,
however, in this work we limit us to
the assumption that the UIE bands
are of vibrational emission from
carbonaceous compounds.}
}
Apparently, whether the UIE carriers contain
pure aromatic rings or mixed aromatic rings
and aliphatic chains, and whether they are pure
hydrocarbon compounds or contain impurities,
are essential for understanding their chemical
structures and identities.


In the post-Spitzer era, these two categories
appear to ``converge'' \citep[see][]{Kwok2022}:
while amorphous solids with a mixed aromatic
and aliphatic composition are now believed to
be nano-sized, similar to PAHs or PAH clusters
\citep{Kwok2011, Kwok2013},
the definition of astronomical PAHs has also been
evolving from the original chemical definition of
planar, pure carbon and hydrogen ring molecules,
to a collection of heterosubstituted aromatic
hydrocarbon molecules with side groups
\citep[see][]{Yang2017}, which was described by
\citet{Allamandola2021} as ``continuous
advancements made with the interstellar PAH model.''

%
It is apparent that the UIE carriers are most likely
a carbonaceous substance whose structure is more
complicated than previously assumed
\citep[e.g., pure PAH molecules; see][]{Kwok2022}.
Indeed, despite four decades of work,
not a single specific PAH molecule has yet been
convincingly identified through its vibrational
spectra in the IR.\footnote{%
  We note that, recently, a dozen small specific PAH
  molecules have actually been identified in space.
  But these identifications were made through their
  rotational spectra in the radio
  \citep[see][and references therein]{McCarthy2026},
  not through their vibrational spectra in the IR.
  }
Also, \citet{Tokunaga2025} argued
that the PAH hypothesis as an explanation for
the UIE bands has a number of problems,
e.g., the exact wavelength of the astronomical
3.3$\mum$ band is inconsistent with PAH molecules
\citep{Tokunaga2021}.

The aromatic nature of the interstellar 3.3$\mum$
band was originally suggested by \citet{Knacke1977}
and \citet{Duley1981}. Observationally,
the aromatic C--H stretching band at 3.3$\mum$
is often accompanied by a weak satellite band
at 3.4$\mum$, often ascribed to aliphatic C--H
stretch
\citep[for their assignment see][]{Yang2017, Tokunaga2021, Bernstein2024}.
It is also interesting
to note that the aliphatic nature of the 3.4$\mum$
band was also originally suggested
by \citet{Duley1981}.

In addition, some UIE sources also exhibit
a weak feature at 6.85$\mum$ and even an
additional weak feature at 7.25$\mum$
\citep[e.g., see][]{Sloan2014, Yang2016}.
These two features are also aliphatic
in nature, arising from aliphatic C--H deformation.
Furthermore, some UIE sources also show
weak emission at $\simali$4.4$\mum$
and $\simali$4.6--4.7$\mum$
\citep[e.g., see][]{Verstraete1996, Peeters2004,
Boulanger2011, Doney2016, Onaka2022, Boersma2023,
Peeters2024, Pereira-santaella2024, Draine2025, Misselt2025},
This indicates that the UIE carriers
are sometimes also deuterated.
Upon deuteration, the UIE carriers
shift the aromatic C--H stretch
at 3.3$\mum$ to $\simali$4.4$\mum$
for the aromatic C--D stretch.
Similarly, the aliphatic C--H stretch
at 3.4$\mum$ shifts to $\simali$4.65$\mum$
for the aliphatic C--D stretch
\citep[see][]{Allamandola2021, Yang2023}.

Based on the {\it Integral Field Unit} (IFU) spectra
obtained with the {\it Near InfraRed Spectrograph}
(NIRSpec) and the {\it Mid Infrared Instrument} (MIRI)
aboard the {\it James Webb Space Telescope} (JWST),
here, we report a widespread detection of
the 3.4, 4.65, and 6.85$\mum$ emission features
in the 30 Doradus star-forming complex
in the Large Magellanic Cloud (LMC),
indicating that the aromatic molecules in 30 Dor
are deuterated and have an appreciable amount
of aliphatic content. \citet{Mori2012} had obtained
the 2.55--13.4$\mum$ slit spectra of the diffuse emission
toward nine positions in the LMC, with the IR camera
on board AKARI. Both the 3.3 and 3.4$\mum$ emission
bands were also clearly detected.

This paper is organized as follows.
We describe in \S\ref{sec:data}
the observations and data reduction.
In \S\ref{sec:analysis}, we analyze the observed spectra
and derive the power emitted in each of the major bands
at 3.3, 3.4, 4.65, 6.2, 6.85, 7.7, and 11.3$\mum$.
In \S\ref{sec:discussion}, we discuss
the aliphatic and deuterated spectral
signals and their implications.
Finally, the major results are summarized
in \S\ref{sec:summary}.
This paper is largely concerned with
the observational results.
In a subsequent paper, we will quantify
the aliphacity and deuteration
of the UIE carriers in the context of
the PAH hypothesis.

\section{Observations and Data Reduction} \label{sec:data}
The JWST spectra of 30 Doradus analyzed in this work
are drawn from the Early Release Observations (ERO)
Program No.\,2729, including the NIRSpec and MIRI
IFU spectroscopy. The NIRSpec IFU
($\simali$2.87--5.27$\mum$; $R$\,$\simali$2700)
and MIRI/MRS IFU ($\simali$4.90--27.9$\mum$;
$R$\,$\simali$3700 in Channel 1 to $R$\,$\simali$1800
in Channel 4) pointings cover the young star S7A
\citep{Walborn2013} and its surrounding dust cloud,
located at the northeast of the super star cluster (SSC) R136.

We download the uncalibrated NIRSpec and MIRI data
from the {\it Mikulski Archive for Space Telescopes} (MAST).
The data reduction is performed with
the JWST Science Calibration Pipeline (version 1.19.1),
using the \texttt{jwst\_1413.pmap} reference file.
The pipeline is run through all three stages
(\texttt{Detector1}, \texttt{Spec2}, and \texttt{Spec3}).
For the NIRSpec data, we enabled
the \texttt{clean\_flicker\_noise} step in Stage 1
to correct the $1/f$ flicker noise,
while for the MIRI data, we applied
the \texttt{residual\_fringe} step in Stage 2
to mitigate residual fringes.
All remaining pipeline steps are executed
with their default configurations.
The final pipeline products consist of
15 spectral cubes: three NIRSpec IFU cubes
separated by grism and twelve MIRI MRS cubes
separated by band.

We find that the NIRSpec IFU cubes
are spatially offset by 0$\farcs$38
from the HST F160W image obtained
from the {\it Hubble Tarantula Treasury Project}
(HTTP).\footnote{%
  \url{https://archive.stsci.edu/hlsp/http}
   }
Since the HST observation has been astrometrically
calibrated \cite[see][]{Sabbi2013, Sabbi2016},
and the NIRSpec and MIRI IFU cubes exhibit
a mutual offset of 0$\farcs$32, we therefore
align all cubes with respect to the HST F160W
astrometric reference frame by applying
World Coordinate System (WCS) corrections.

\begin{figure}[h!]
\centering
\includegraphics[width=0.9\textwidth]{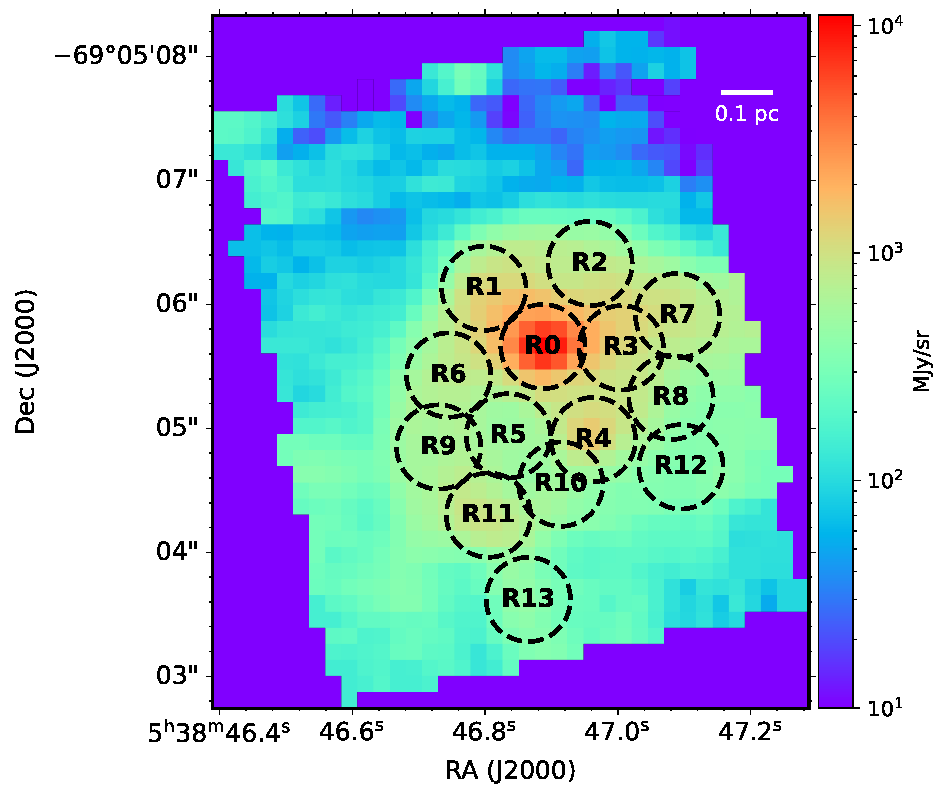}
\vspace{-0.5cm}
\caption{\label{fig:30Dor}
JWST's MIRI IFU slice at 6.2$\mum$
for the 30 Doradus star-forming complex
in the LMC. R0--R13 represent the apertures
adopted for spectral extraction.}
\vspace{-0.3cm}
\end{figure}

Based on the MIRI IFU slice at 6.2$\mum$,
as illustrated in Figure~\ref{fig:30Dor},
we define 14 regions
labelled R0, R1, R2, ..., and R13,
with R0 centering on the star S7A.
%
Each circular aperture has a radius of
0$\farcs$34, corresponding to twice
the pixel size of channel 2.\footnote{%
We use this channel as reference
because it covers the longest wavelength
out to the UIE feature of interest at 11.3$\mum$.
}
An additional one-dimensional residual fringe
correction (\texttt{fit\_residual\_fringes\_1d})
is applied to all the MIRI spectra to mitigate
the effect of the fringes on the spectra.
We find that small discontinuities exist
out to the UIE feature of interest at 11.3$\mum$.
wavelength overlaps between the adjacent segments.
Following \citet{Zhang2025}, we multiplicatively
scale and stitch the spectra for the cubes
from NIRSpec G235H to MIRI Channel 3 Long,
using the MIRI Channel 2 Short segment as the reference.
The average scaling factors for NIRSpec and MIRI
are 0.88 and 0.98, respectively, which are consistent
with those reported in the literature
\citep{Misselt2025, Chown2025}.
For all the regions, the spectra are extracted,
and prominent UIE emission bands are detected.
We note here that since $R0$ coincides with
the star S7A, its UIE features are severely
overwhelmed by the very strong continuum.
We therefore exclude this region from further analysis.
%


\section{Spectral Analysis} \label{sec:analysis}
\begin{figure}
\centering
\includegraphics[width=0.49\textwidth]{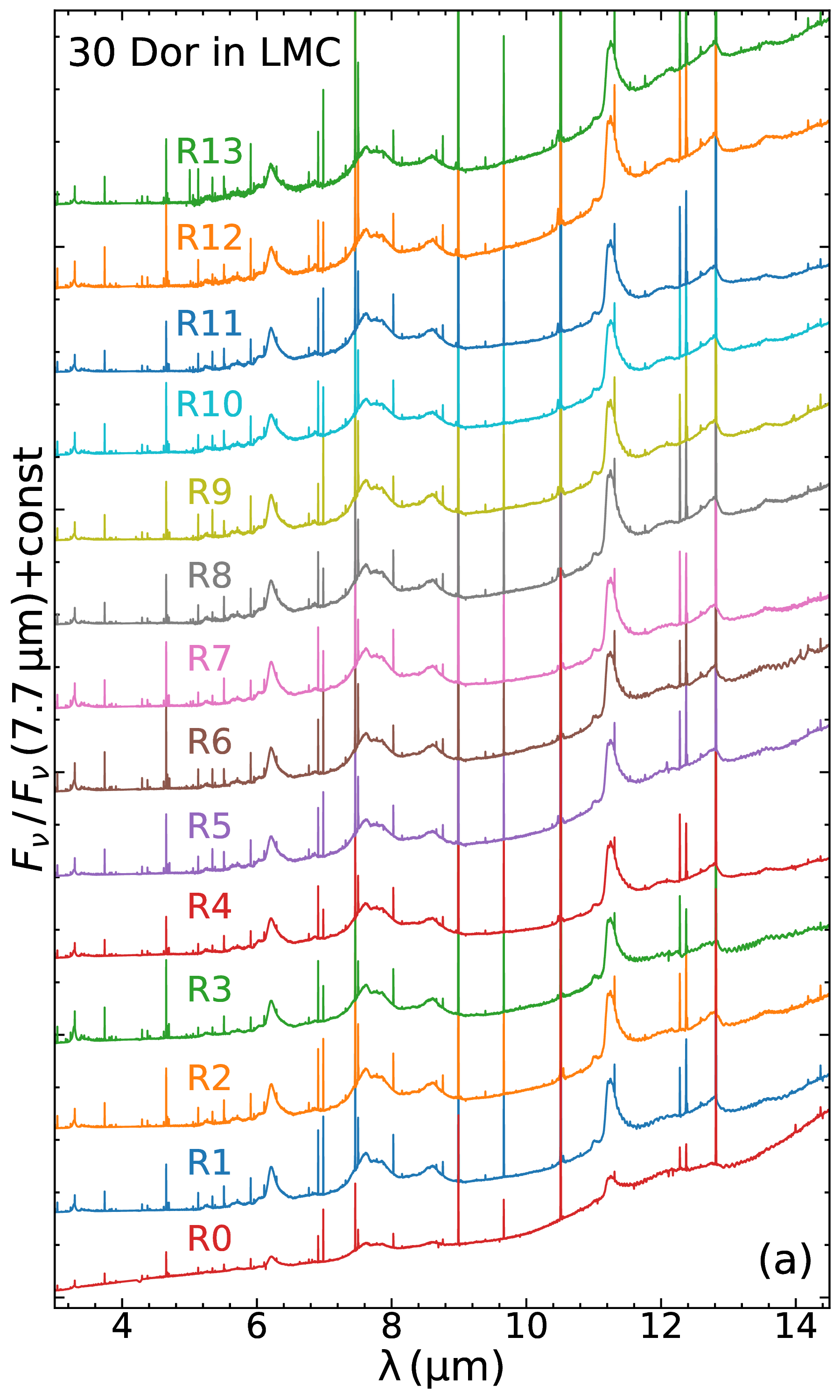}
\includegraphics[width=0.49\textwidth]{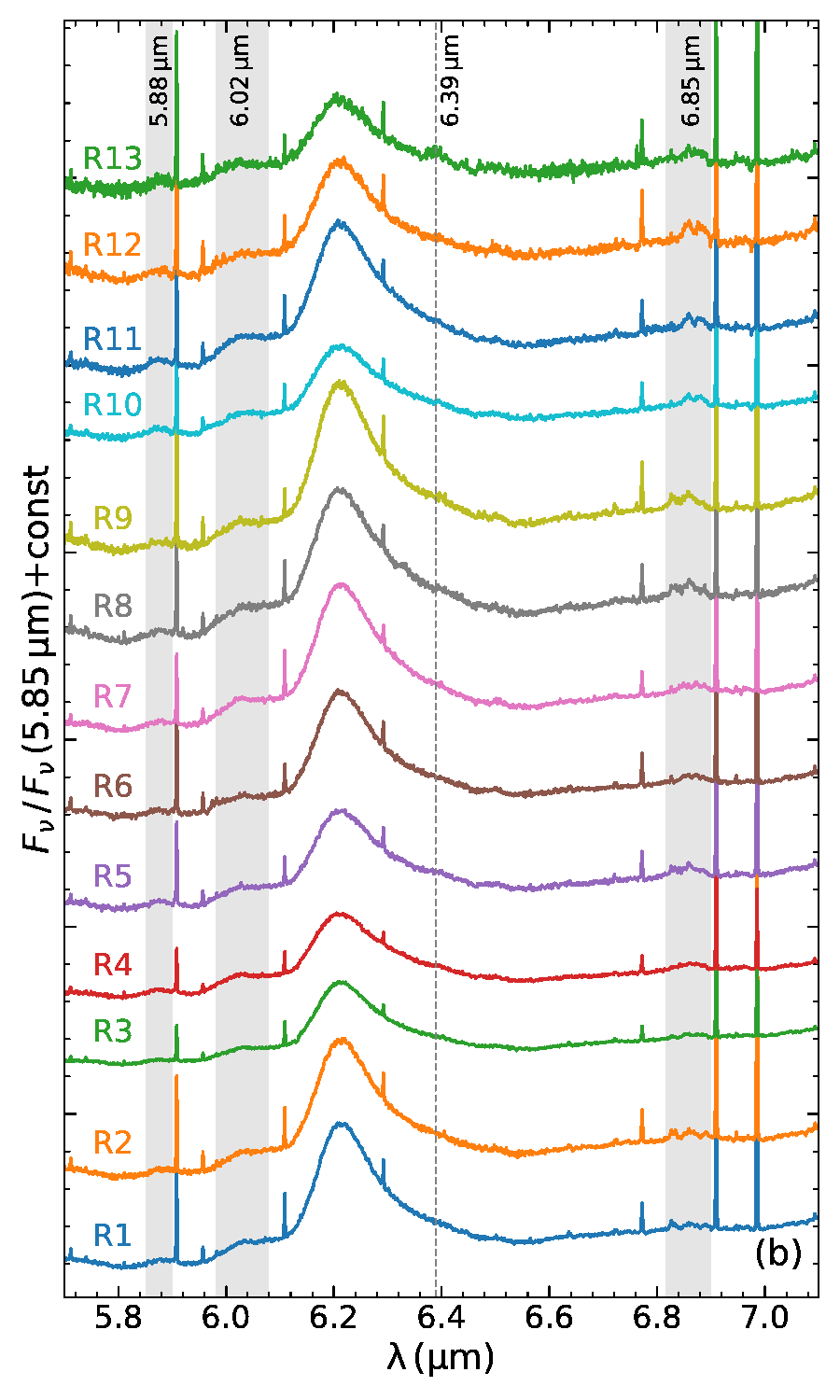}
\vspace{-0.5cm}
\caption{\label{fig:specplot}
Left panel (a):
The 3.0--14.5$\mum$ spectra of Regions 0--13
(from bottom to top), normalized at 7.7$\mum$.
The spectra are shifted vertically for clarity.
Right panel: Zoomed-in view
in the 5.7--7.1$\mum$ wavelength range
for Regions 1--13 (from bottom to top),
normalized at 5.85$\mum$.
The spectra have been vertically shifted for clarity.
Shaded belts highlight the weak 5.88 and 6.02$\mum$ features,
as well as the 6.85$\mum$ complex.
The vertical dashed line indicates
the tentative, weak subfeature at 6.39$\mum$.
}
\vspace{-0.3cm}
\end{figure}

We show in Figure~\ref{fig:specplot}a the JWST/NIRSpec
and MIRI spectra extracted for all the selected 14 regions.
The major UIE emission bands are prominent
in all these regions, along with many gaseous
emission lines, including the recombination lines
of atomic hydrogen (HI).
In addition, the aliphatic C--H features at 3.4 and
6.85$\mum$, as well as the 4.65$\mum$ aliphatic
C--D stretch, are also clearly detected.
We employ the PAHFIT software \citep{Smith2007}
to decompose the observed spectra to determine
the power emitted in each band in each region.
%
The Pf$\delta$ (9$\rightarrow$5)
hydrogen recombination line at 3.297$\mum$
coincides with the 3.3$\mum$ C--H band,
and the Pf$\beta$ (7$\rightarrow$5)
hydrogen line at 4.653$\mum$
coincides with the 4.65$\mum$ C--D band
\citep{Geballe1985, Hora2026}.
They are included in the PAHFIT software,
and their contributions to the power emitted
in the 3.3 and 4.65$\mum$ bands,
respectively, are subtracted.

Building upon the framework of \citet{Smith2007}
and \citet{Lai2020}, we refine the 3.4$\mum$ complex
in our fitting scheme by including three weak satellite
bands at 3.46, 3.52, and 3.56$\mum$.
As shown in Figure~\ref{fig:specplot}b,
the 30 Dor spectra exhibit two weak features at 5.88
and 6.02$\mum$, which are also detected
in the Orion Bar \citep{Chown2024}.
Therefore, these two features
are also included in the fitting scheme.
In addition, the pronounced red wing of
the 6.2$\mum$ band requires two separate profiles.
A tentative component appears to be present
at 6.39$\mum$ in the spectrum of R13,
but due to its low signal-to-noise ratio (SNR),
it is not included in the fitting scheme.

The 6.85$\mum$ aliphatic C--H deformation
band is treated as a blended complex,
consisting of three individual components
at $\simali$6.83, 6.86, and 6.88$\mum$.
For the 11.3$\mum$ UIE complex,
we follow the configuration of \citet{Donnan2023}
and fit it with five components.
We also identify a weak feature at 11.02$\mum$,
which we model as a separate component.
Our modifications to the spectral features
considered in PAHFIT are summarized
in Table~\ref{tab:modifications}.
In addition, we further supplement
the fitting using the JWST line list templates
adopted in the CAFE software
\citep{2025ascl.soft01001D}.\footnote{%
  \url{https://github.com/GOALS-survey/CAFE}
}

\begin{figure*}[h!]
\centering
\includegraphics[width=0.55\textwidth]{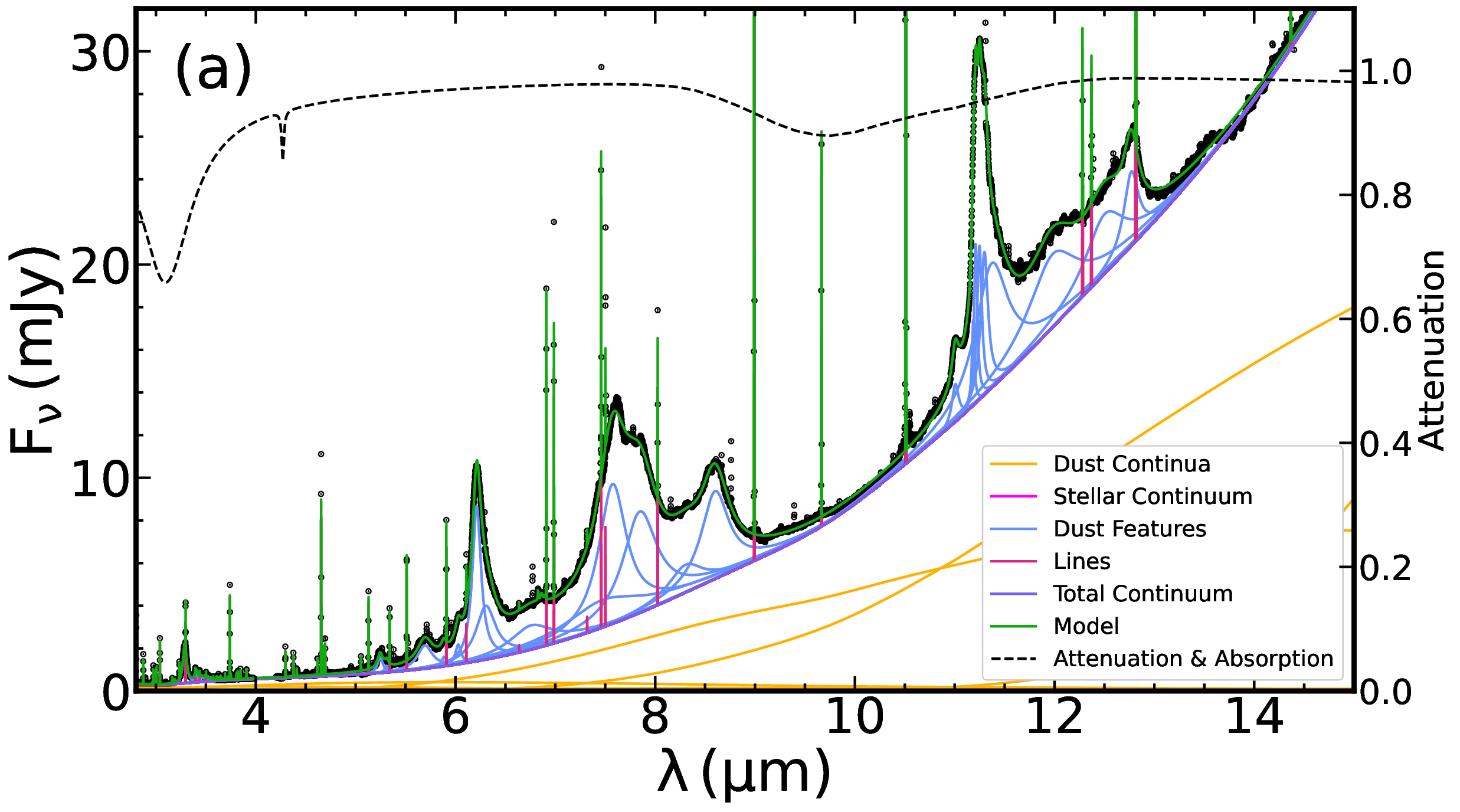}
\includegraphics[width=0.41\textwidth]{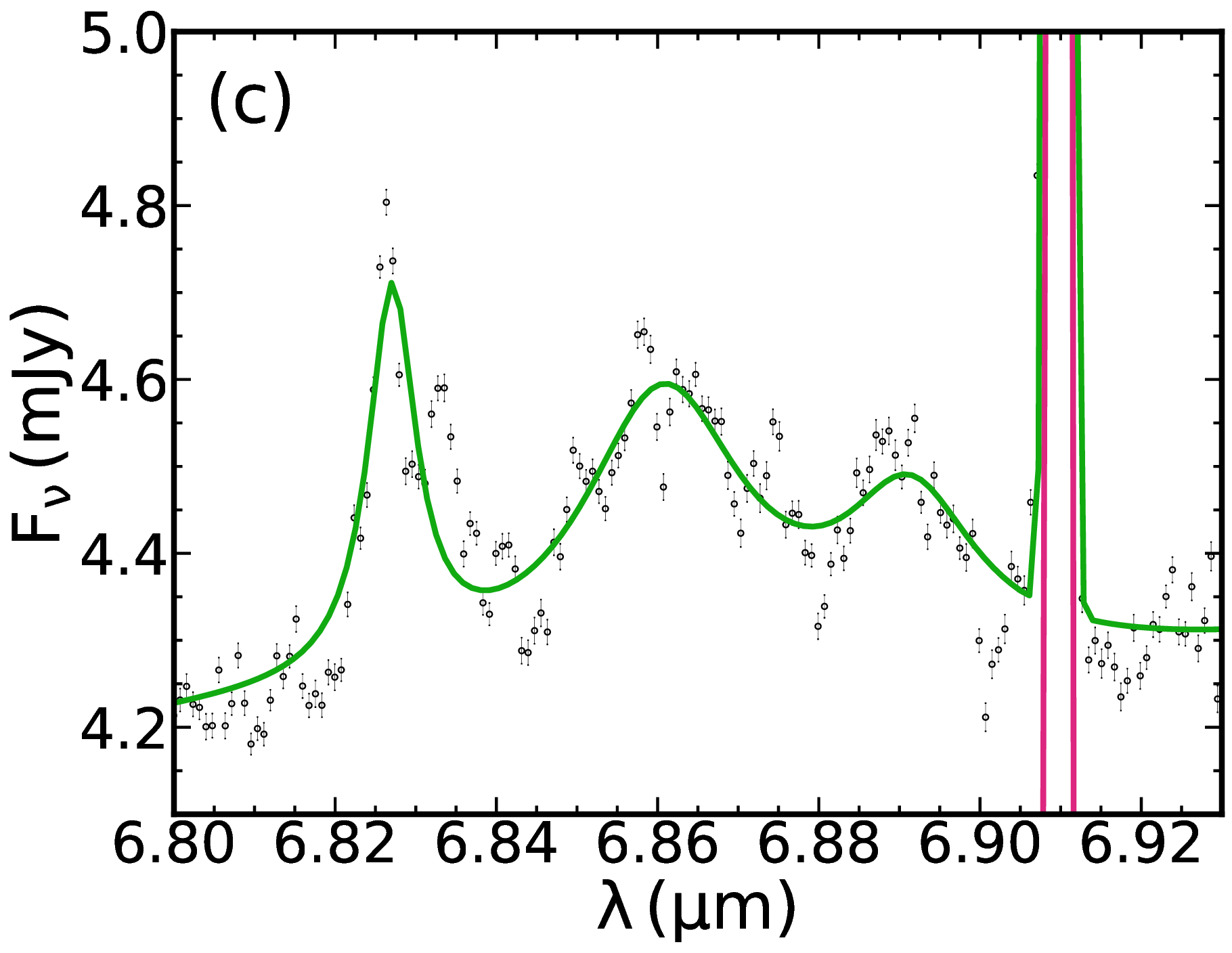}
\includegraphics[width=0.55\textwidth]{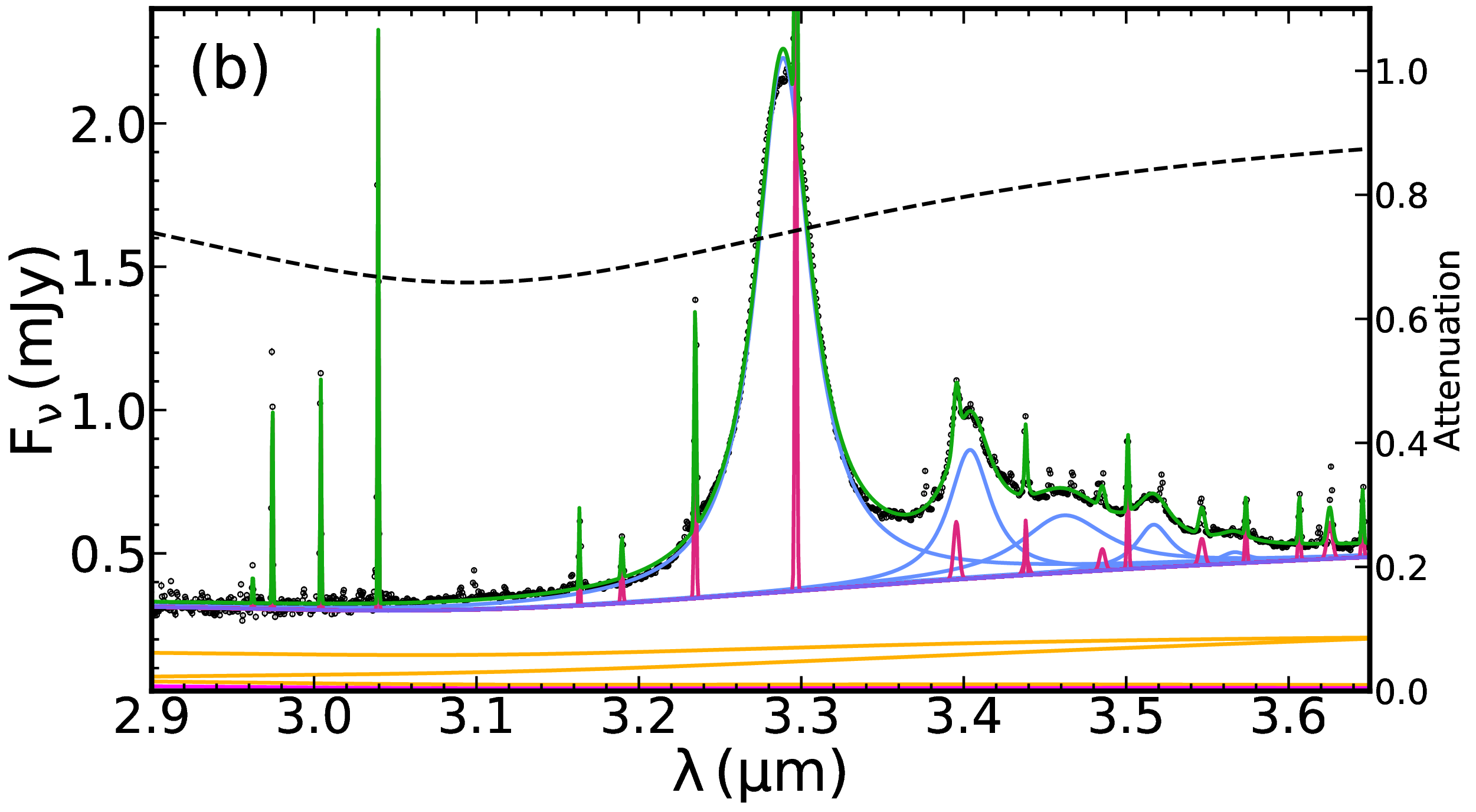}
\includegraphics[width=0.41\textwidth]{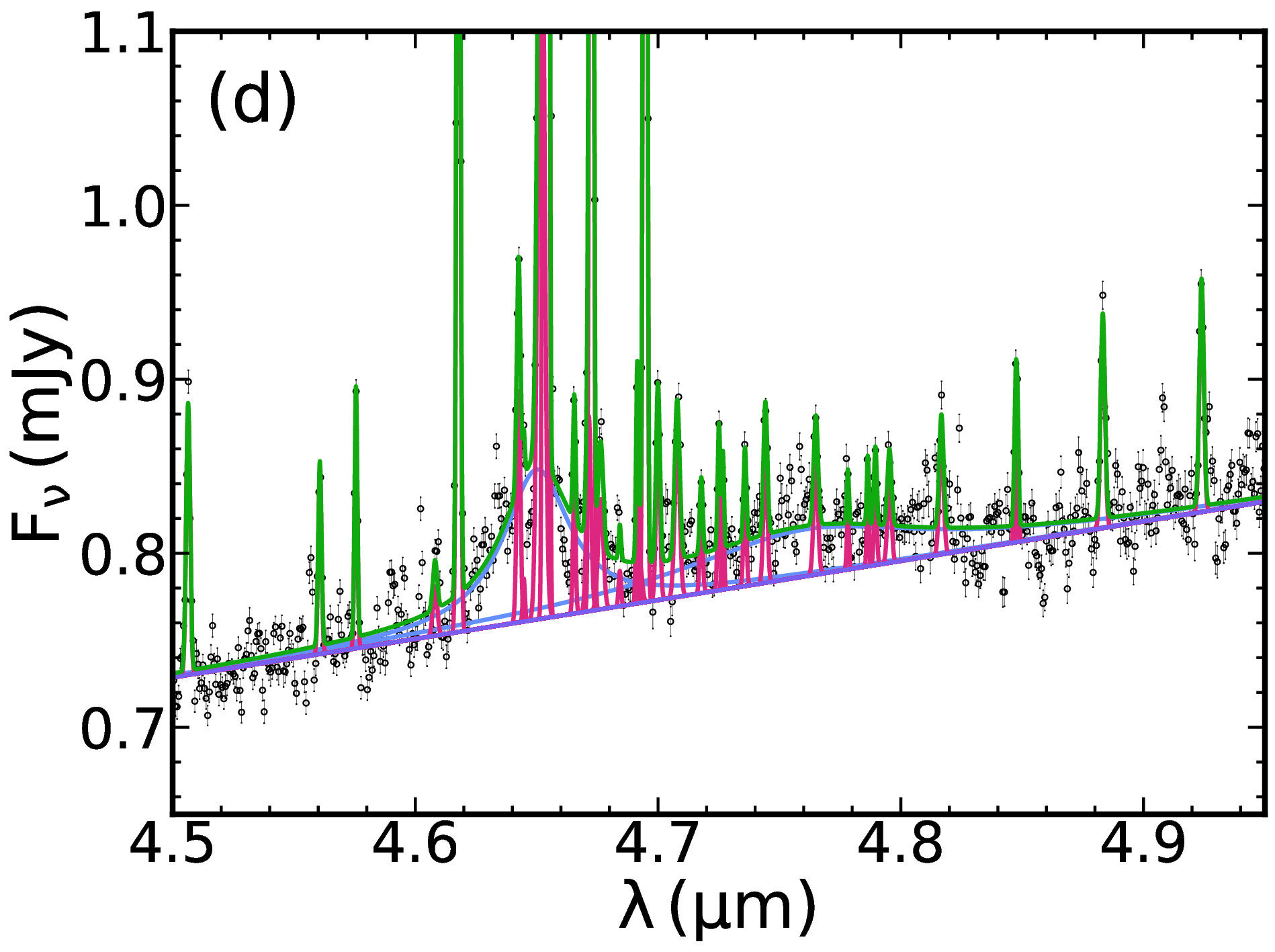}
\caption{\label{fig:specfit_plot}
(a) The JWST/NIRSpec and MIRI spectrum of Region 1
is decomposed with PAHFIT.
(b) Zoomed-in view of the aromatic C--H stretches
around 3.3$\mum$ and the aliphatic C--H stretches
around 3.4$\mum$.
(c) Zoomed-in view of the aliphatic C--H
deformation around 6.85$\mum$.
For the first time, the 6.85$\mum$ complex
is resolved into three sub-features
at $\simali$6.83, 6.86, and 6.88$\mum$.
(d) Zoomed-in view of the aliphatic C--D
stretches around 4.65$\mum$.
}
\vspace{-0.1cm}
\end{figure*}

For illustration, we show in
Figure~\ref{fig:specfit_plot}
the PAHFIT results for the R1 region,
including the overall UIE bands
(Figure~\ref{fig:specfit_plot}a)
as well as the aliphatic C--H stretches
at 3.40, 3.46, 3.52, and 3.56$\mum$
(Figure~\ref{fig:specfit_plot}b)
and the aliphatic C--H deformation bands
at 6.83, 6.86, and 6.88$\mum$
(Figure~\ref{fig:specfit_plot}c).
%
%

Given the exceptionally weak nature of
the 4.65$\mum$ aliphatic C--D stretching
feature, we do not include it in the global PAHFIT
fitting scheme. Instead, we perform a local fit
in which the continuum is modeled as a combination
of a power law, emission lines with Gaussian profiles
and UIE features with Drude profiles.
The 4.65$\mum$ band and sometimes an even weaker band
at 4.75$\mum$ are modeled as two independent Drude profiles
for the aliphatic C--D stretches.
%
Again, for illustration, we show
in Figure~\ref{fig:specfit_plot}d
for the fit to the 4.65 and 4.75$\mum$ components.

\begin{figure}[h!]
\centering
\includegraphics[width=0.48\textwidth]{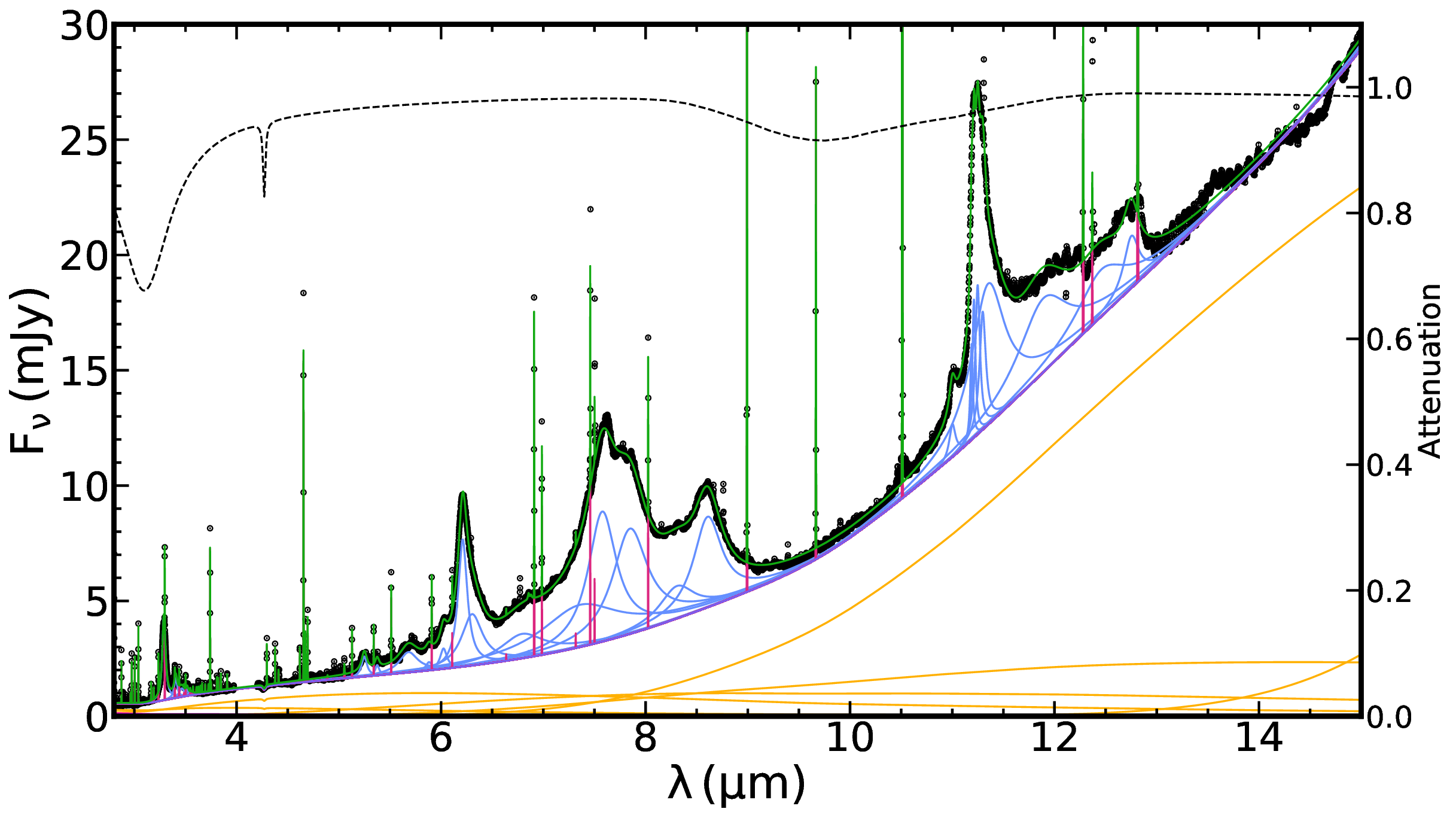}
\includegraphics[width=0.48\textwidth]{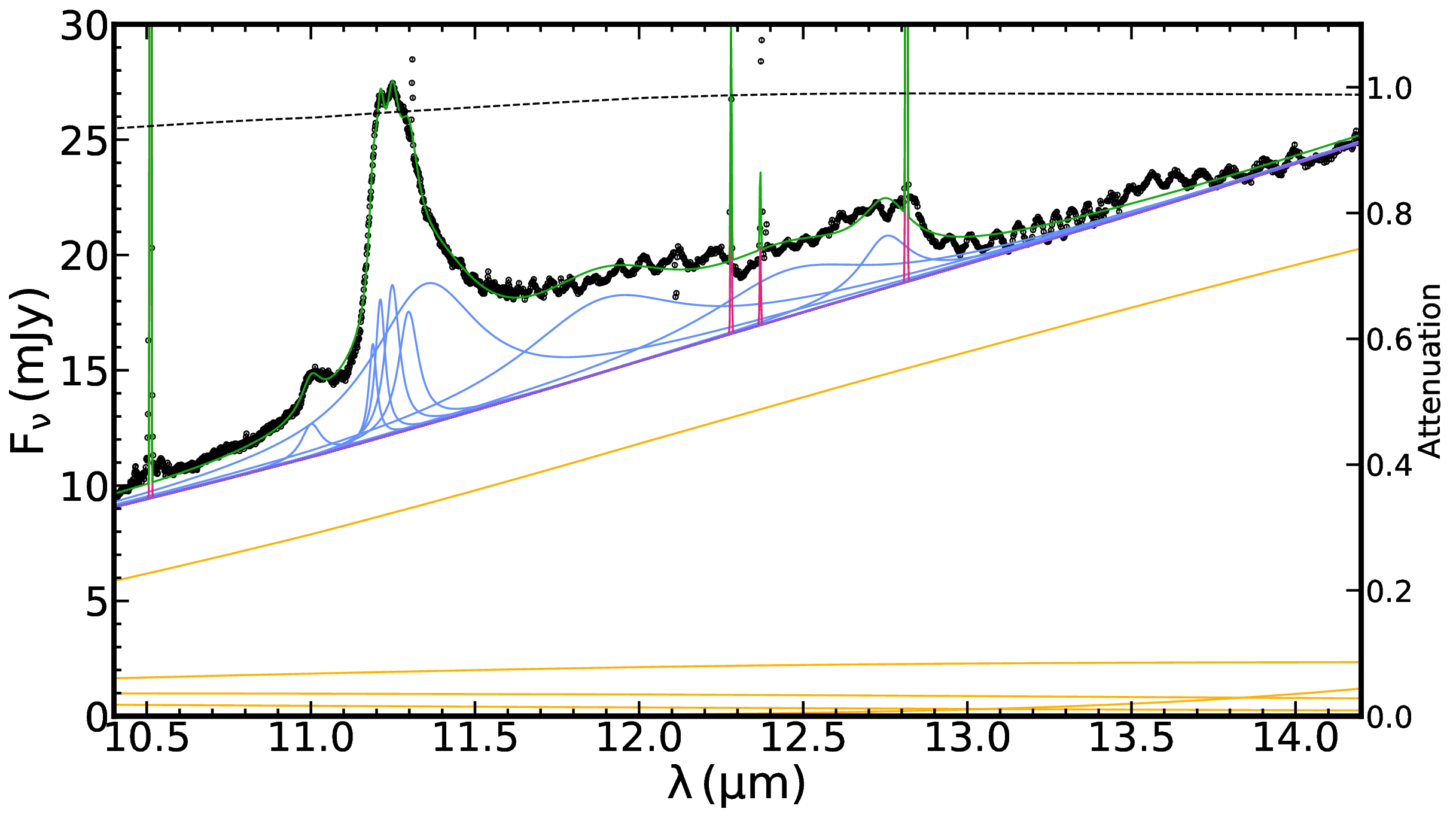}
\caption{\label{fig:R3}
PAHFIT analysis of Region R3.
The left panel is for the full-spectrum,
while the right panel provides a zoomed-in
view around the 11.3$\mum$ feature.
}
\end{figure}

The PAHFIT spectral decomposition for
the R3 region seems unsatisfactory,
as illustrated in Figure~\ref{fig:R3}.
At wavelengths longward of $\simali$11.5$\mum$,
the spectrum exhibits pronounced fringing.
This leads to a suboptimal fit,
making it difficult to decompose
the 12.7$\mum$ band. This would
compress the 11.3$\mum$ feature
and lead to an underestimation of
its intensity $I_{11.3}$.
Furthermore, the fringing precludes
robust constraints on the continuum
at the longest wavelengths.
Since PAHFIT employs a global continuum,
these inaccuracies can bias the inferred
intensities of other features as well.
For these reasons, we will not exclude
R3 for correlation analysis.

Finally, the fitting results for all these regions
are summarized in Table~\ref{tab:fitresult}.
The intensities for the major UIE bands,
as well as the aliphatic features at 3.4, 4.65,
and 6.85$\mum$ are listed.

\section{Discussion: Aliphacity and Deuteration}\label{sec:discussion}

\subsection{Aliphatic C--H Stretches at 3.4$\mum$}\label{sec:3.4um}
\begin{figure*}[h!]
\centering
\includegraphics[width=0.32\textwidth]{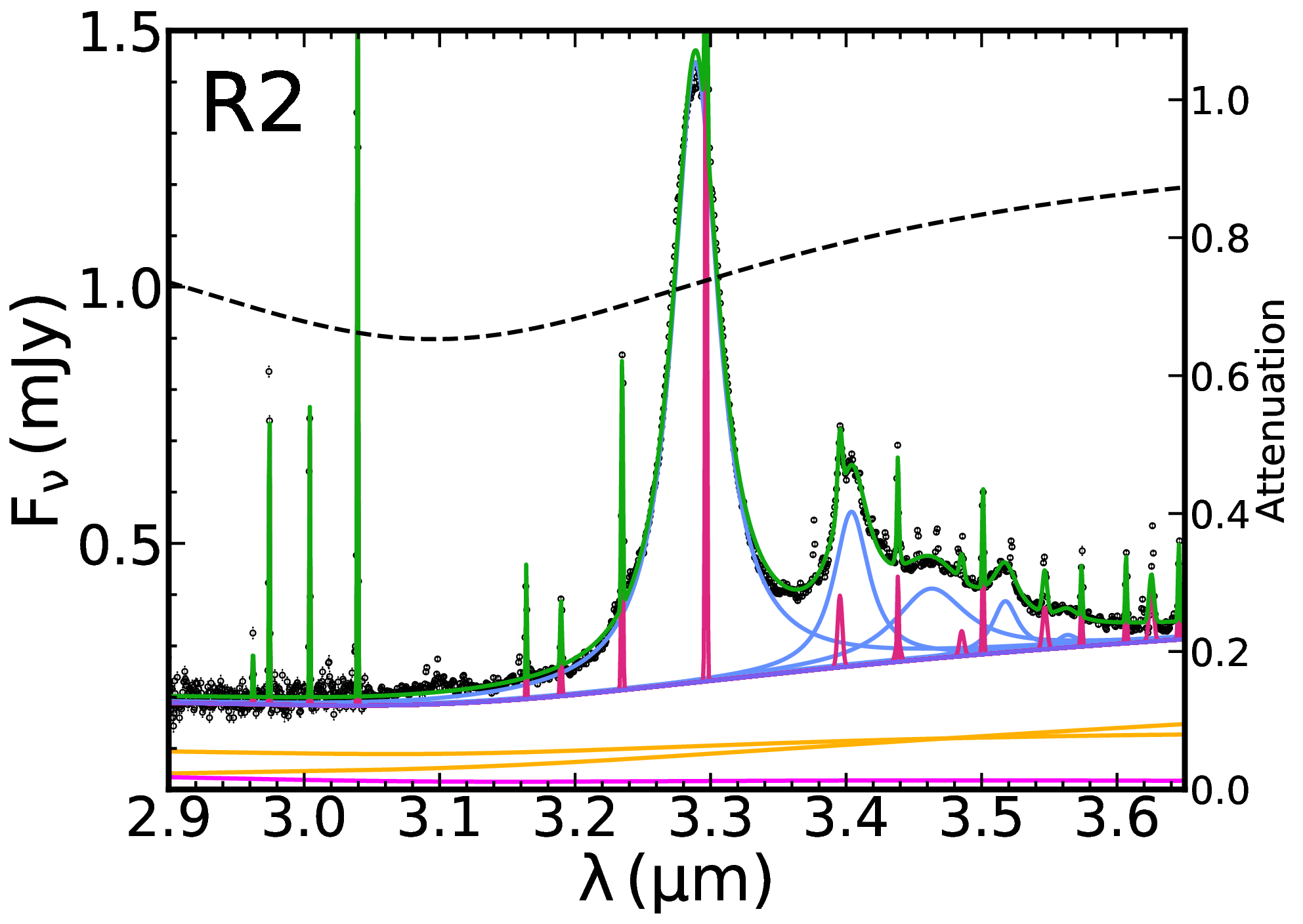}
\includegraphics[width=0.32\textwidth]{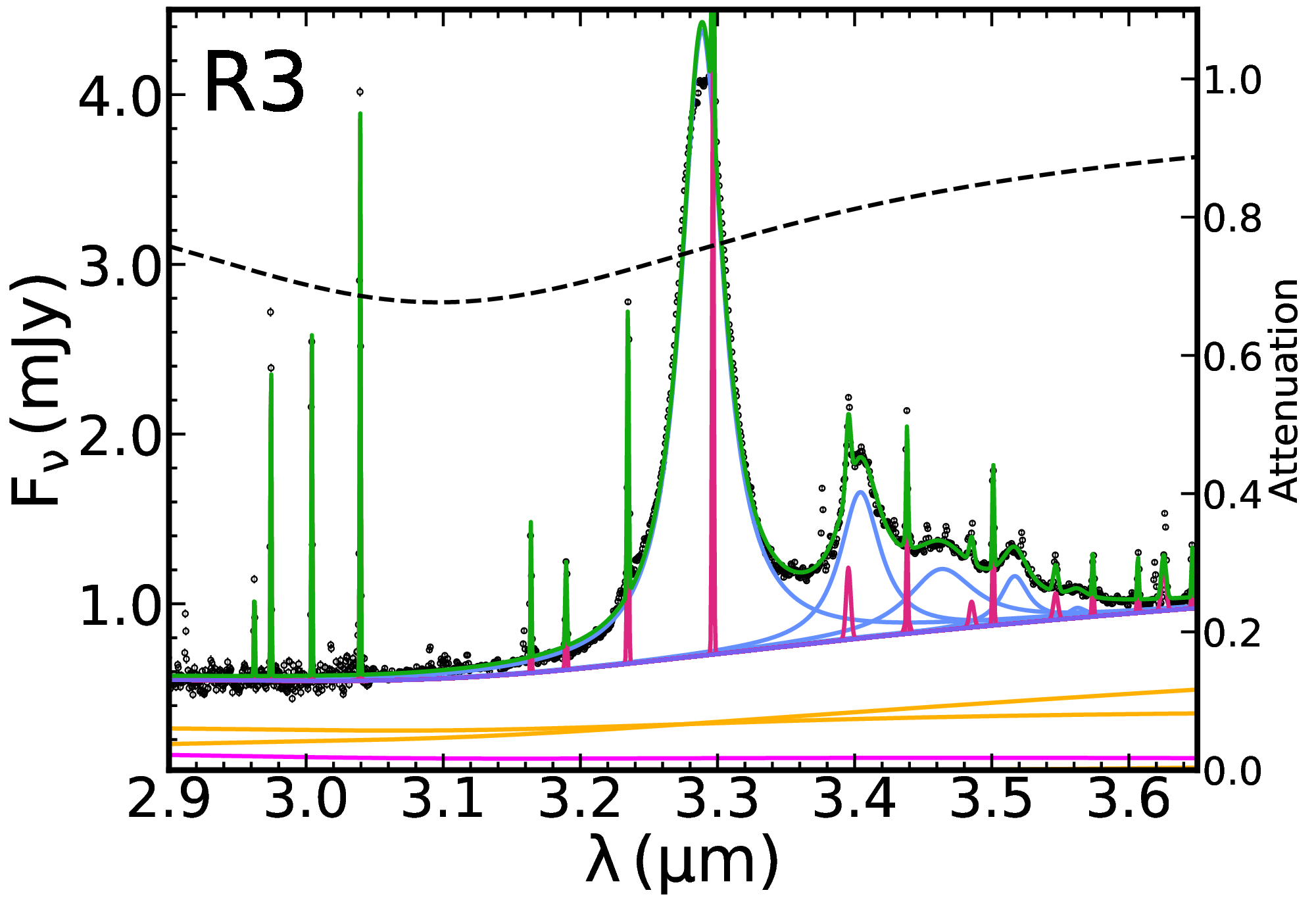}
\includegraphics[width=0.32\textwidth]{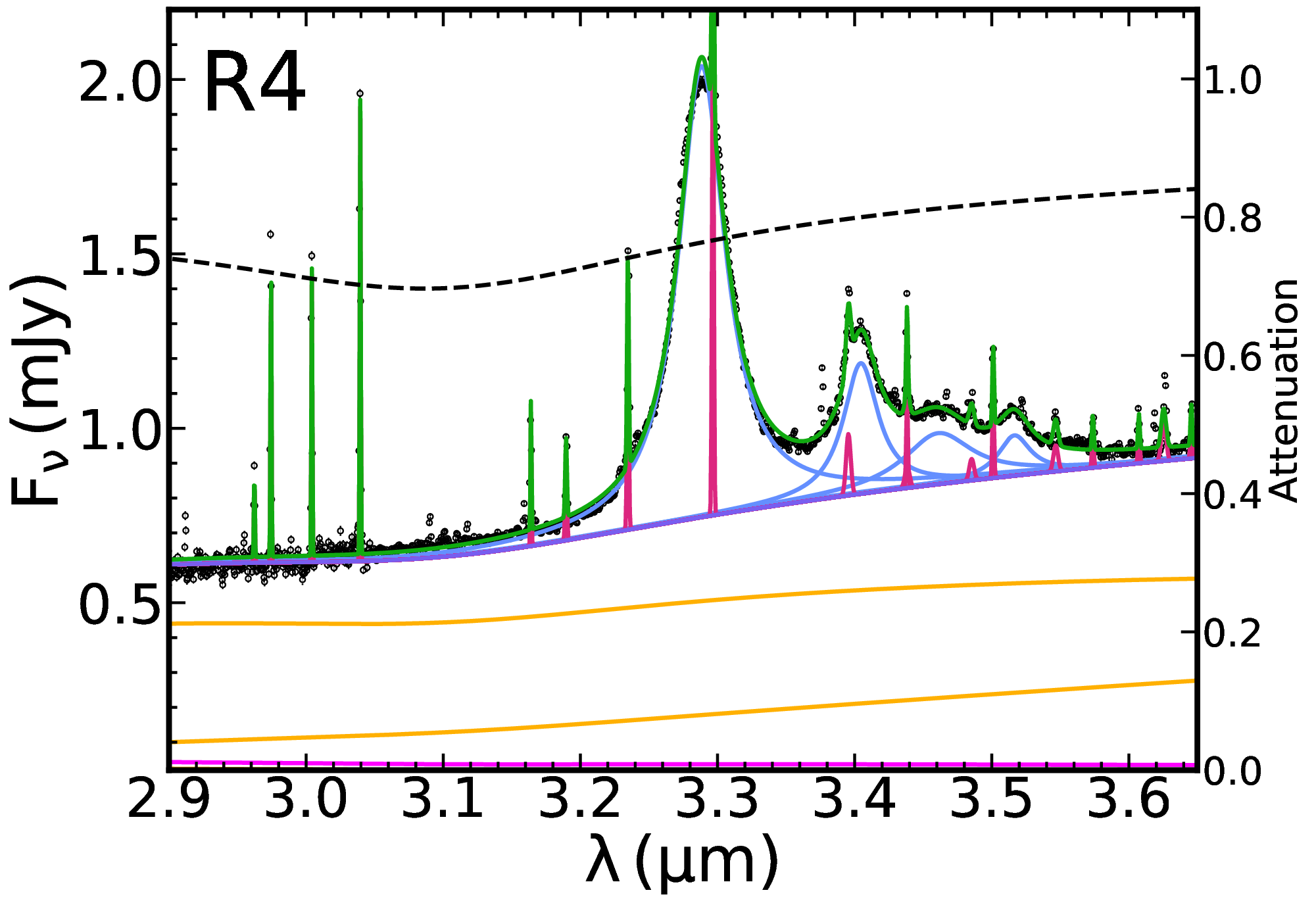}
\includegraphics[width=0.32\textwidth]{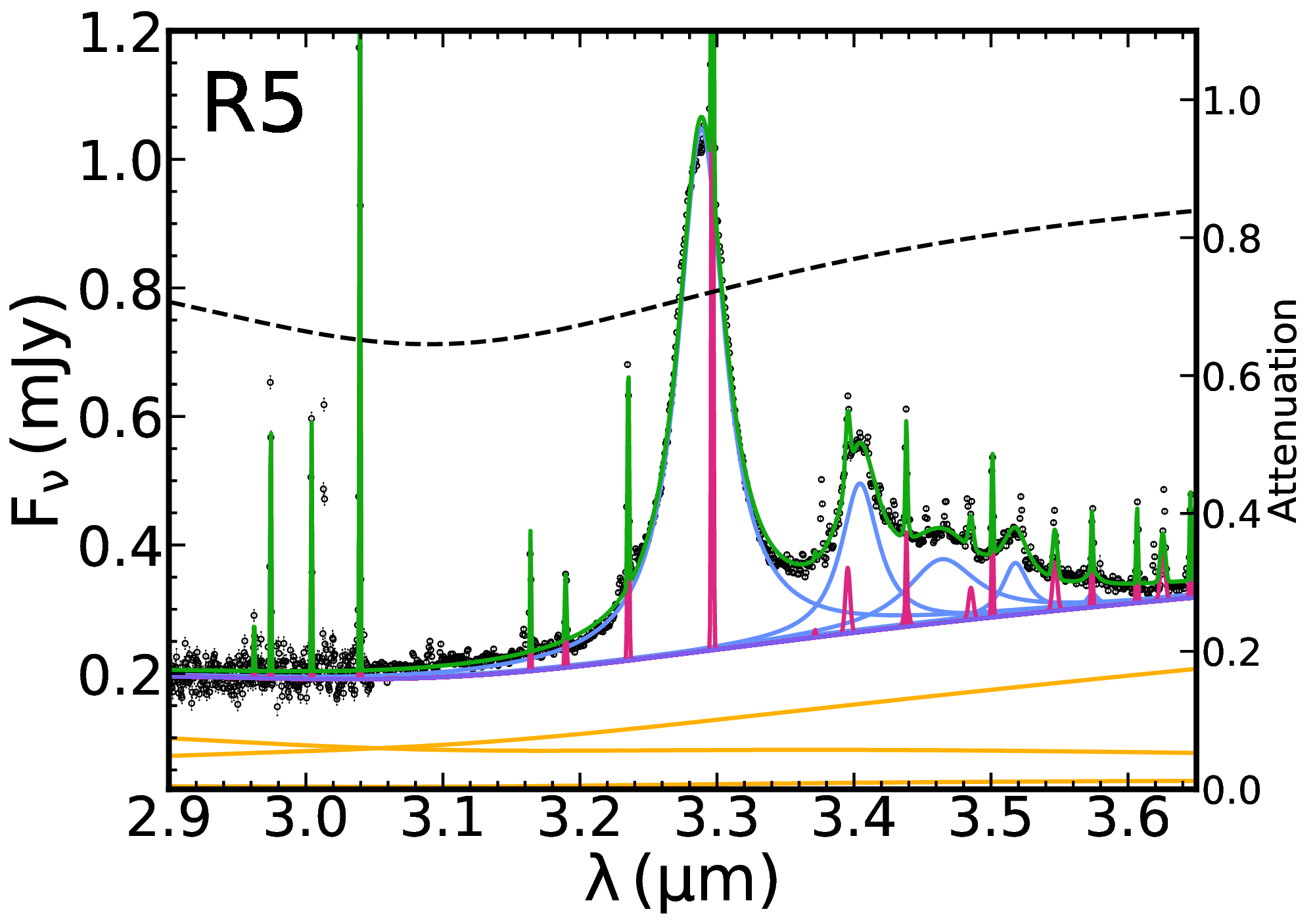}
\includegraphics[width=0.32\textwidth]{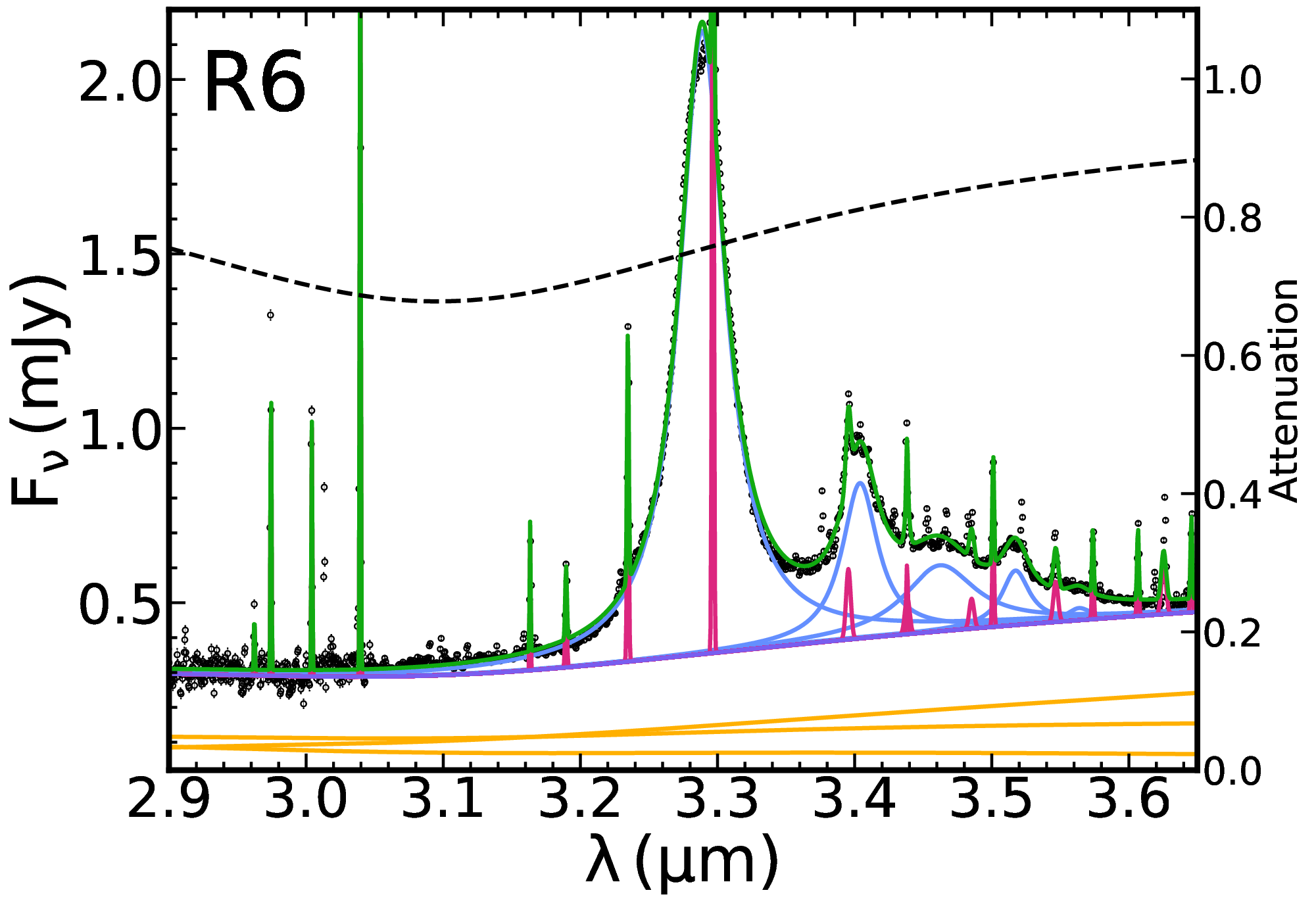}
\includegraphics[width=0.32\textwidth]{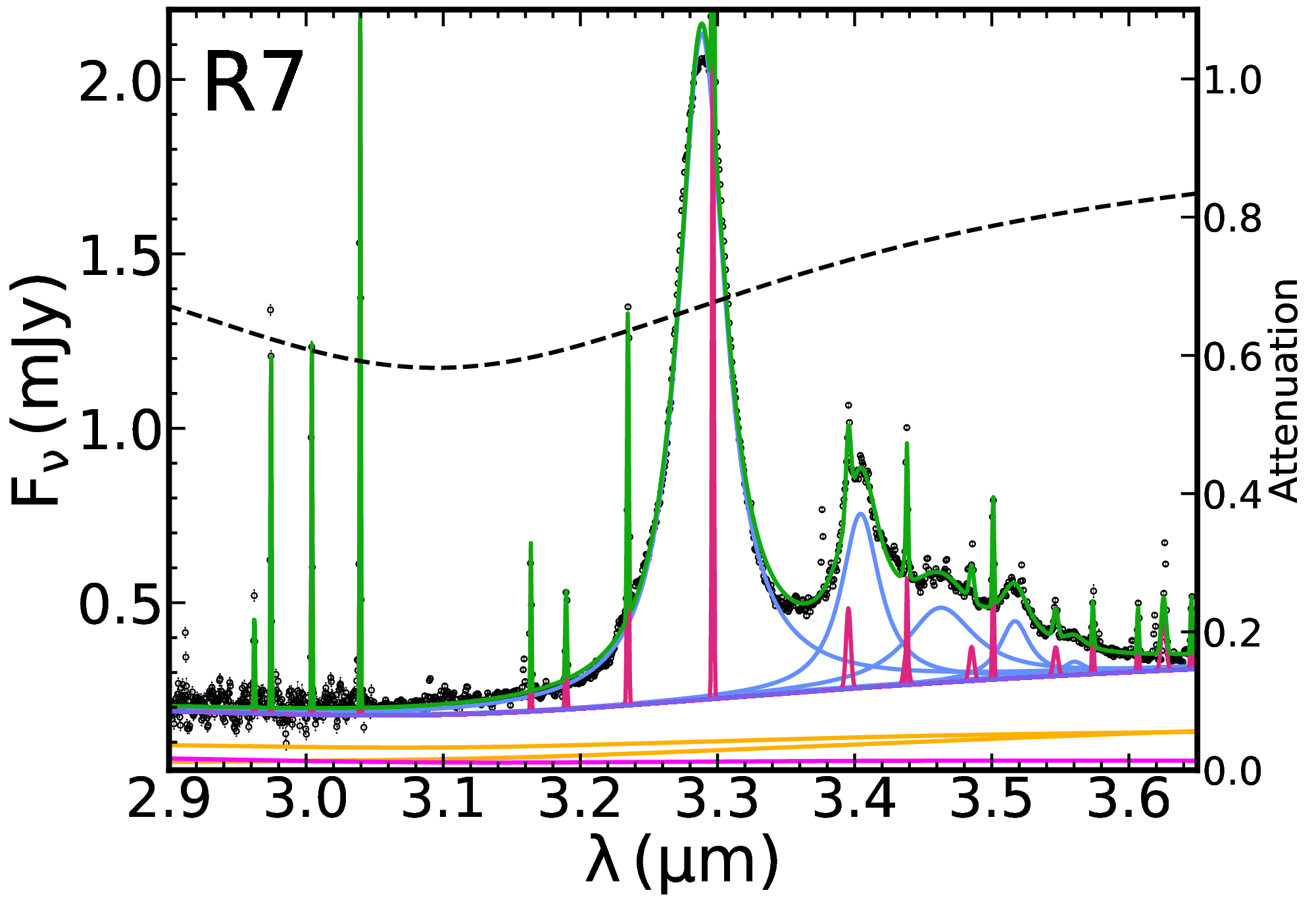}
\includegraphics[width=0.32\textwidth]{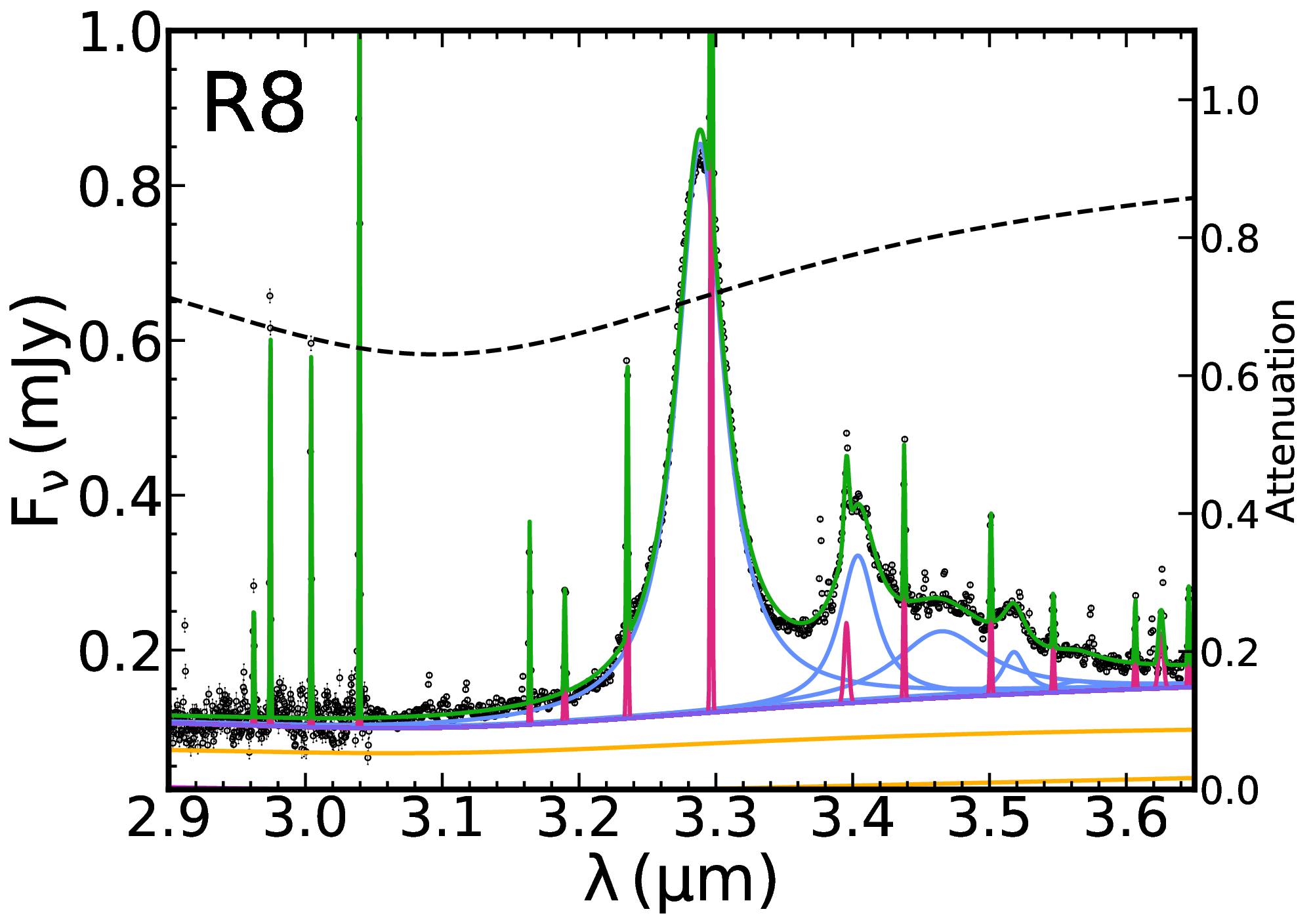}
\includegraphics[width=0.32\textwidth]{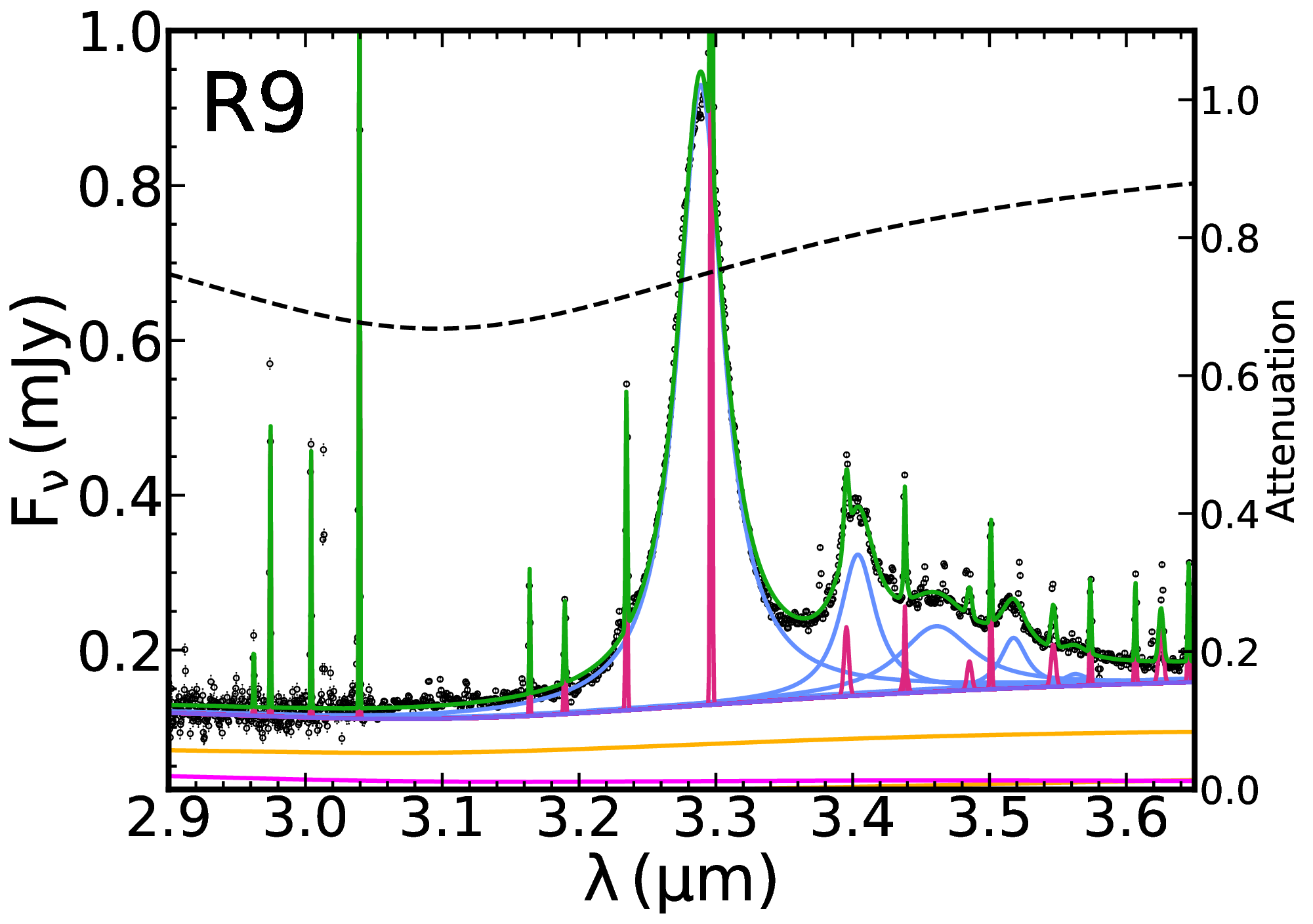}
\includegraphics[width=0.32\textwidth]{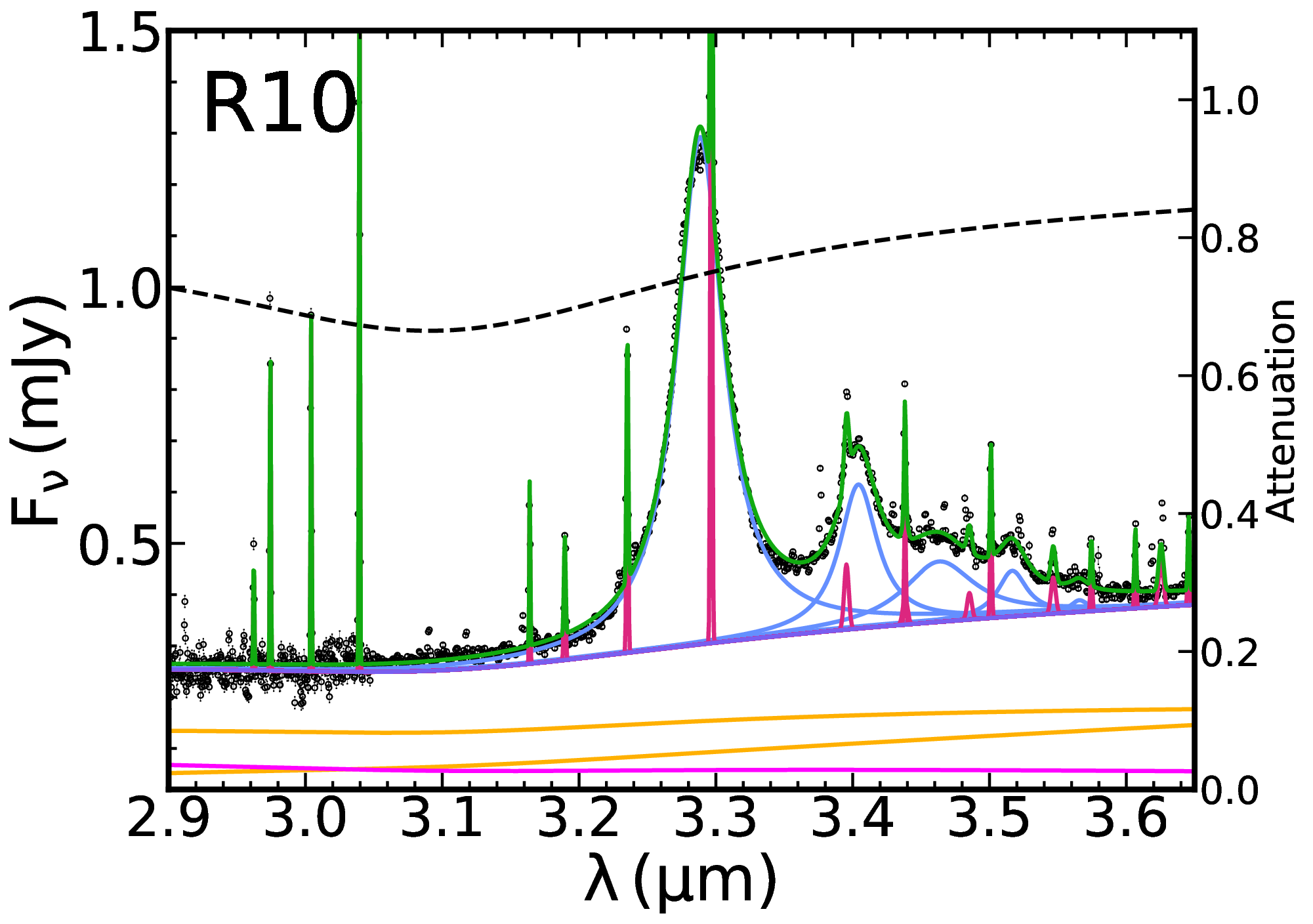}
\includegraphics[width=0.32\textwidth]{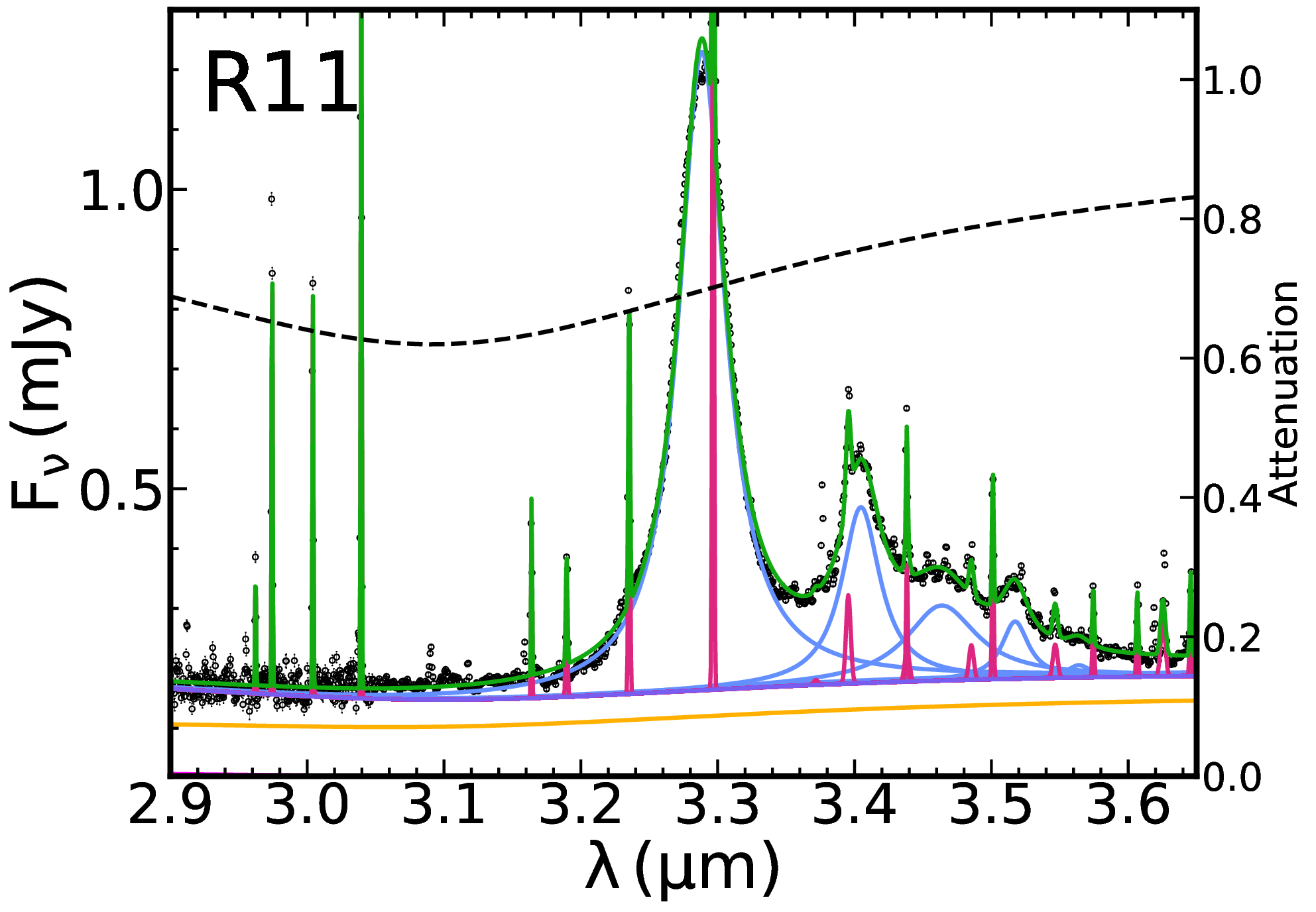}
\includegraphics[width=0.32\textwidth]{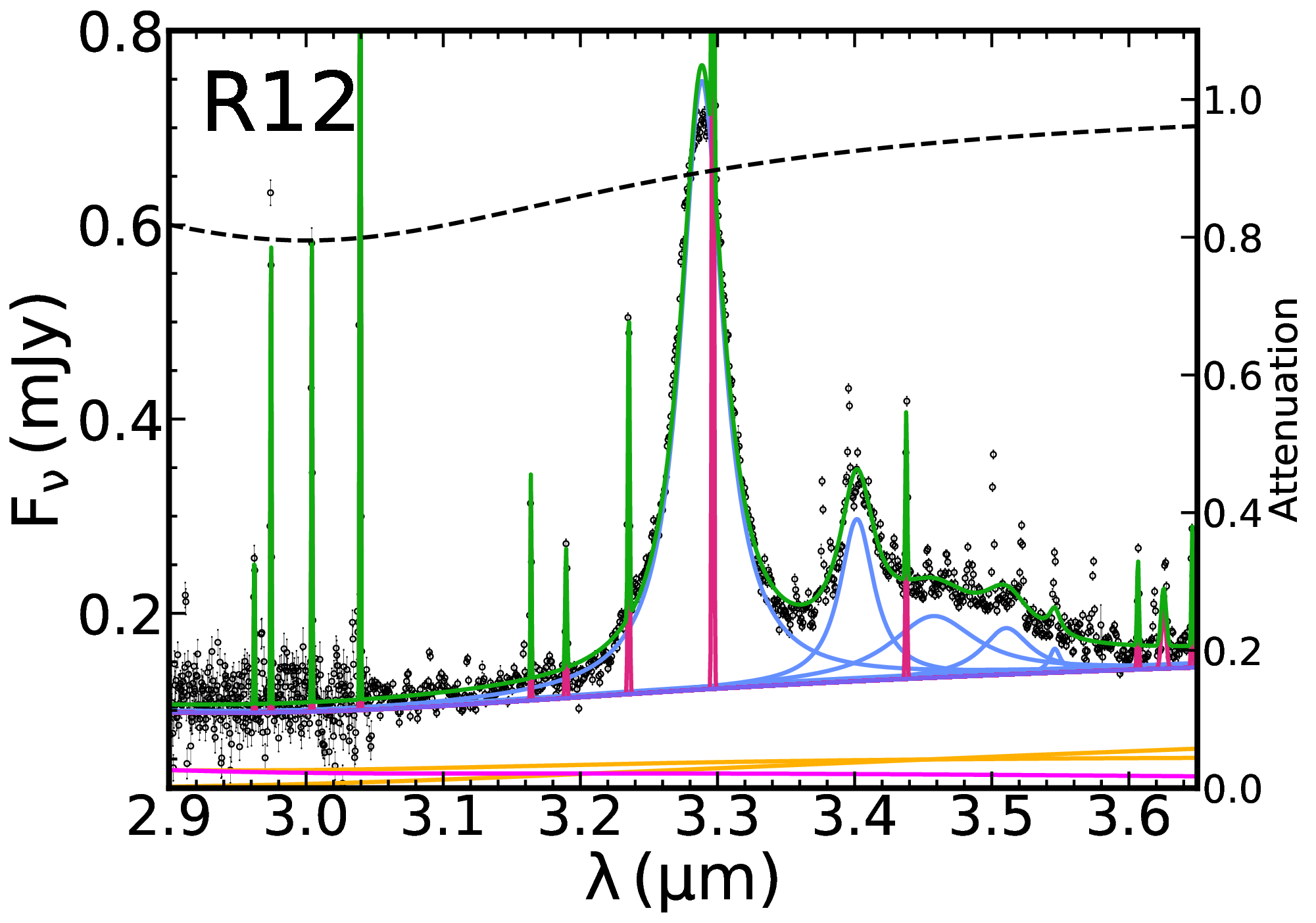}
\includegraphics[width=0.32\textwidth]{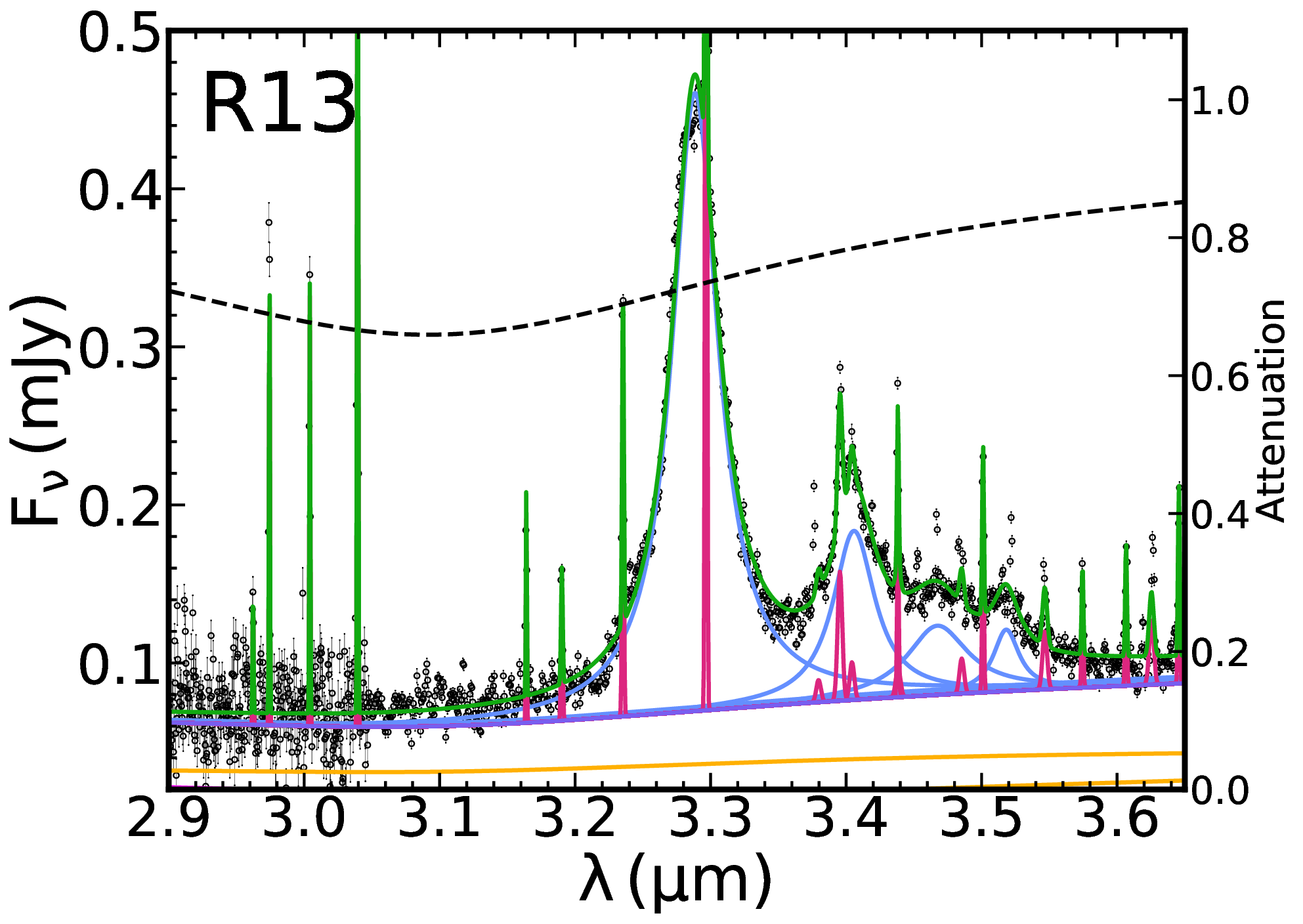}
\vspace{-0.2cm}
\caption{\label{fig:33fit}
Zoomed-in view of the PAHFIT
decomposition around 3.3$\mum$
for the R2--R13 regions.
}
\vspace{-0.1cm}
\end{figure*}

Figure~\ref{fig:33fit} shows the PAHFIT analysis
of the 3.3$\mum$ aromatic C--H stretch
as well as the aliphatic C--H stretches
at $\simali$3.40, 3.46, 3.52, and 3.56$\mum$
for all the selected regions in 30 Dor, except R1.
The spectra of all these regions resemble
each other and display a pronounced
3.3$\mum$ emission band, accompanied
by a series of weaker bands at 3.40, 3.46, 3.52,
and 3.56$\mum$, which show a tendency to
decrease in strength with increasing wavelength.
Similar spectra have been reported in previous
studies, both in the pre-JWST era
\citep[e.g., see][]{Nagata1988, Geballe1992, Joblin1996,
Sloan1997, Kondo2012, 
Yamagishi2012, Inami2018, Lai2020} and in the JWST era
\citep[e.g., see][]{Boersma2023, Kondo2024,
Schroetter2024, Peeters2024, Lyu2025, Misselt2025,
Mckinney2026, Mentzer2026}.
In Table~\ref{tab:fit_parameters}, we present
the best-fit parameters for the individual subfeatures
of the 3.4$\mum$ complex, including the peak
wavelengths, widths and power emitted.
While the peak wavelengths of these subfeatures
are consistent with that of \citet{Mori2012},
their widths are somewhat different.
For instance, the 3.41$\mum$ band derived here
($\gamma_{3.41}\approx0.033\mum$)
is considerably narrower than that of
Mori et al.\ (2012; $\approx0.060\mum$).
The 3.51 and 3.56$\mum$ subfeatures
($\gamma_{3.51}\approx0.025\mum$,
$\gamma_{3.56}\approx0.020\mum$)
are also somewhat narrower than that of
Mori et al.\ (2012;
$\gamma_{3.51}\approx0.034\mum$,
$\gamma_{3.56}\approx0.034\mum$).
On the opposite, the 3.46$\mum$ subfeature
($\approx0.063\mum$)
is substantially broader than that
of Mori et al.\ (2012; $\approx0.034\mum$).
Nevertheless, the width of the 3.3$\mum$
band ($\gamma_{3.3}\approx0.041\mum$)
is closely similar to that of
Mori et al.\ (2012; $\approx0.042\mum$).
We note that the wavelengths, widths
and relative strengths of these subfeatures
are not invariant among different sources
\citep[e.g., see][]{Nagata1988, Geballe1994,
Beintema1996, Tokunaga1997}.

\begin{figure}[h!]
\centering
\includegraphics[width=0.9\textwidth]{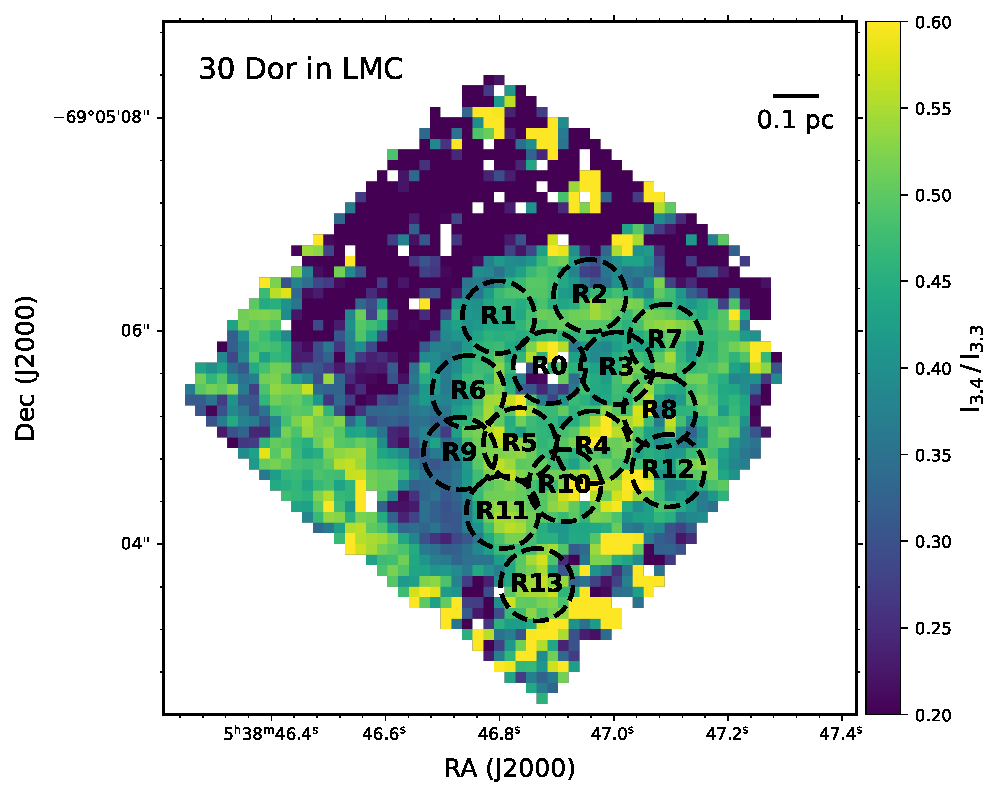}
\vspace{-0.3cm}
\caption{\label{fig:map1}
Pixel-by-pixel map of
the aliphatic to aromatic C--H band ratio
($I_{3.4}/I_{3.3}$), with the apertures used
for regional spectral extraction overlaid.
}
\vspace{-0.3cm}
\end{figure}

We show in Figure~\ref{fig:map1} the spatial
variation of $I_{3.4}/I_{3.3}$, the intensity ratio
of the 3.4$\mum$ feature to the 3.3$\mum$ feature.
Apparently, the cloud complex is rich in aromatics
and aliphatics.
There appears to be a gradual inward increase
in $I_{3.4}/I_{3.3}$ from the cloud boundaries
(most notably in the R6 and R9 regions).
This phenomenon, where $I_{3.4}/I_{3.3}$ shows
a dependence on regional shielding of UV irradiation,
resembles that observed in PDRs.
In PDRs such as NGC\,7023 and the Orion Bar,
$I_{3.4}/I_{3.3}$ gradually increases deeper into
the cloud, suggesting that the aliphatic structures
associated with the aromatic molecules
have likely undergone photochemical processing
\citep{Peeters2024, Misselt2025}.

We note that all spatially resolved observations
of PDRs consistently show the UIE emission to
peak in the atomic HI layer, between the ionized
gas and the molecular cloud,
rather than in the H$_2$-emitting gas itself
\citep[e.g., see][]{Sellgren1990, Habart2024}.
\citet{Zagury2021} found that the UIE bands
systematically cluster around hydrogen transitions.
More broadly, \citet{Zagury2023} reported that nebular
emission features from the visible to the IR likewise
tend to cluster around transitions of the hydrogen series.

\subsection{Aliphatic C--H Deformation:
                    The 6.85$\mum$ Complex} \label{sec:6.85um}
\begin{figure*}[h!]
\centering
\includegraphics[width=0.32\textwidth]{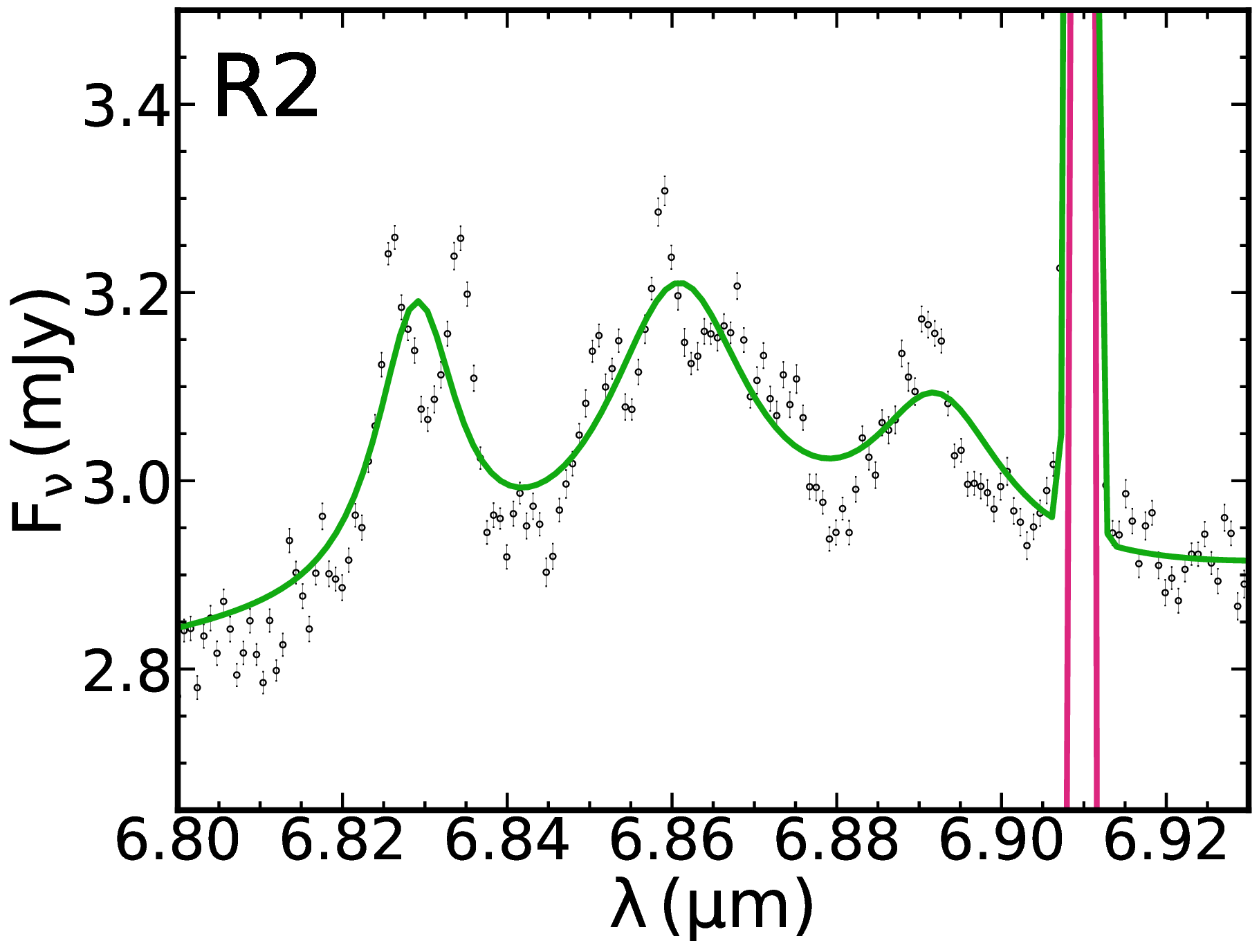}
\includegraphics[width=0.32\textwidth]{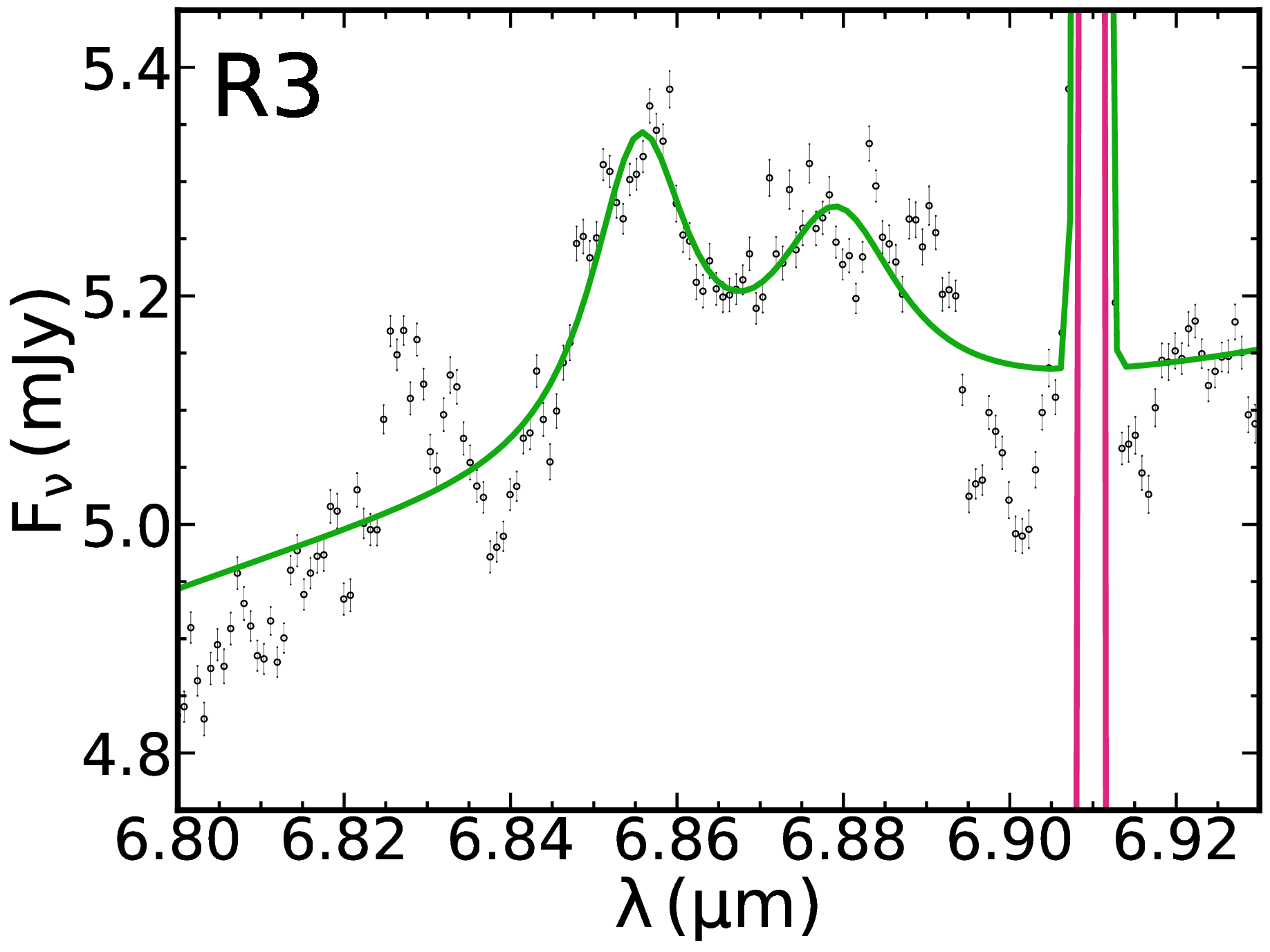}
\includegraphics[width=0.32\textwidth]{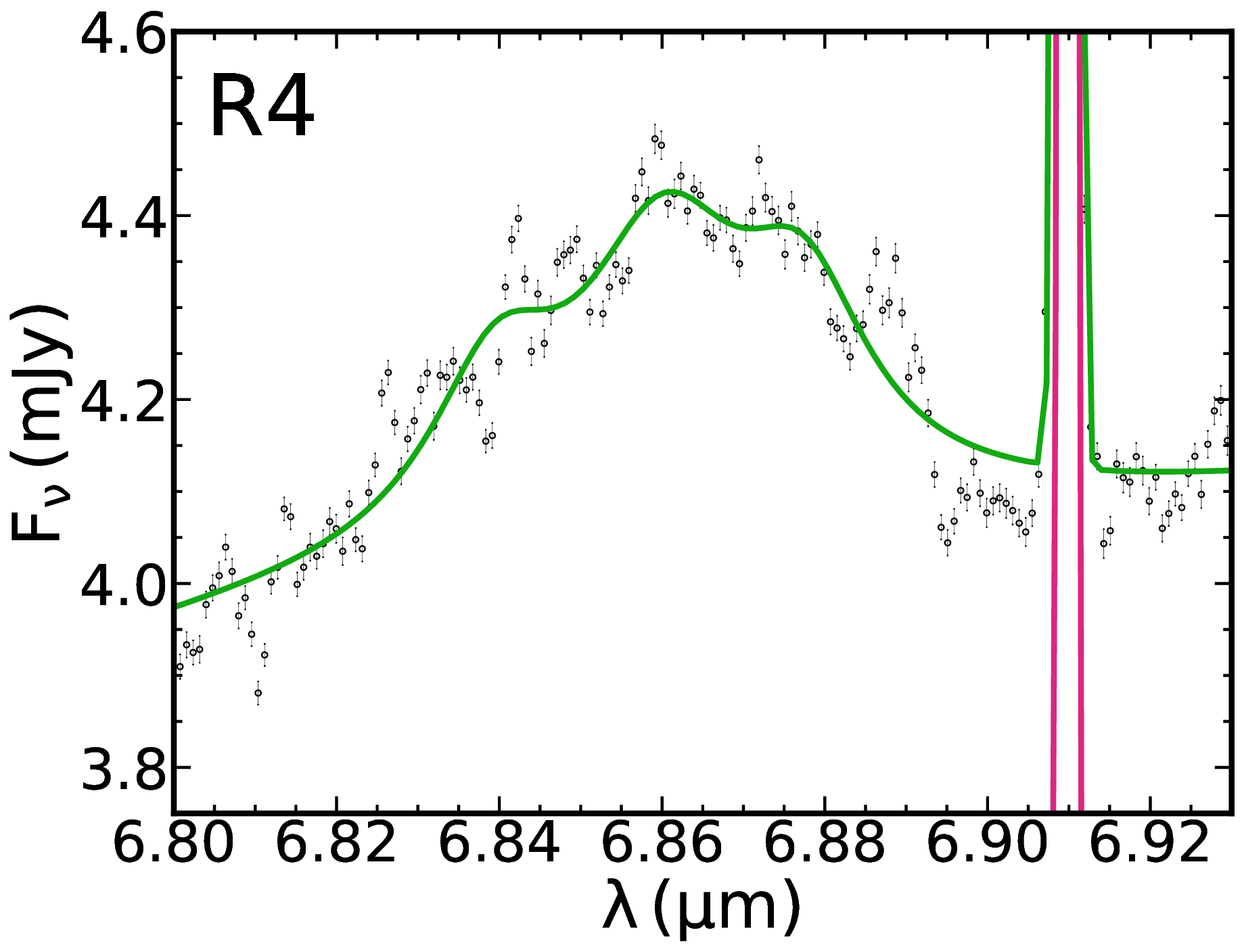}
\includegraphics[width=0.32\textwidth]{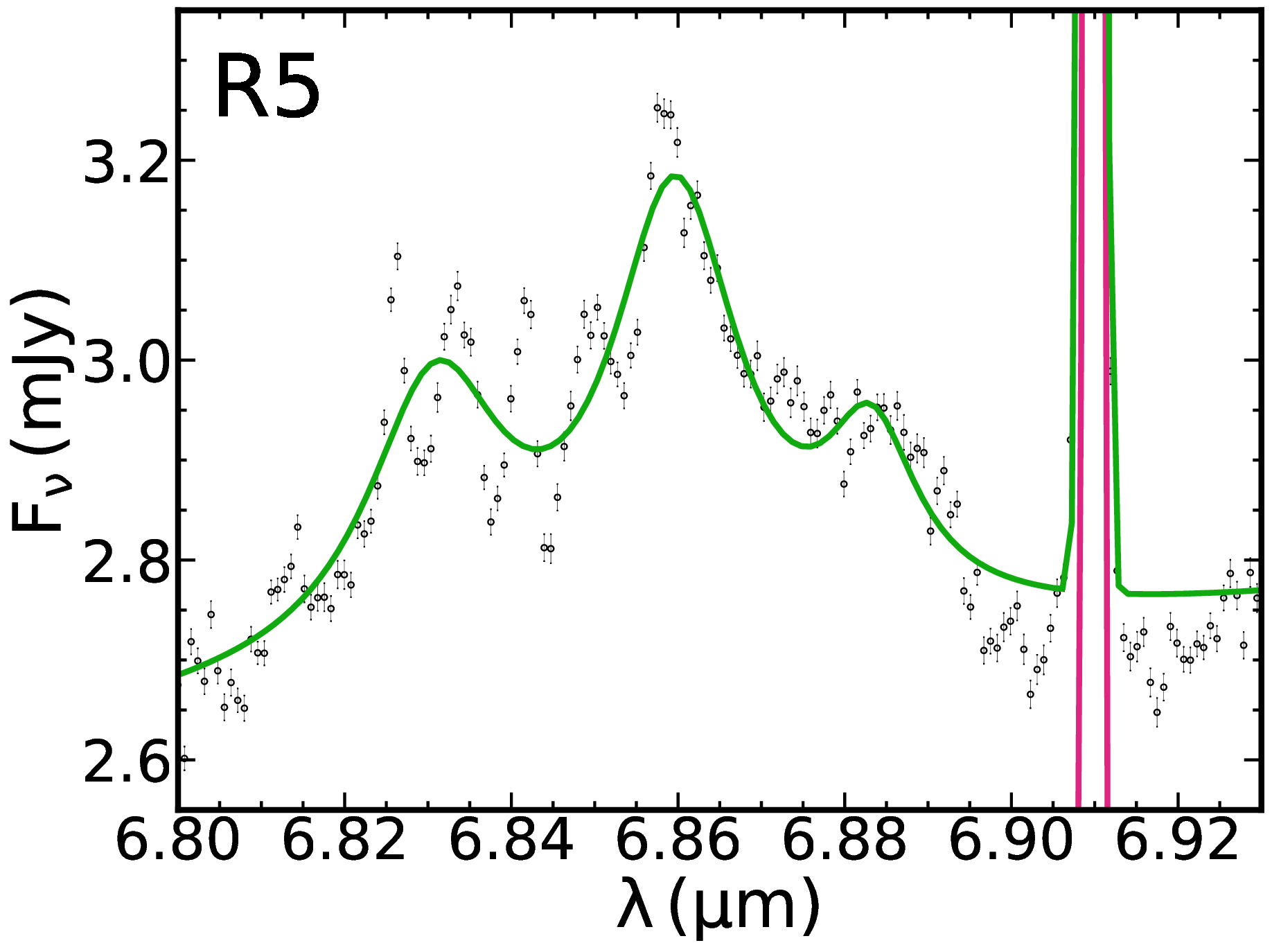}
\includegraphics[width=0.32\textwidth]{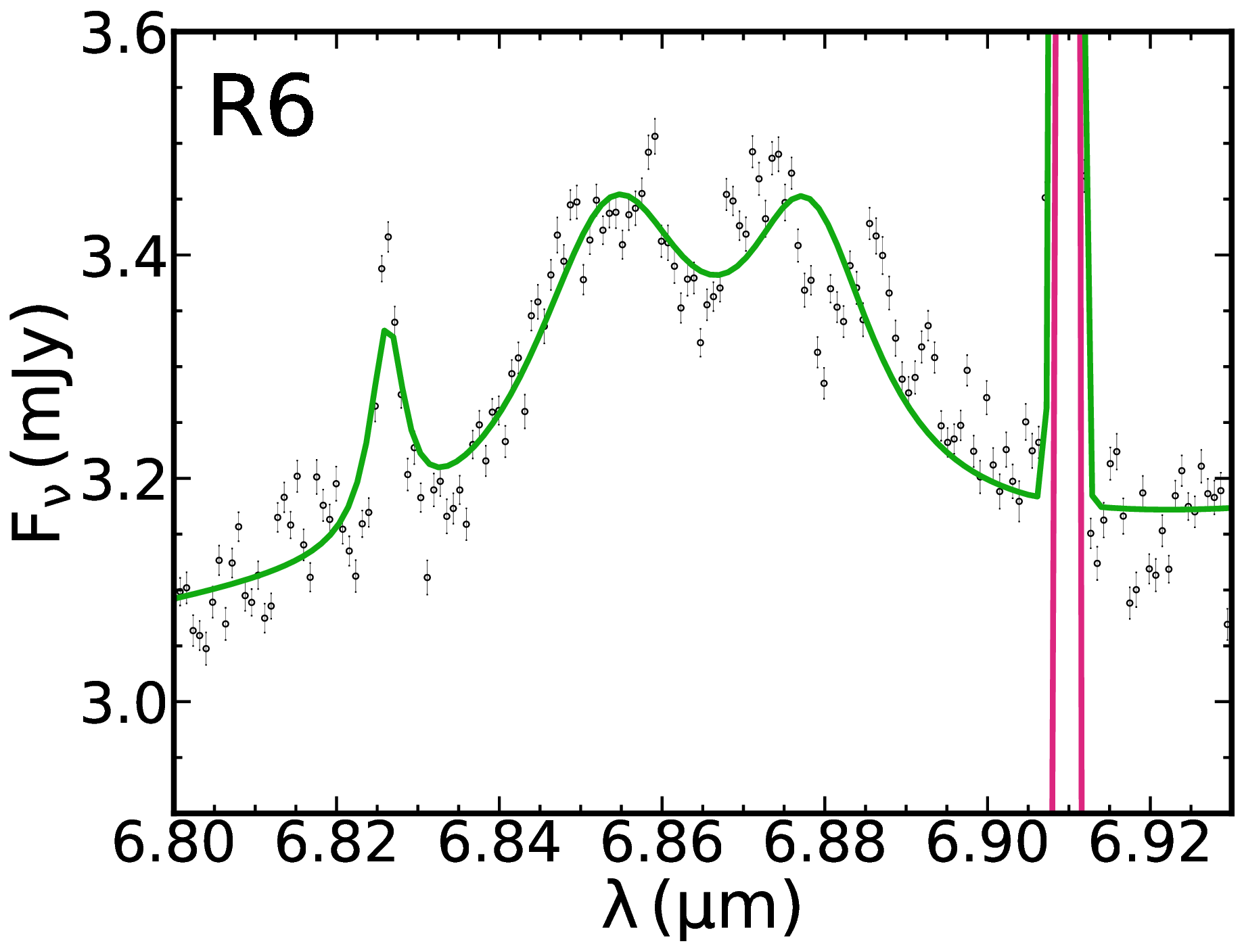}
\includegraphics[width=0.32\textwidth]{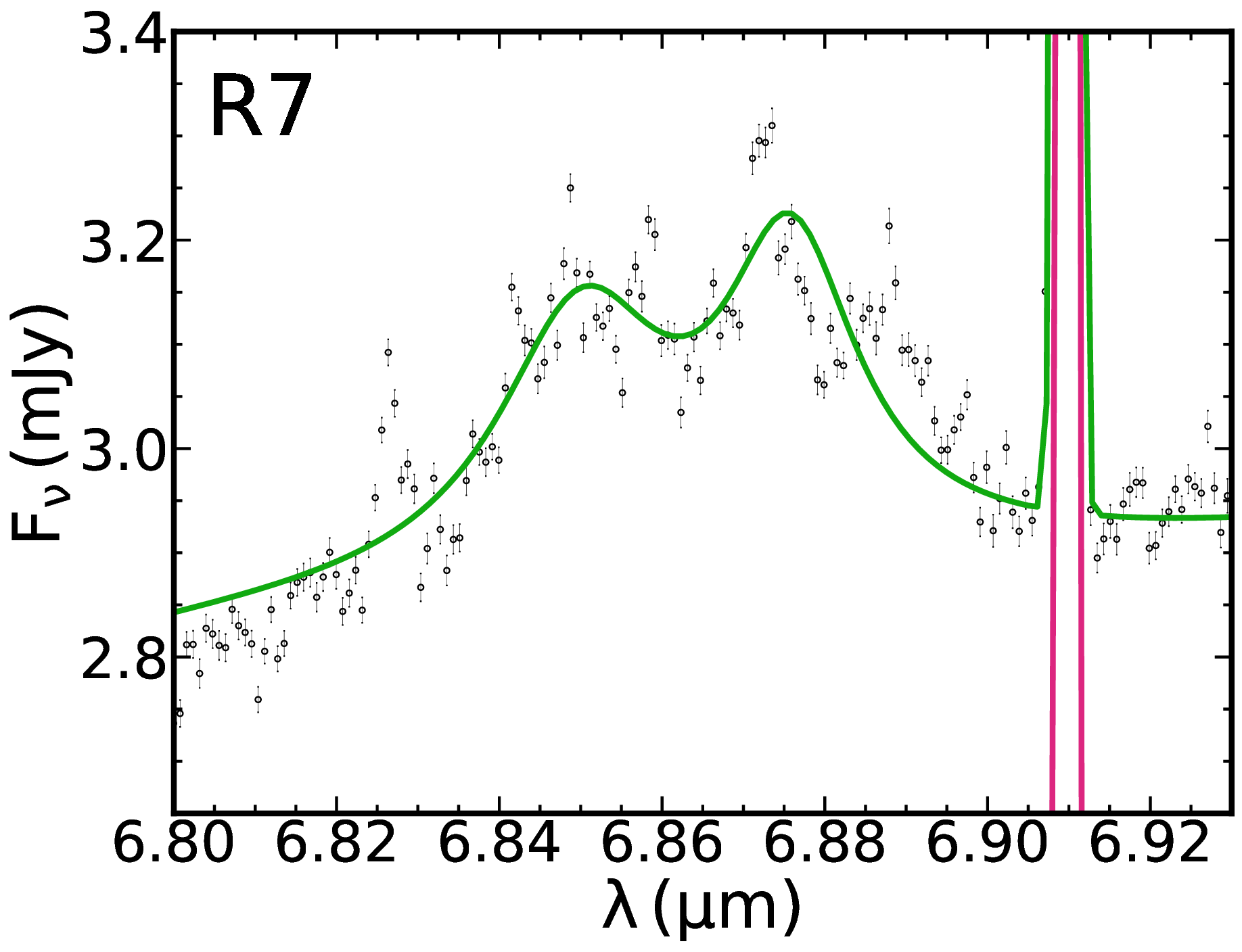}
\includegraphics[width=0.32\textwidth]{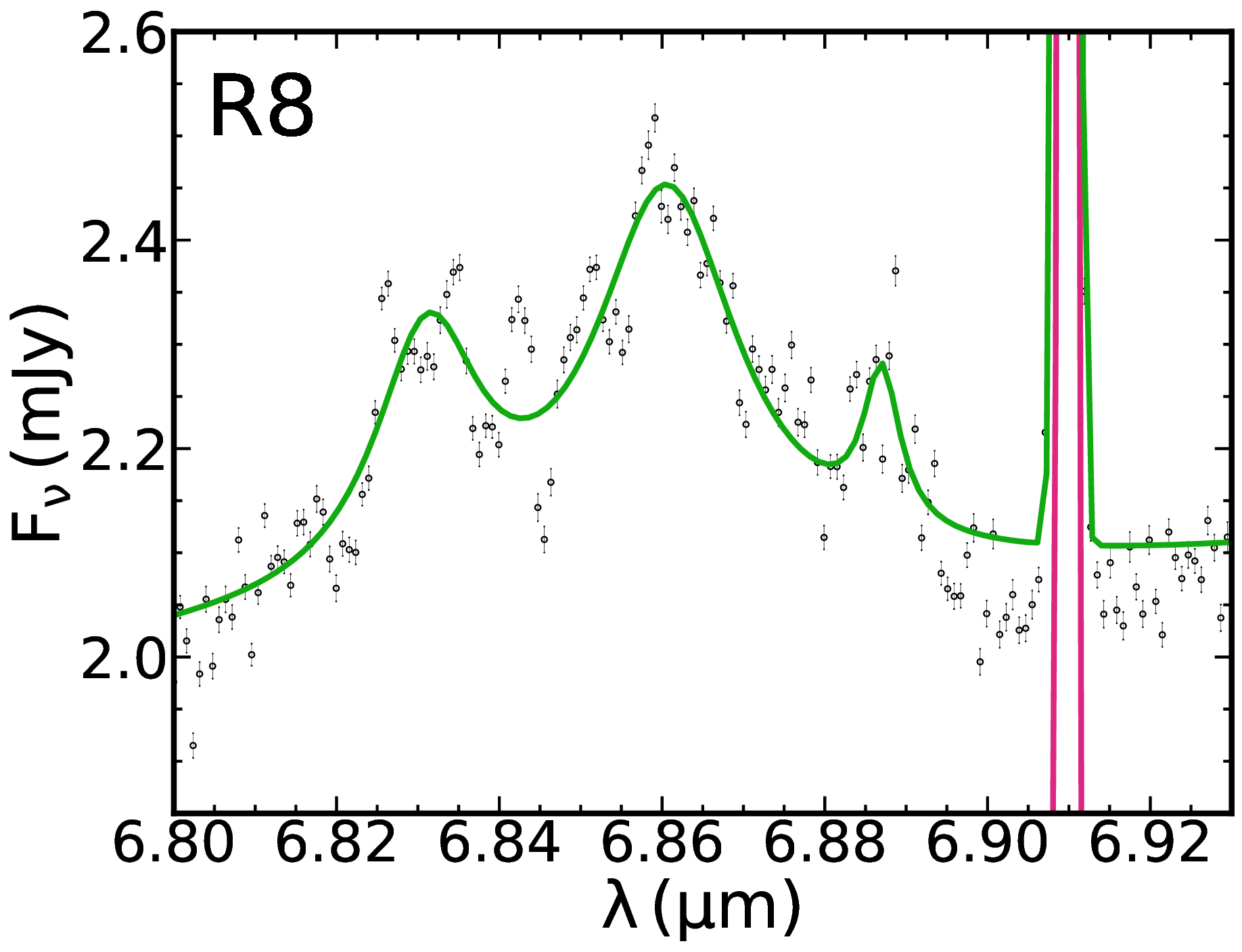}
\includegraphics[width=0.32\textwidth]{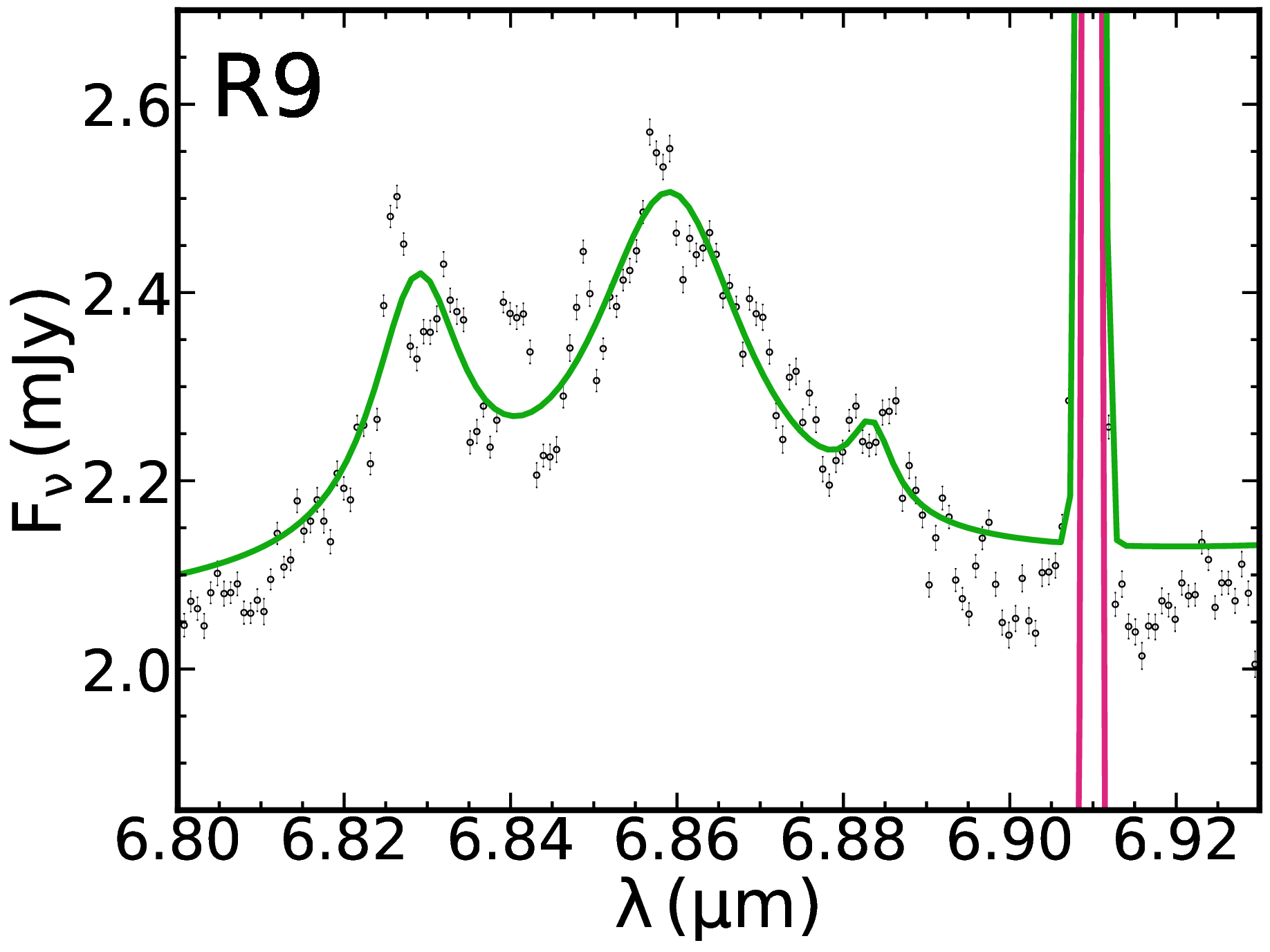}
\includegraphics[width=0.32\textwidth]{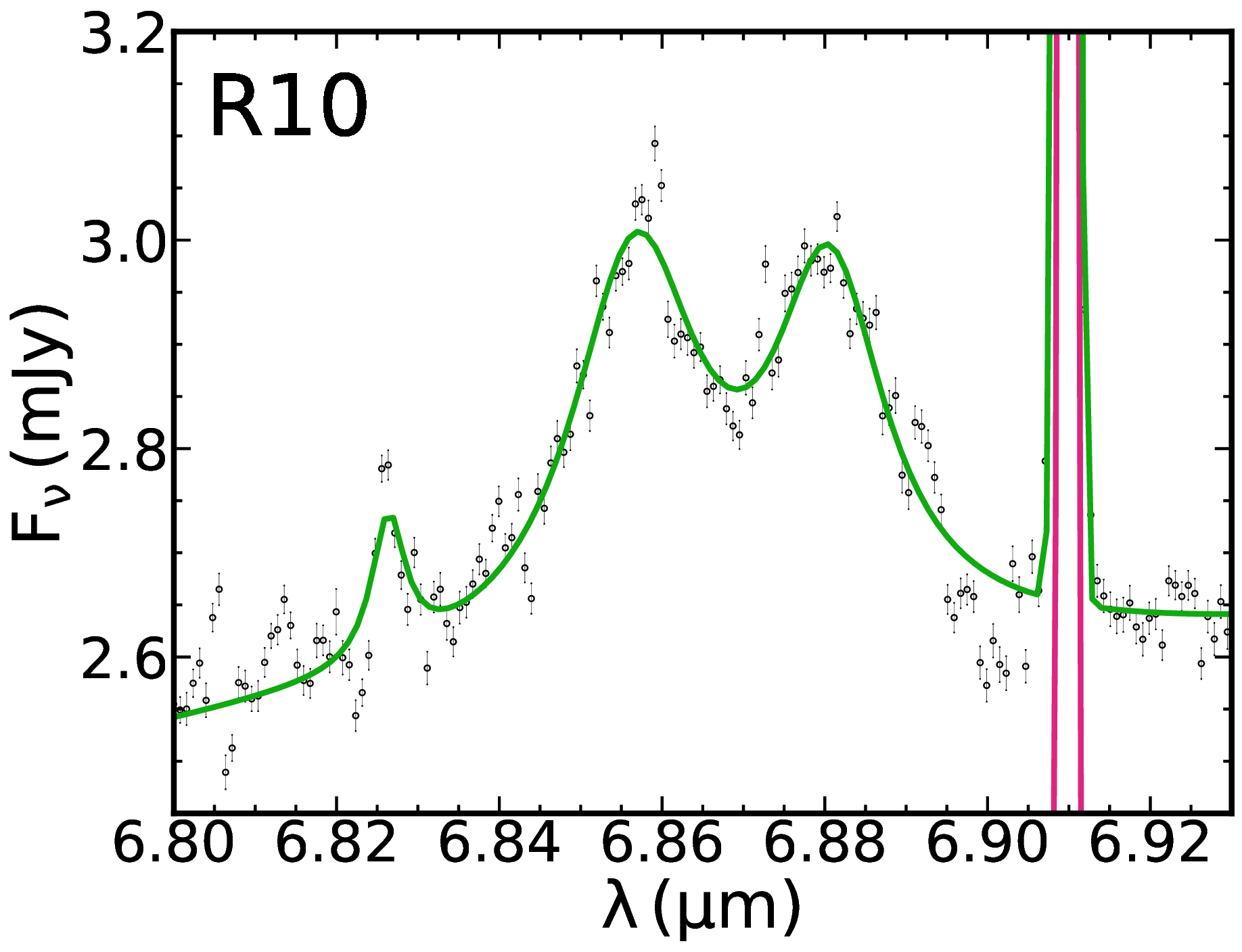}
\includegraphics[width=0.32\textwidth]{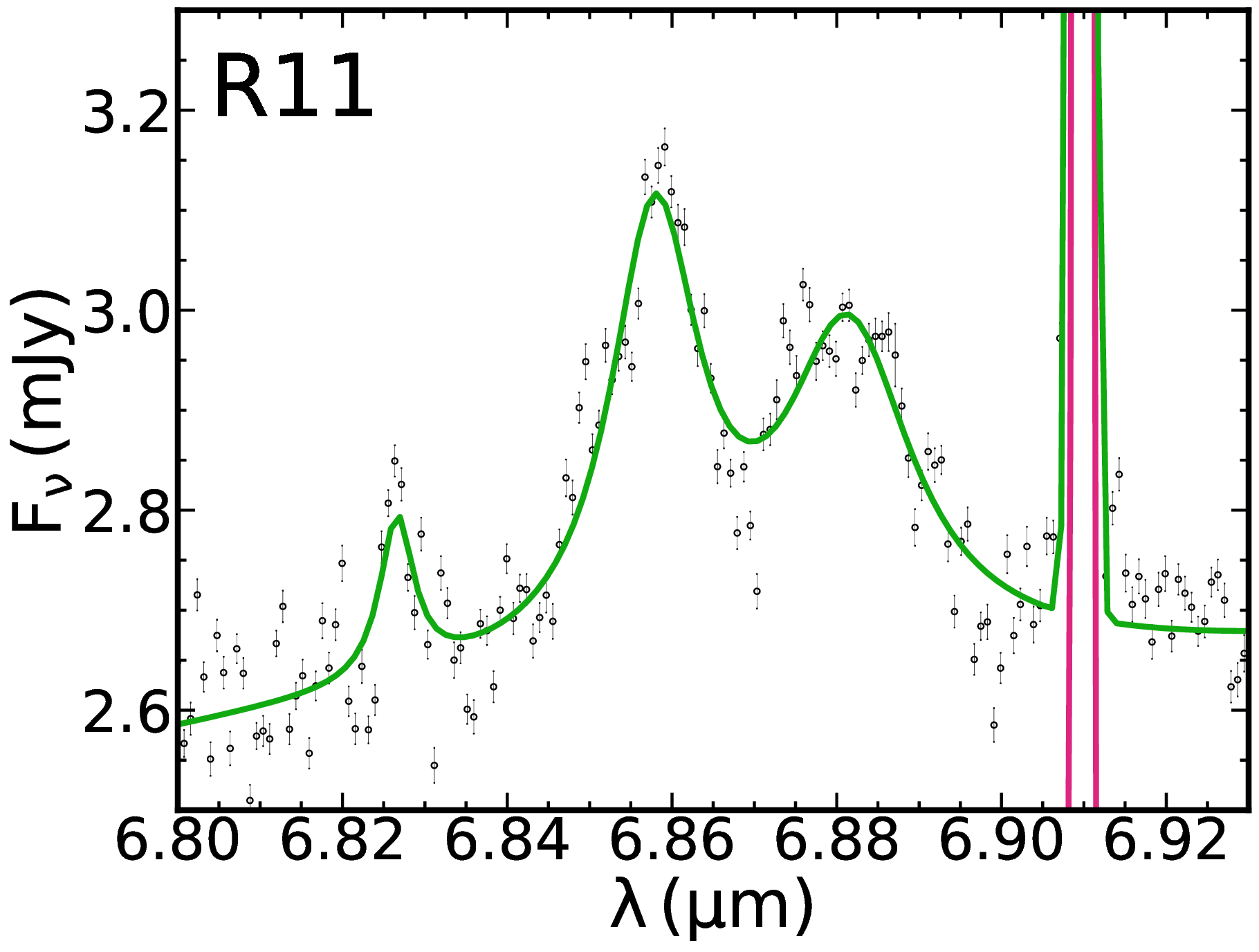}
\includegraphics[width=0.32\textwidth]{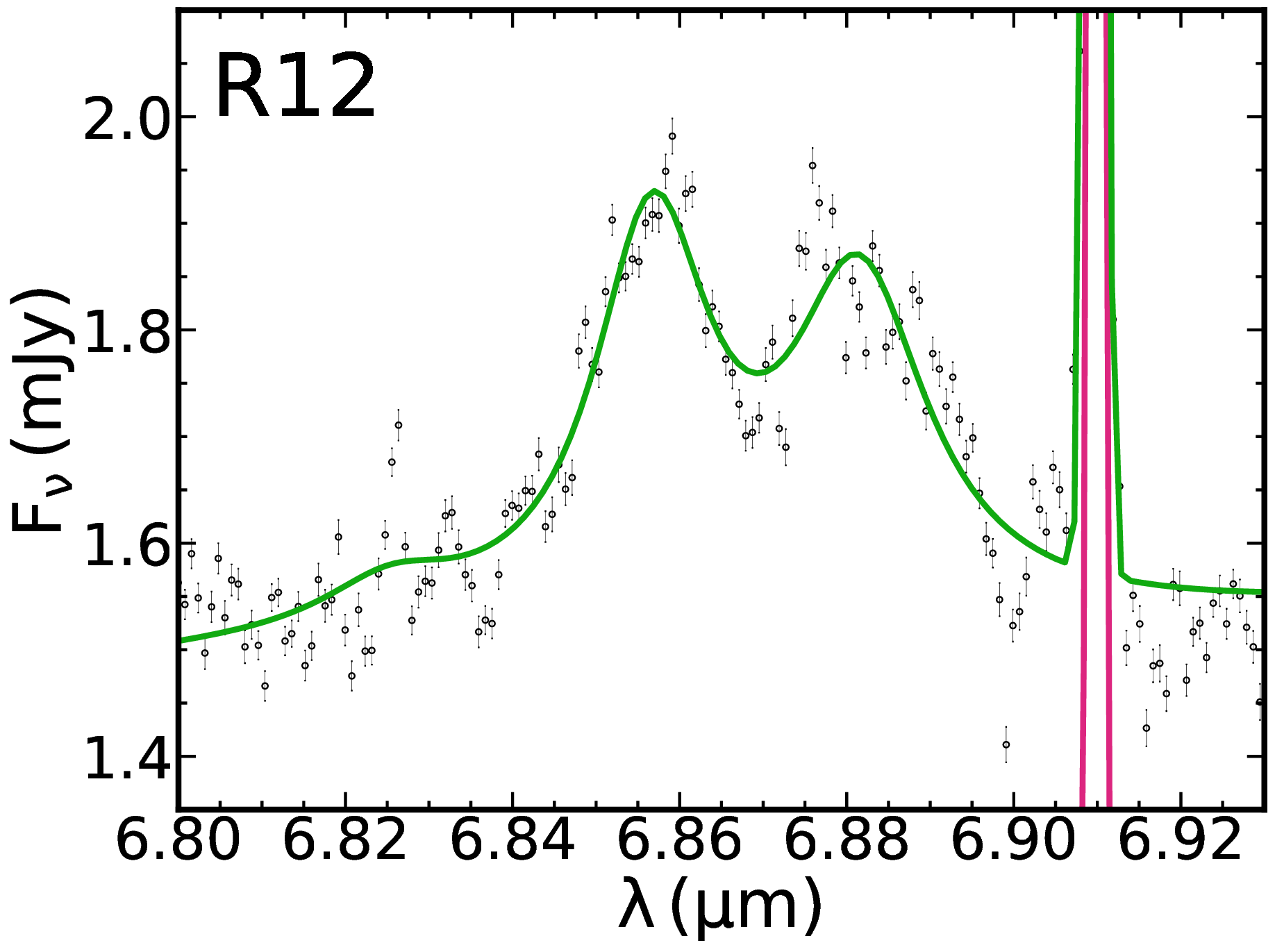}
\includegraphics[width=0.32\textwidth]{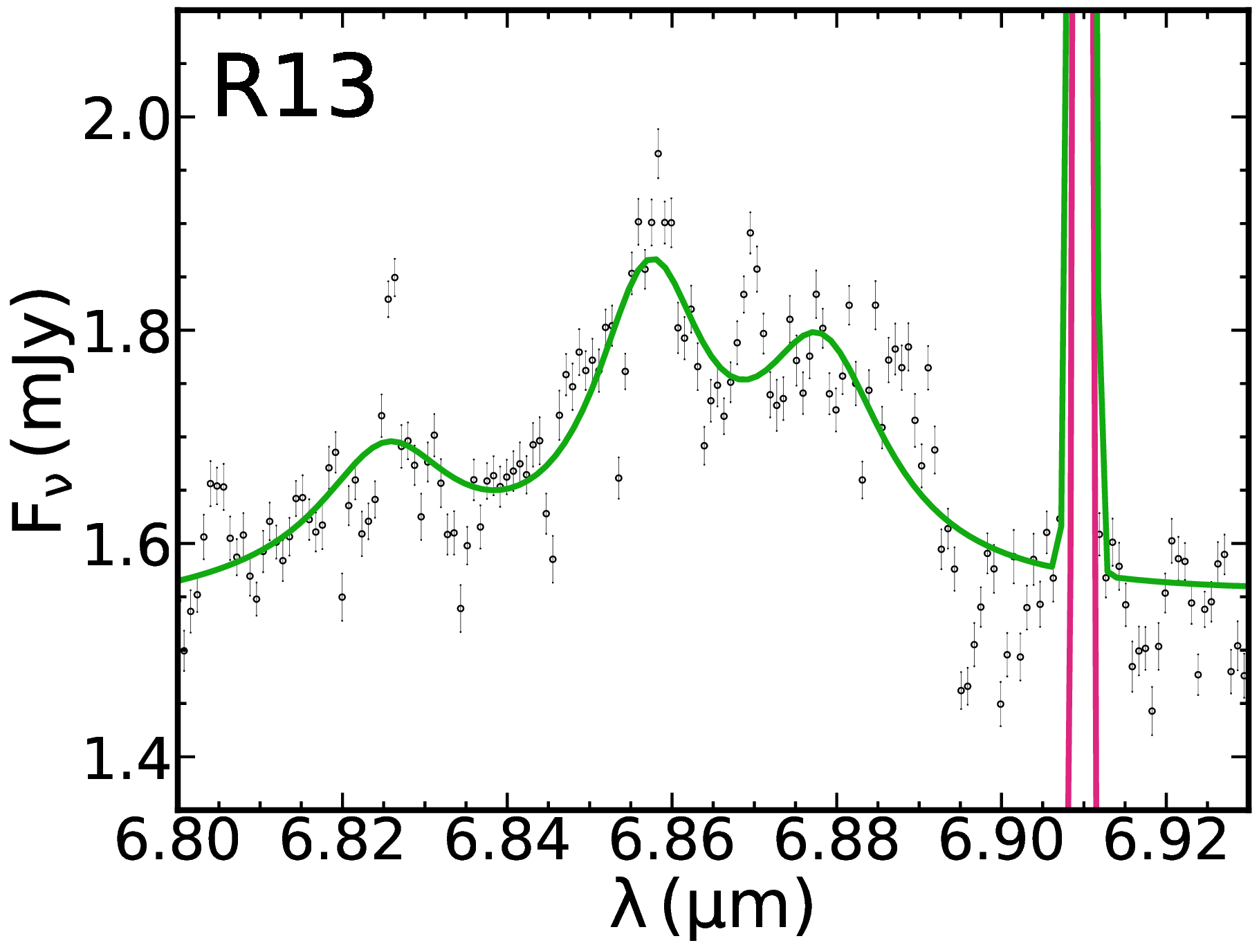}
\vspace{-0.2cm}
\caption{\label{fig:685fit}
Zoomed-in view of the PAHFIT
decomposition around 6.85$\mum$
for the R2--R13 regions.
The 6.85$\mum$ complex is composed
of three sub-features
at $\simali$6.83, 6.86, and 6.88$\mum$.
}
\vspace{-0.1cm}
\end{figure*}

In addition to the 3.4$\mum$ band,
aliphatic C--H stretches also emit at
the 6.85 and 7.25$\mum$ deformation bands,
with the former often appreciably stronger
than the latter one \citep[e.g., see][]{Sloan2014, Yang2016}.
Figure~\ref{fig:685fit} presents a zoomed-in
view of the 6.85$\mum$ complex
for the R2--R13 regions.
It is apparent that the 6.85$\mum$ feature
is typically composed of three weak subfeatures
at $\simali$6.83, 6.86, and 6.88$\mum$.
Table~\ref{tab:fit_parameters} also
summarizes the best-fit parameters
(i.e., peak wavelengths, widths, and intensities)
as well as the relative intensities of these subfeatures.
While the peak wavelengths and widths
of these subfeatures remain more or less
stable, their relative intensities
vary from one region to another.
Overall, the 6.858$\mum$ subfeature
is rather robust, whereas the 6.83 and 6.88$\mum$
subfeatures exhibit significant variations
across different regions. The reason for this
variation is unclear, but it may involve subtle
differences in their originating molecular structures
(e.g., methyl and methylene groups).
We should note that atomic hydrogen has
a recombination line at 6.826$\mum$
(i.e., the HI Humphreys 6$\rightarrow$5 transition).
As the 6.83$\mum$ subfeature seen here is
considerably broader than other HI lines,
it is unlikely that the 6.826$\mum$ HI line
dominates the 6.83$\mum$ subfeature.

\begin{figure}[h!]
\centering
\includegraphics[width=0.46\textwidth]{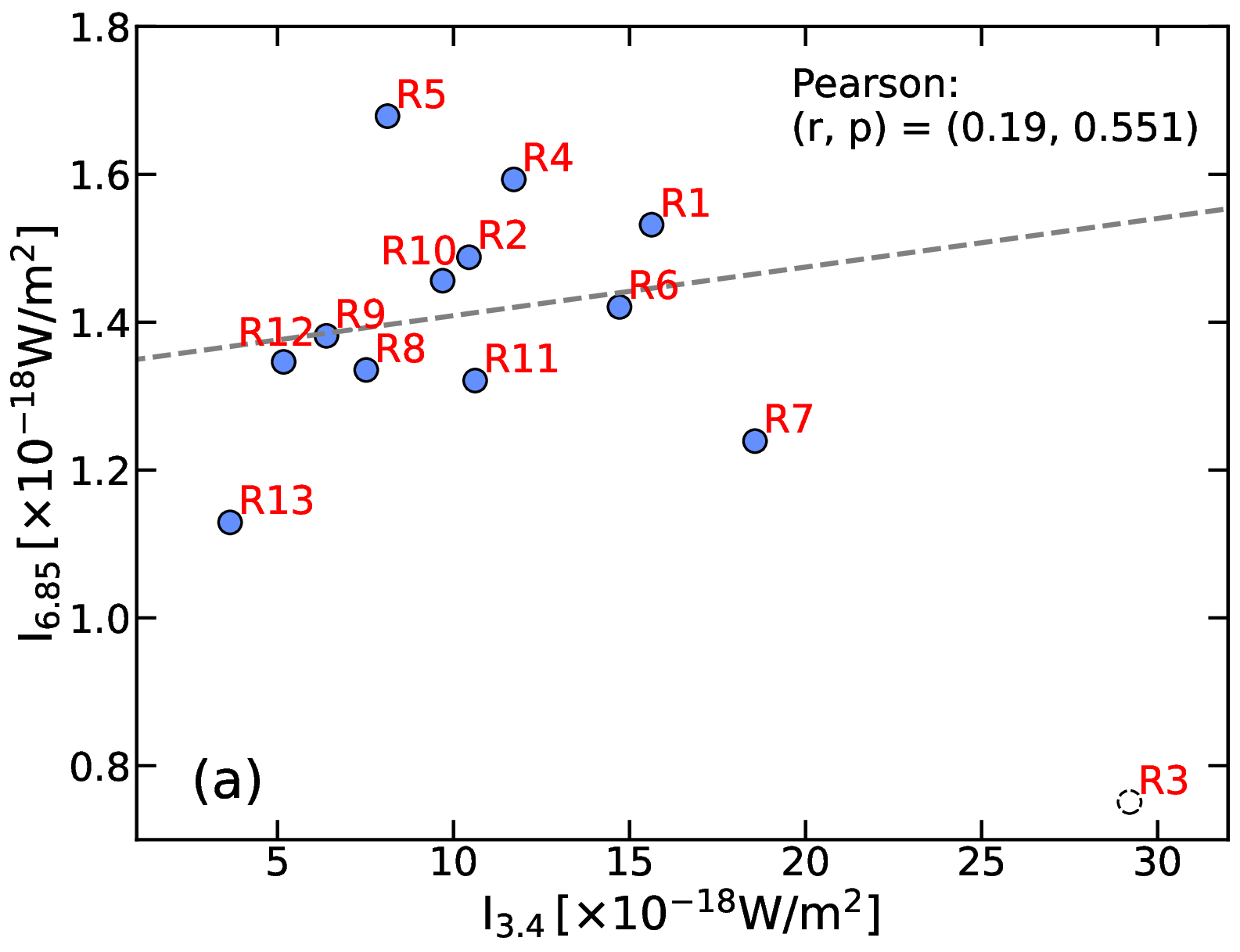}
\includegraphics[width=0.48\textwidth]{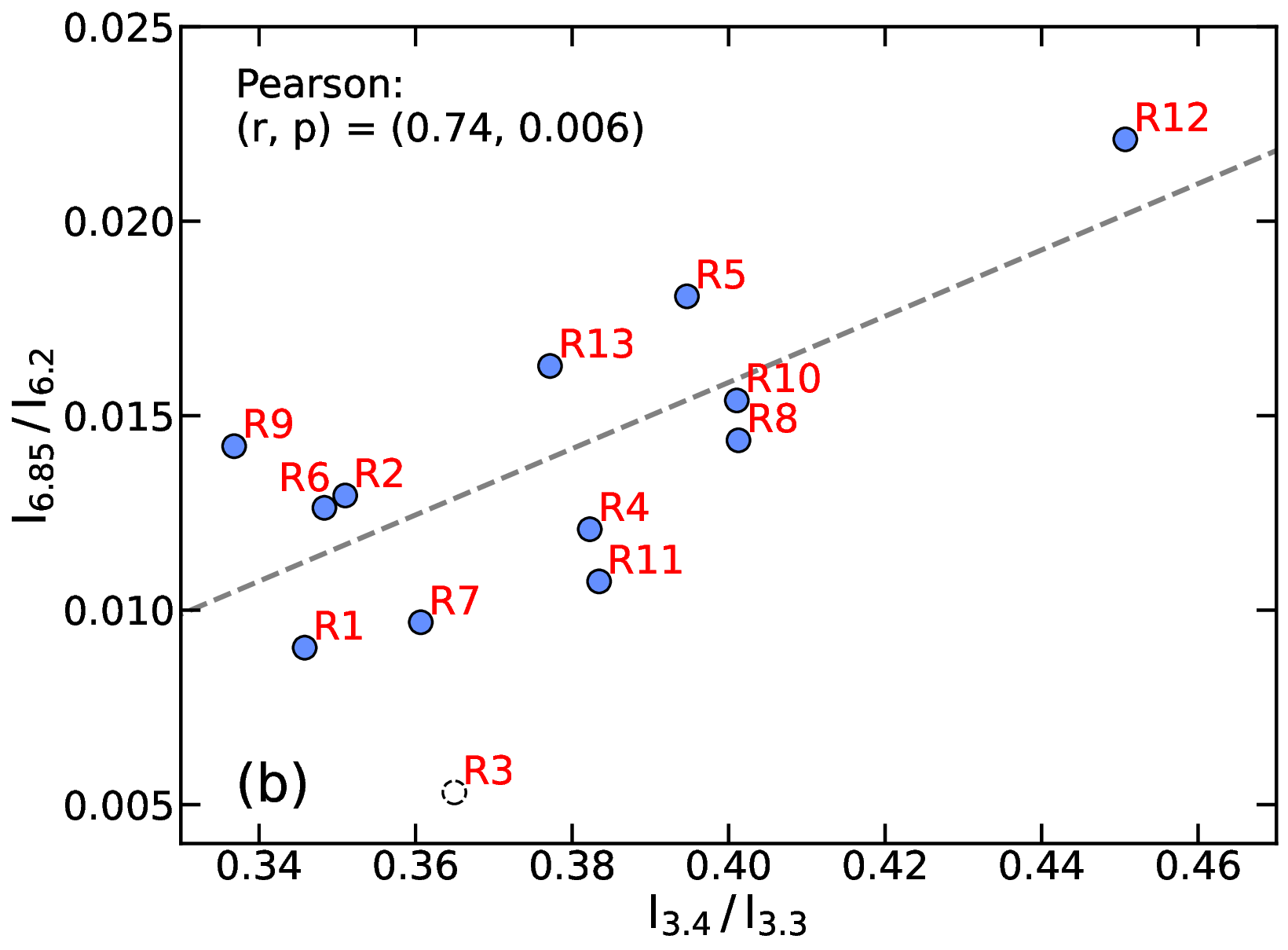}
\vspace{-0.3cm}
\caption{\label{fig:I34_I685}
Left panel (a): Relation between the 3.4$\mum$
aliphatic C--H stretch and the 6.85$\mum$
aliphatic C--H deformation.
Right panel (b): $I_{3.4}/I_{3.3}$ vs. $I_{6.85}/I_{6.2}$.
The results for R3 are indicated by
an open circle due to its low quality spectrum
and this region has been excluded
from the correlation analysis.
While $I_{3.4}$ is not correlated with $I_{6.85}$,
$I_{3.4}/I_{3.3}$ is positively correlated
with $I_{6.85}/I_{6.2}$, confirming
the aliphatic nature of the 6.85$\mum$ band.
That the 3.4$\mum$-emitter and
the 6.85$\mum$-emitter are different in size
explains why $I_{3.4}$ is not correlated with $I_{6.85}$.
}
\vspace{-0.3cm}
\end{figure}

Figure~\ref{fig:I34_I685}a examines
the correlation between the intensities
of the 3.4$\mum$ and 6.85$\mum$ features.
Surprisingly, a Pearson correlation analysis finds
a correlation coefficient of $r\approx0.19$ and
a statistical significance of $p\approx0.15$,
indicating no correlation.
This is somewhat surprising, as both features
arise from aliphatic C--H stretches.
However, one should keep in mind that the 3.4
and 6.85$\mum$ features are not emitted by
the same molecules. \citet{Yang2016} have
shown that the 6.85$\mum$-feature emitter
is larger in size than the 3.4$\mum$-feature emitter.
To demonstrate this,
we show in Figure~\ref{fig:I34_I685}b
the relation between $I_{3.4}/I_{3.3}$
and $I_{6.85}/I_{6.2}$, both of which are
indicative of the aliphatic fractions
of the UIE emitters \citep[see][]{Li2012, Yang2016}.
A strong positive correlation is found,
confirming the aliphatic nature of
the 6.85$\mum$ band.

\subsection{Aliphatic C--D Stretches:
                    The 4.65$\mum$ Complex} \label{sec:4.65um}
The aromatic C--D emission band
at $\simali$4.4$\mum$
as well as the aliphatic C--D band
at $\simali$4.6--4.7$\mum$ have been
tentatively detected in several objects
in the pre-JWST era \citep{Verstraete1996, Peeters2004,
Boulanger2011, Doney2016, Onaka2022}.
This band was not seen in the AKARI
spectra of the LMC obtained by \citet{Mori2012}.
More recently, due to its superb sensitivity,
JWST clearly detected the aliphatic C--D emission
at $\simali$4.65$\mum$ both in Galactic and
extragalactic sources,
including the M17 photodissociation region \citep[PDR;][]{Boersma2023},
the Orion Bar \citep{Peeters2024},
the luminous IR galaxy merger
NGC\,3256 \citep{Pereira-santaella2024},
the Whirlpool galaxy M51 \citep{Draine2025},
the Horsehead nebula and the reflection nebula
NGC\,7023 \citep{Misselt2025}.
However, the detection of the 4.4$\mum$
aromatic C--D band is less certain.
The emission complex around 4.4$\mum$
actually peaks at $\simali$4.35$\mum$,
which overlaps with or is even dominated
by C--N stretches \citep{Esposito2025, Chen2026}.
As a matter of fact, neither M17 nor M51 shows
appreciable aromatic C--D emission at 4.4$\mum$.

\begin{figure*}[h!]
\centering
\includegraphics[width=0.32\textwidth]{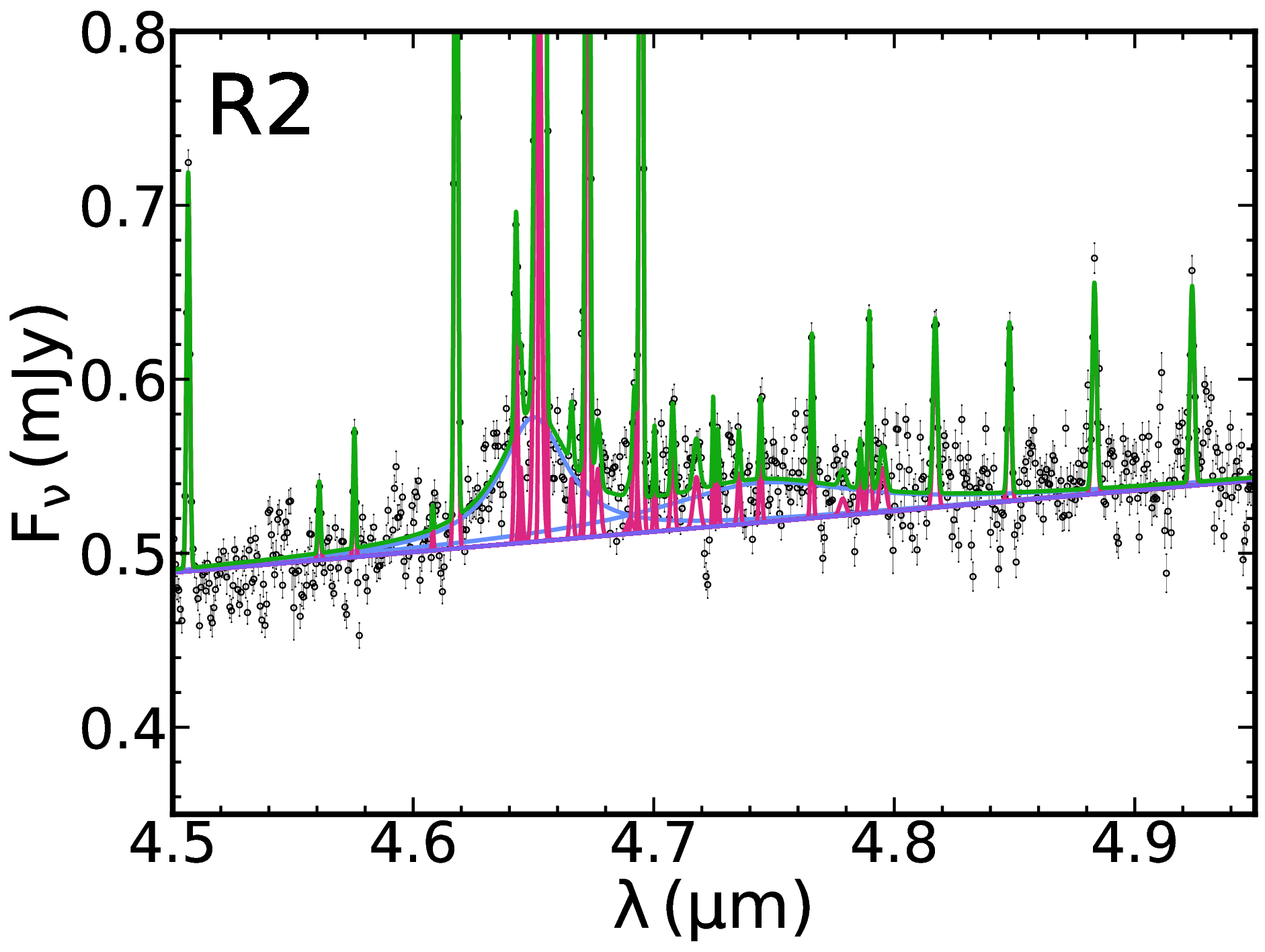}
\includegraphics[width=0.32\textwidth]{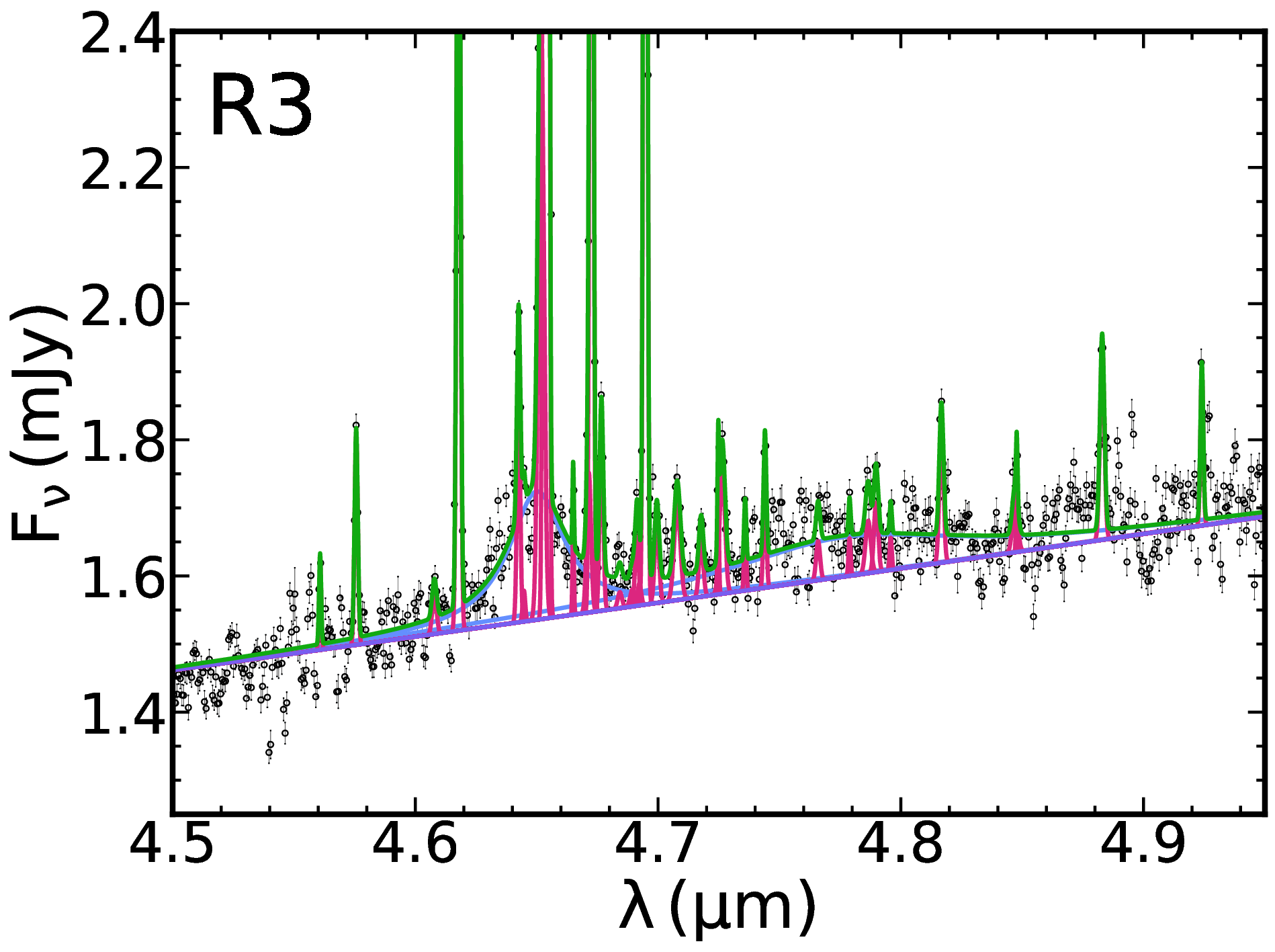}
\includegraphics[width=0.32\textwidth]{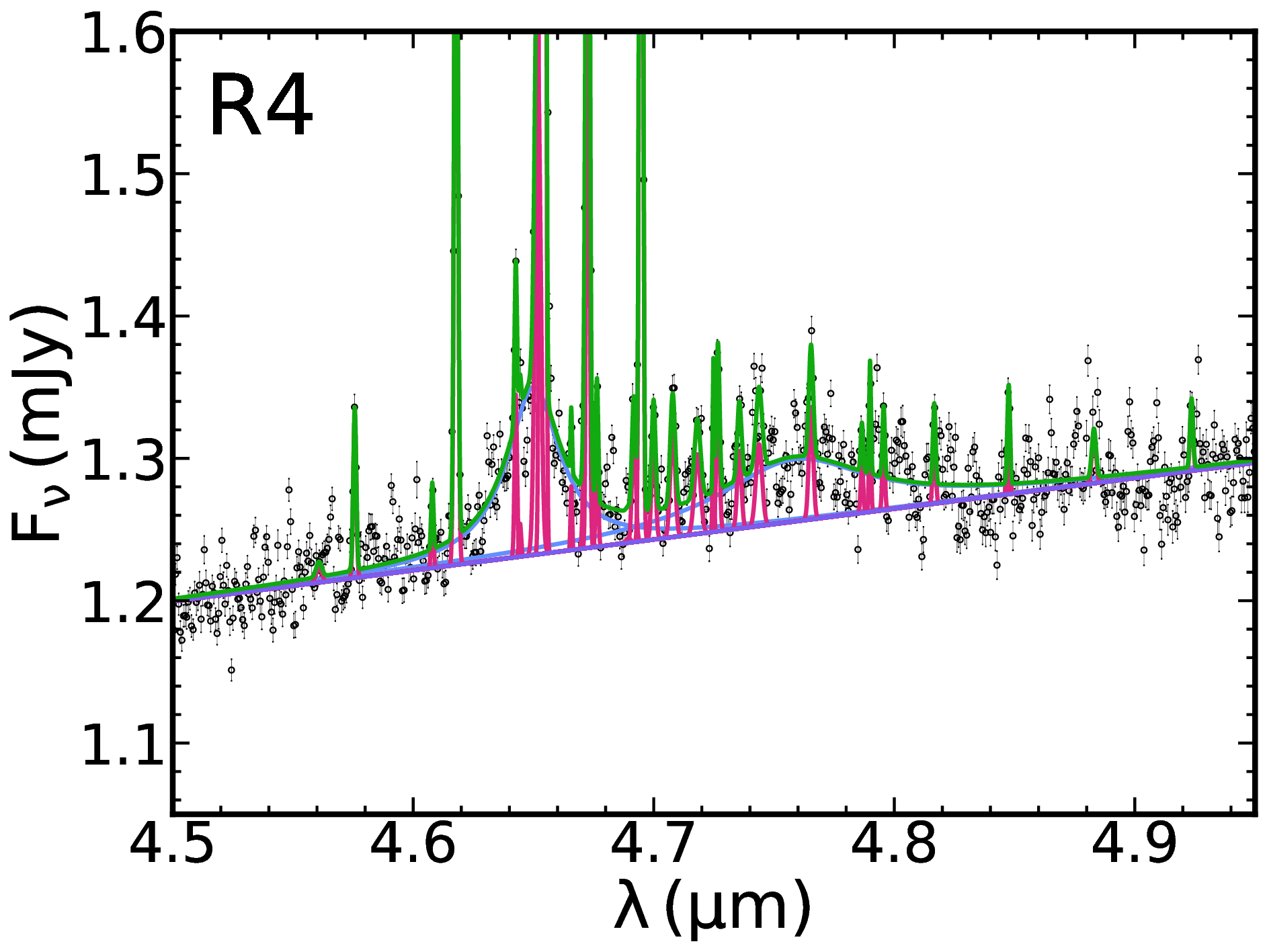}
\includegraphics[width=0.32\textwidth]{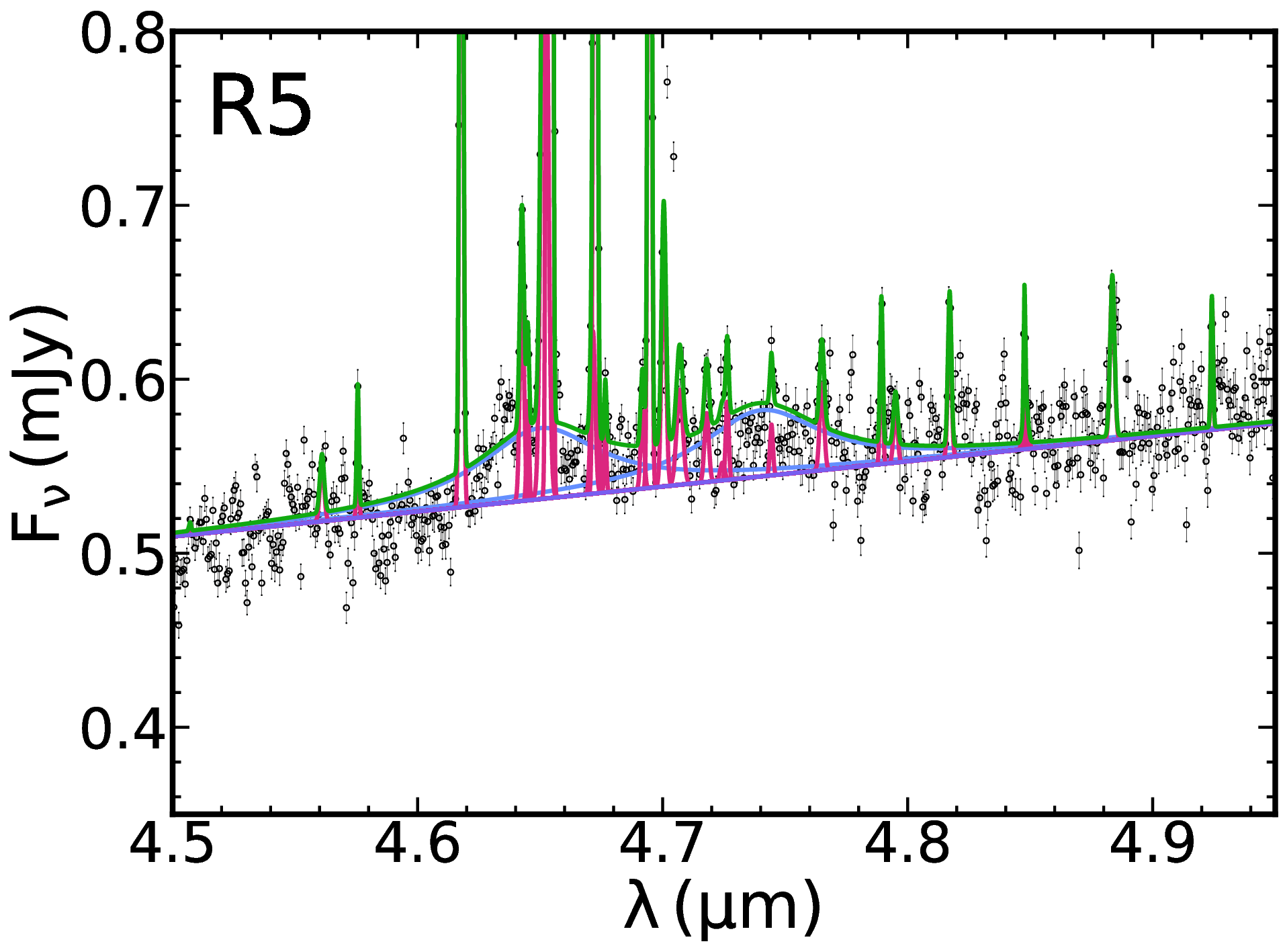}
\includegraphics[width=0.32\textwidth]{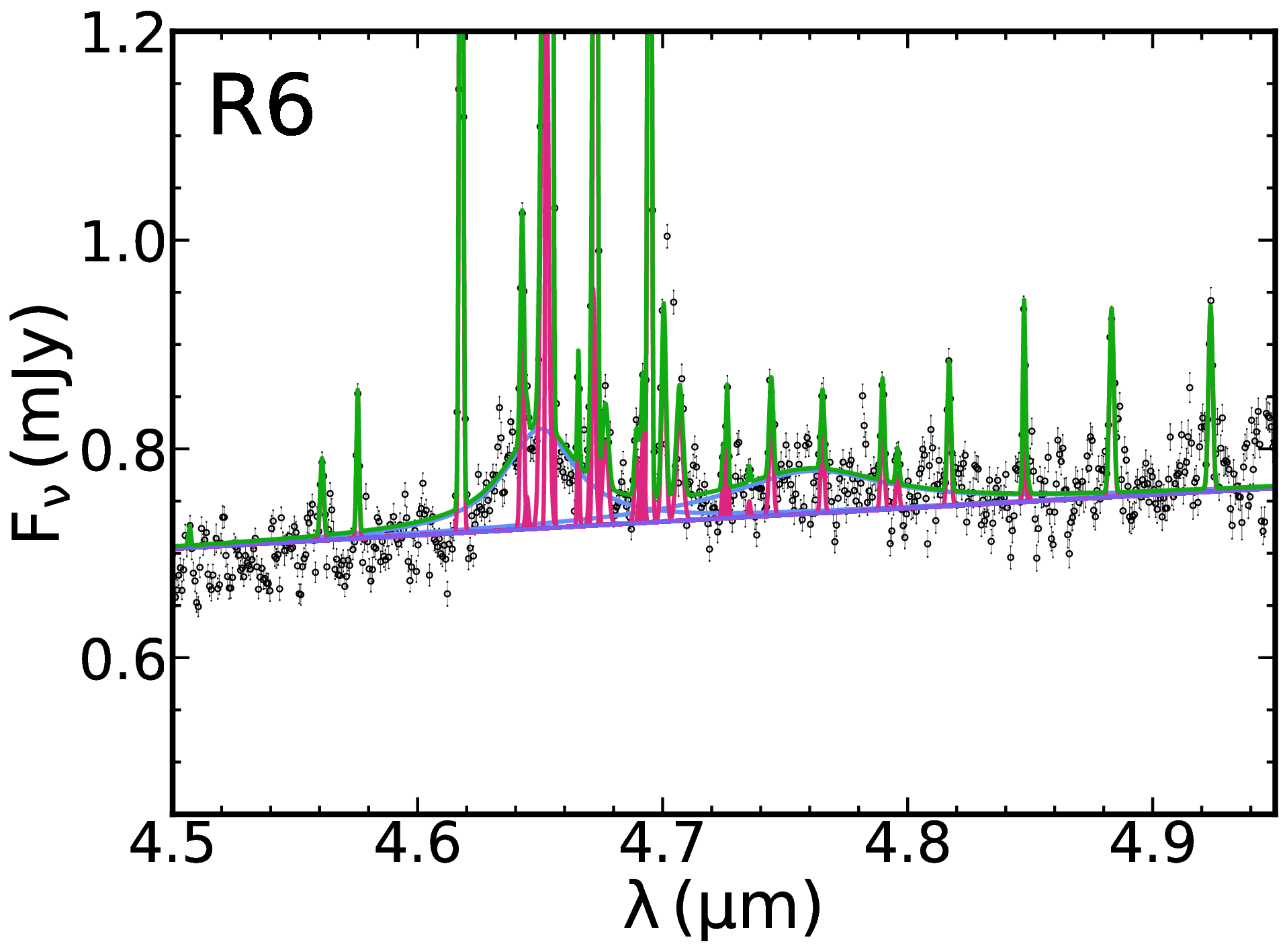}
\includegraphics[width=0.32\textwidth]{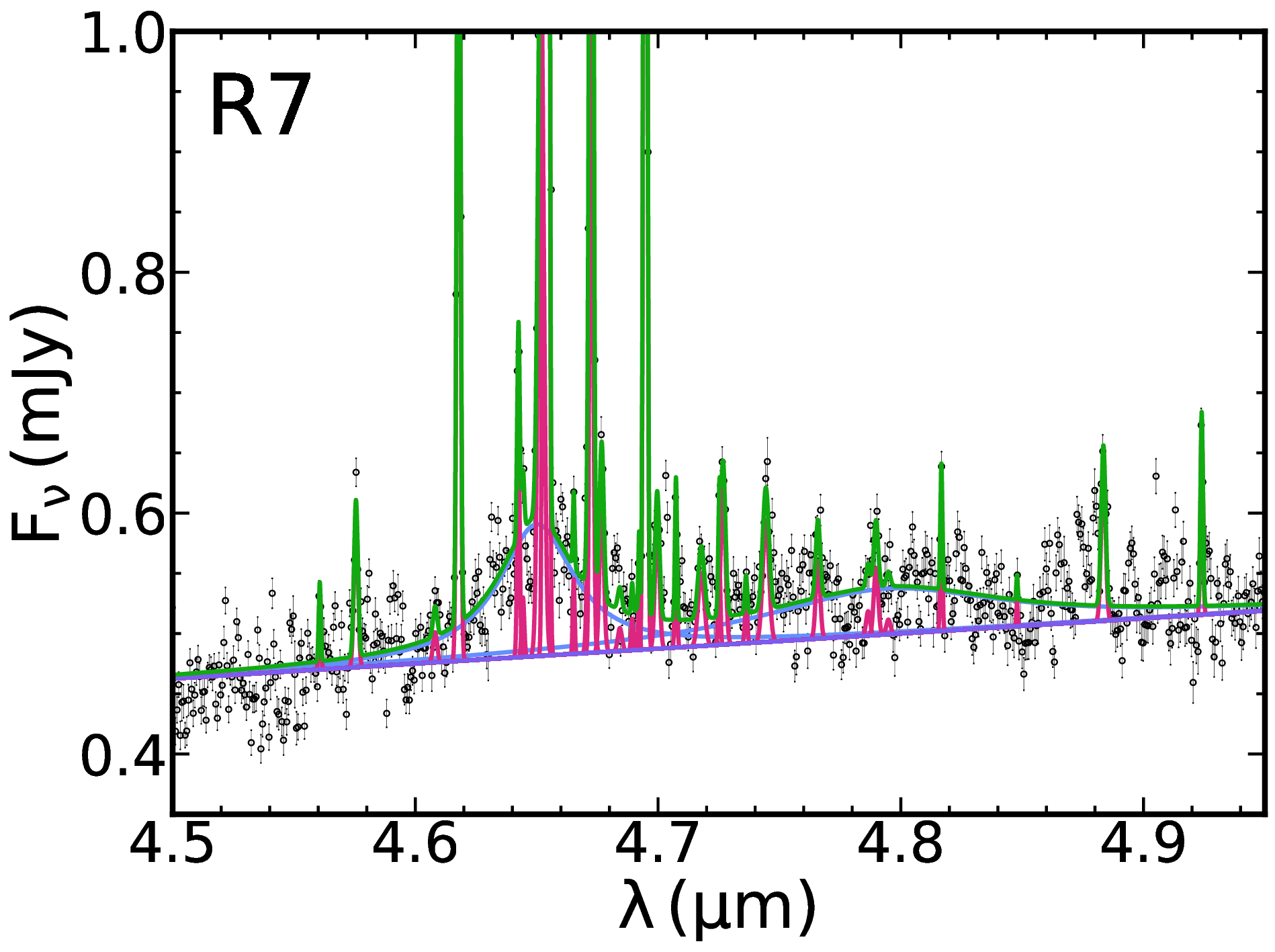}
\includegraphics[width=0.32\textwidth]{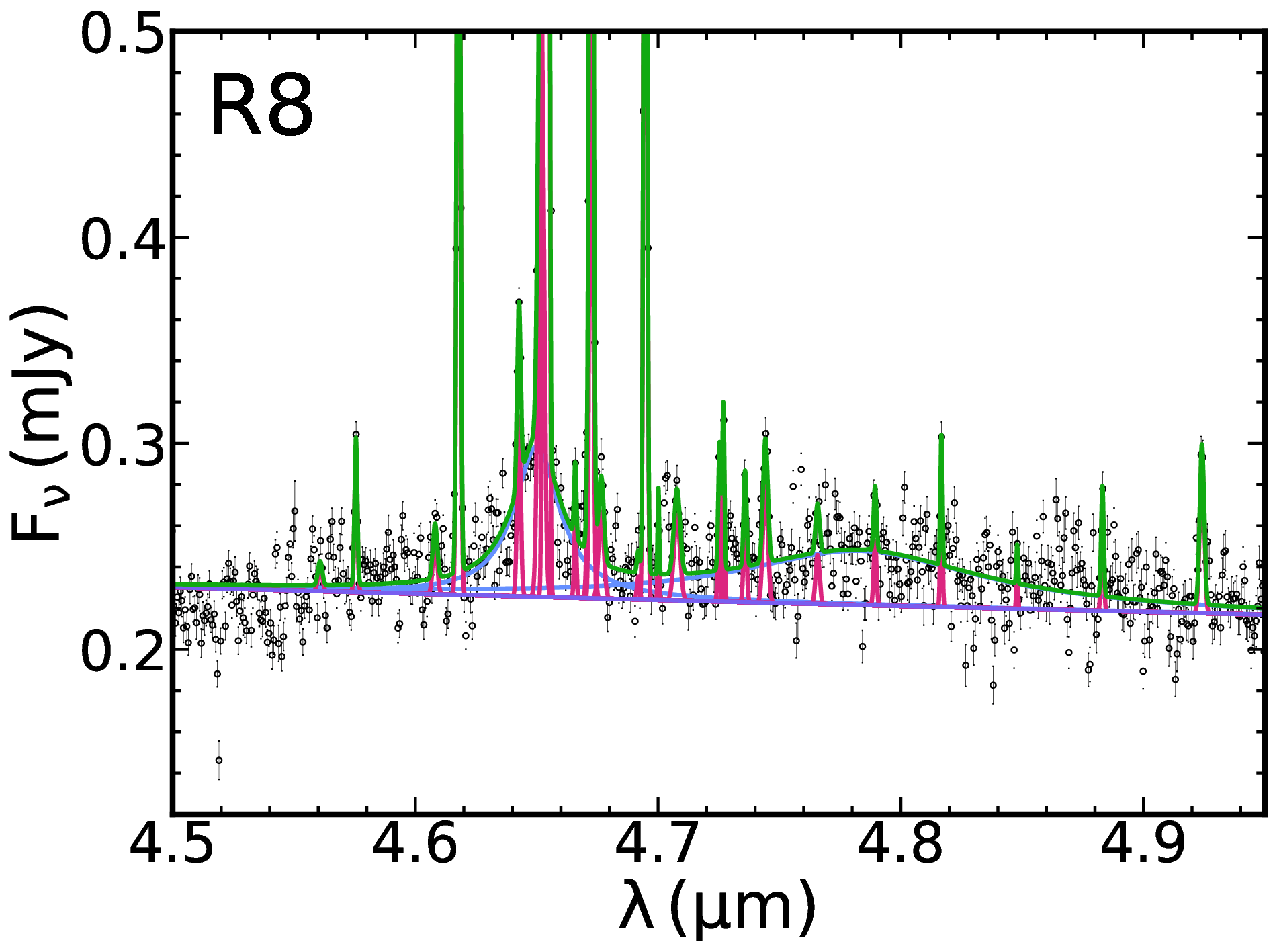}
\includegraphics[width=0.32\textwidth]{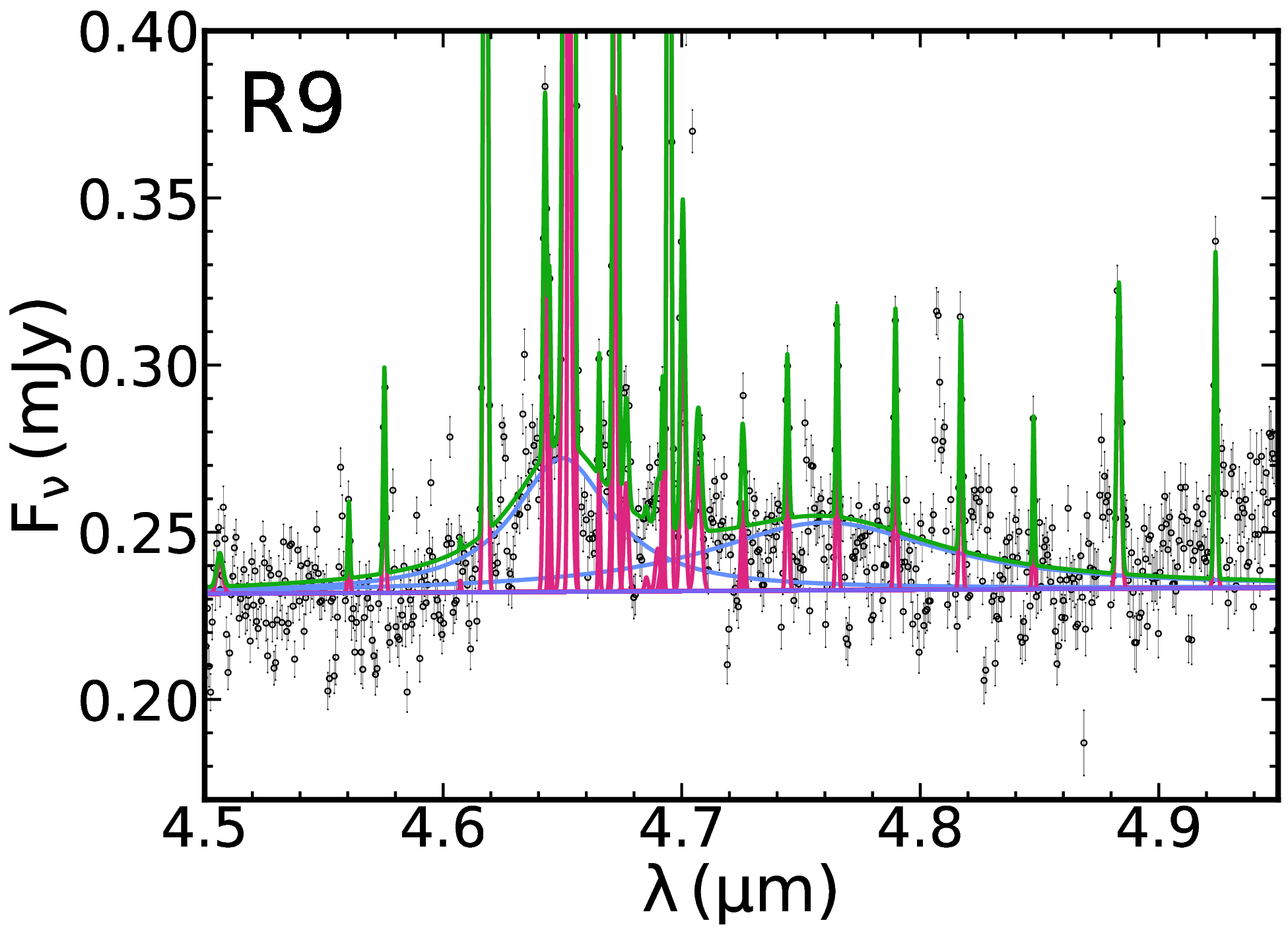}
\includegraphics[width=0.32\textwidth]{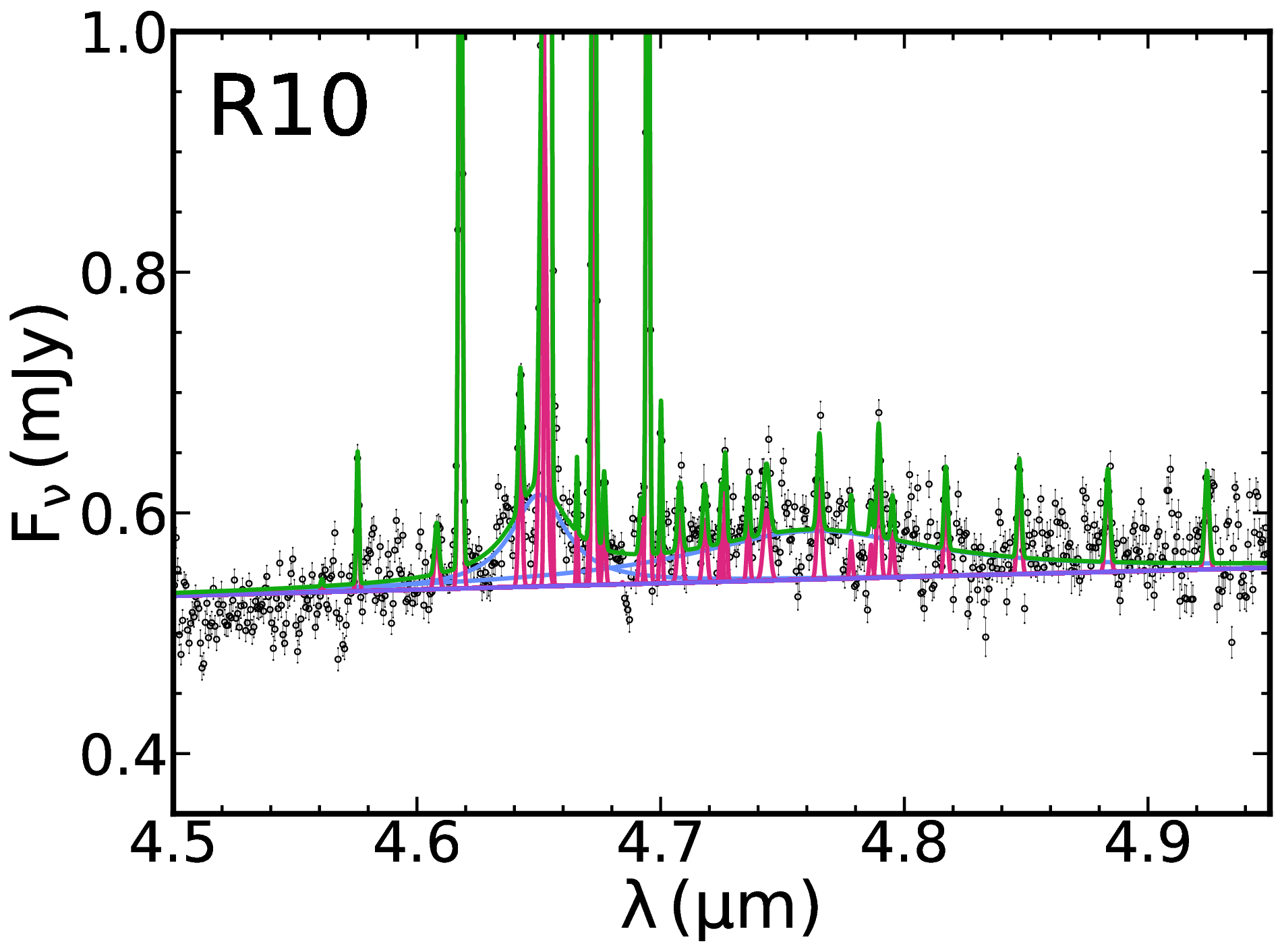}
\includegraphics[width=0.32\textwidth]{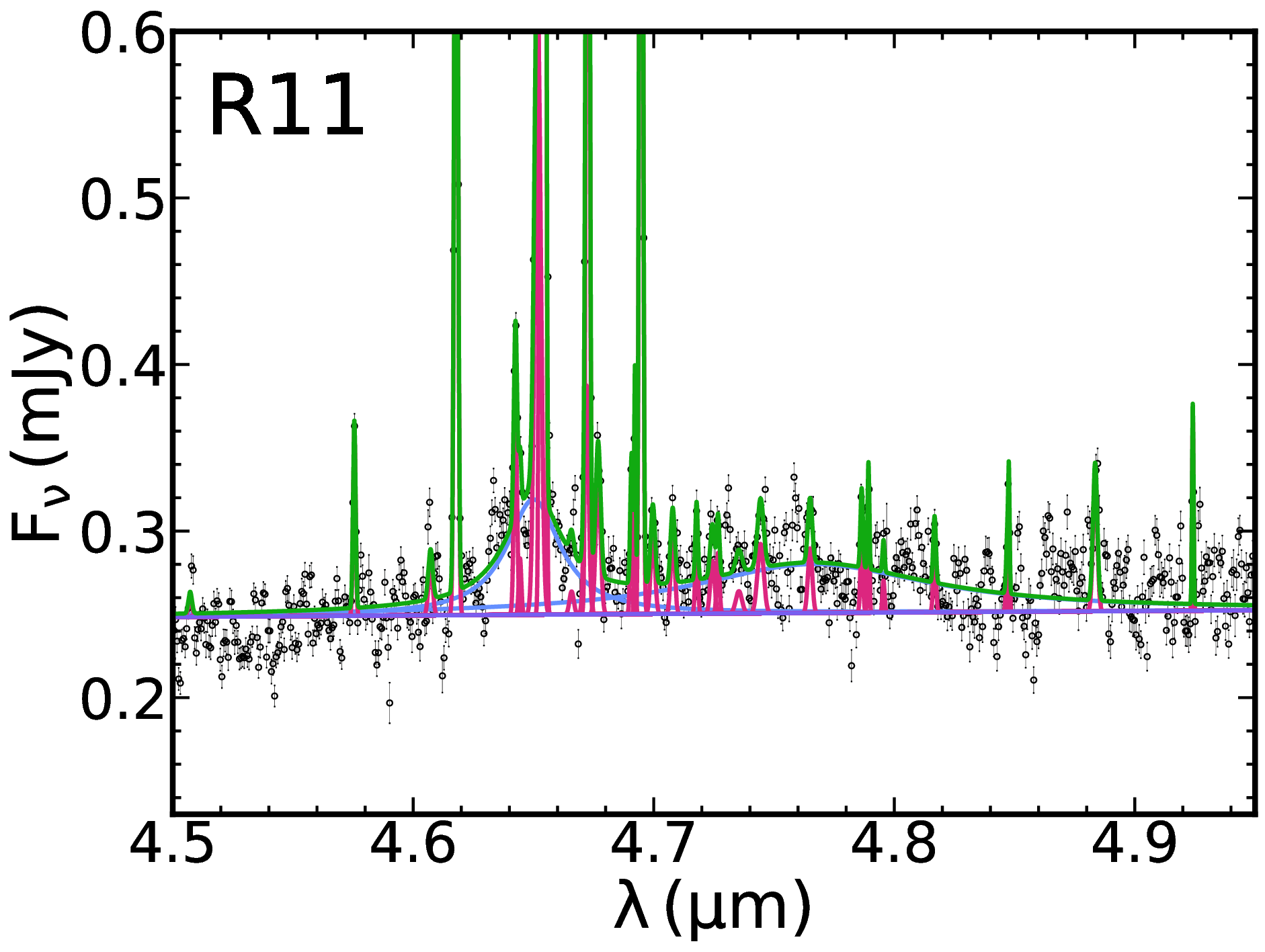}
\includegraphics[width=0.32\textwidth]{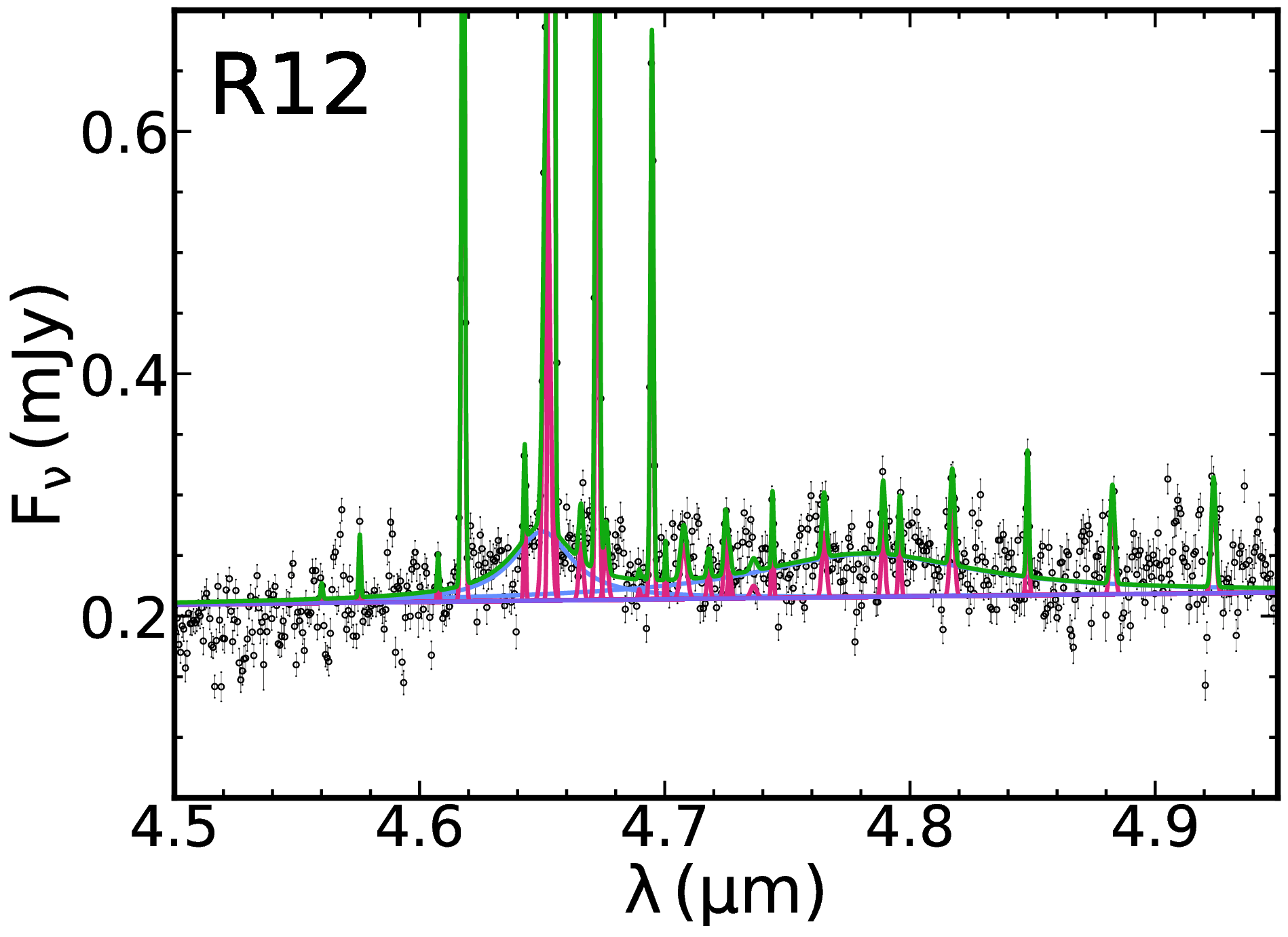}
\includegraphics[width=0.32\textwidth]{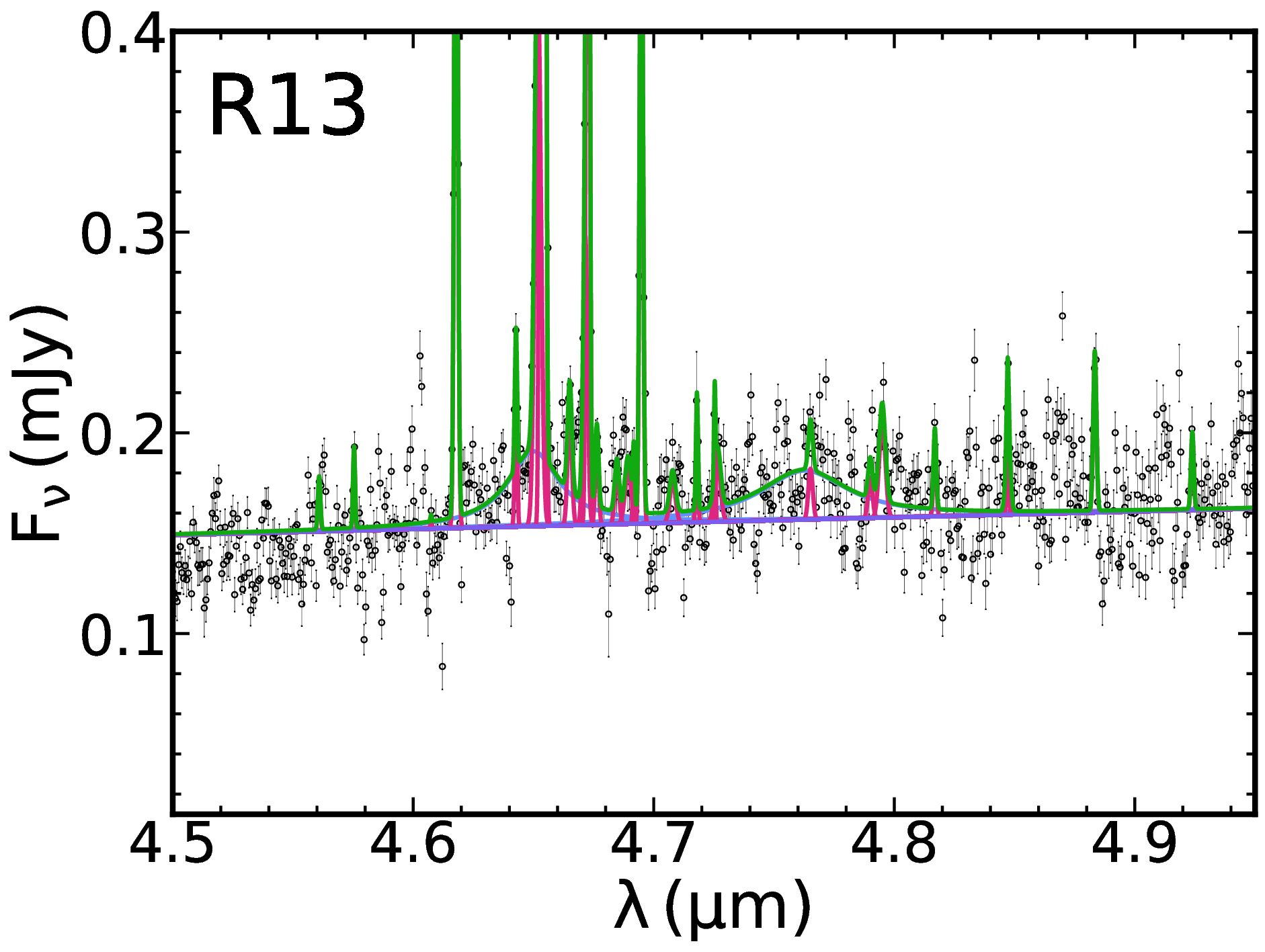}
\vspace{-0.2cm}
\caption{\label{fig:465fit}
Local spectral decomposition
of the 4.65 and 4.75$\mum$ aliphatic
C--D bands for Regions R2--R13.
}
\vspace{-0.1cm}
\end{figure*}

For 30 Dor, as illustrated in Figure~\ref{fig:465fit},
the 4.4$\mum$ feature is difficult to identify
due to the wavelength coverage gaps of NIRSpec.
Nevertheless, the 4.65$\mum$ feature is clearly detected.
Again, Table~\ref{tab:fit_parameters}
also summarizes the best-fit widths and intensities,
with the central wavelength fixed at 4.65$\mum$.
On average, we derive the width
to be $\gamma_{4.65}\simali$0.034$\mum$,
which is somewhat broader than
those reported previously,
e.g., $\simali$0.020$\mum$
\citep{Boersma2023} and
$\simali$0.027$\mum$
\citep{Draine2025}.
\citet{Yang2025} argued that the aliphatic C--D stretch
seen in the JWST spectra of the Orion Bar
\citep{Peeters2024} comprises two components
peaking at 4.65 and 4.75$\mum$.
However, due to the low SNR of the 30 Dor data,
the 4.75$\mum$ component is suboptimal.
Consequently, only the intensity of
the 4.65$\mum$ component is adopted
for our subsequent analysis.

\begin{figure}[h!]
\centering
\includegraphics[width=0.6\textwidth]{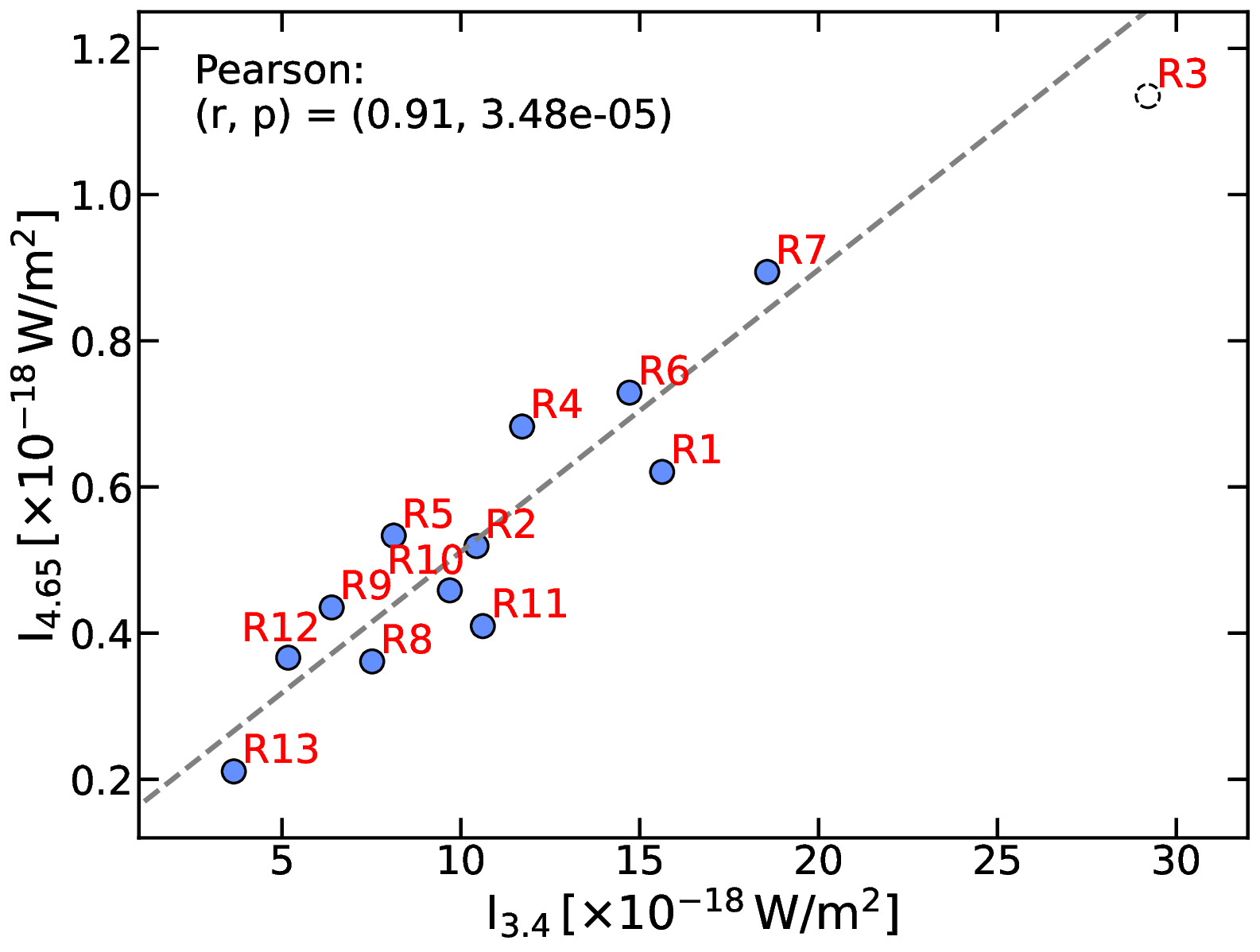}
\vspace{-0.3cm}
\caption{\label{fig:I34_I465}
Relation between the 3.4$\mum$
aliphatic C--H stretch and the 4.65$\mum$
aliphatic C--D stretch.
The results for R3 are indicated by
an open circle due to its low quality spectrum
and this region has been excluded
from the correlation analysis.
}
\vspace{-0.3cm}
\end{figure}

We examine in Figure~\ref{fig:I34_I465}
the relationship between the 3.4 and 4.65$\mum$ features.
With a Pearson correlation coefficient of $r\approx0.91$
($p<0.01$), they exhibit a significant positive correlation.
As the 4.65$\mum$ feature originates from the C--D
stretching mode in deuterated aliphatic group---a mechanism
highly analogous to the aliphatic C--H stretching that produces
the 3.4$\mum$ feature---this correlation perfectly aligns
with physical expectations.

\section{Summary} \label{sec:summary}
Utilizing JWST's NIRSpec and MIRI IFU data,
we have studied the UIE spectra of the 30 Doradus
star-forming complex in the LMC.
Our major results are as follows:
\begin{enumerate}
\item The UIE carriers in 30 Dor have an appreciable
amount of aliphatic contents and are aliphatically
deuterated, as revealed by the widespread detection
of the aromatic 3.3$\mum$ C--H stretches,
the 3.4$\mum$ aliphatic C--H stretches,
and the 6.85$\mum$ aliphatic C--H deformation,
as well as the 4.65$\mum$ aliphatic C--D stretches.
\item For the first time, the 6.85$\mum$ aliphatic C--H
deformation band is resolved into three subfeatures
at $\simali$6.83, 6.86, and 6.88$\mum$.
\item The intensities of the 3.4$\mum$ C--H band
and of the 4.65$\mum$ C--D band are correlated,
supporting the aliphatic nature of the 4.65$\mum$ band.
\end{enumerate}

\acknowledgments
We thank the anonymous referee
for helpful comments and suggestions
that considerably improved the quality
and presentation of this work.
JHP and XJY are supported in part by NSFC\,12333005
and 12122302, CMS-CSST-2021-A09,
and the Innovative Research Group Project
of Natural Science Foundation of Hunan Province
of China No. 2024JJ1008.
This work is based on observations
made with the NASA/ESA/CSA James Webb Space Telescope.
The data were obtained from
the Mikulski Archive for Space Telescopes
at the Space Telescope Science Institute,
which is operated by the Association of Universities
for Research in Astronomy, Inc., under NASA contract
NAS 5-03127 for JWST.
These observations are associated
with program \#2729.
The specific observations used in this paper
can be accessed via the MAST at DOI:
\url{https://doi.org/10.17909/rwfa-2870}.



\begin{thebibliography}{30}
\expandafter\ifx\csname natexlab\endcsname\relax\def\natexlab#1{#1}\fi
\providecommand{\url}[1]{\href{#1}{#1}}
\providecommand{\dodoi}[1]{doi:~\href{http://doi.org/#1}{\nolinkurl{#1}}}
\providecommand{\doeprint}[1]{\href{http://ascl.net/#1}{\nolinkurl{http://ascl.net/#1}}}
\providecommand{\doarXiv}[1]{\href{https://arxiv.org/abs/#1}{\nolinkurl{https://arxiv.org/abs/#1}}}

\bibitem[{Allamandola {et~al.}(1985)Allamandola, Tielens, \& Barker}]{Allamandola1985}
Allamandola, L.~J., Tielens, A. G. G.~M., \& Barker, J.~R. 1985, Astrophysical Journal, Vol. 290, p. L25-L28 (1985), 290, L25, \dodoi{10.1086/184435}


\bibitem[{Allamandola {et~al.}(2021)Allamandola, Boersma, Lee, Bregman, \& Temi}]{Allamandola2021}
Allamandola, L.~J., Boersma, C., Lee, T.~J., Bregman, J.~D., \& Temi, P. 2021, The Astrophysical Journal Letters, 917, L35, \dodoi{10.3847/2041-8213/ac17f0}


\bibitem[{Beintema {et~al.}(1996)}]{Beintema1996}
Beintema, D.~A., {van den Ancker}, M.~E., Molster, F.~J., {et~al.} 1996, A\&A, 315, L369

\bibitem[Bernstein \& Geballe(2024)]{Bernstein2024}
Bernstein, L.~S. \& Geballe, T.~R.\ 2024, \apj, 962, 2, 114, \dodoi{10.3847/1538-4357/ad1245}



\bibitem[{Boersma {et~al.}(2023)Boersma, Allamandola, Esposito, Maragkoudakis, Bregman, Temi, Lee, Fortenberry, \& Peeters}]{Boersma2023}
Boersma, C., Allamandola, L.~J., Esposito, V.~J., {et~al.} 2023, The Astrophysical Journal, 959, 74, \dodoi{10.3847/1538-4357/ad022b}

\bibitem[{Boulanger {et~al.}(2011)Boulanger, Onaka, Pilleri, \& Joblin}]{Boulanger2011}
Boulanger, F., Onaka, T., Pilleri, P., \& Joblin, C. 2011, EAS Publications Series, 46, 399, \dodoi{10.1051/eas/1146041}

\bibitem[{Cataldo {et~al.}(2013)Cataldo, Garc{\'{\i}}a-Hern{\'a}ndez, \& Manchado}]{Cataldo2013}
Cataldo, F., Garc{\'{\i}}a-Hern{\'a}ndez, D.~A., \& Manchado, A.\ 2013, MNRAS, 429, 3025, \dodoi{10.1093/mnras/sts558}

\bibitem[{{Chen} {et~al.}(2026){Chen}, {Li}, \& {Li}}]{Chen2026}
{Chen}, T., {Li}, K.~J., \& {Li}, A. 2026, Astronomy \& Astrophysics, submitted


\bibitem[{Chown {et~al.}(2024)}]{Chown2024}
Chown, R., Sidhu, A., Peeters, E., {et~al.} 2024, Astronomy \& Astrophysics, 685, A75, \dodoi{10.1051/0004-6361/202346662}

\bibitem[{Chown {et~al.}(2025)Chown, Okada, Peeters, Sidhu, Khan, Schefter, Trahin, Canin, Van De~Putte, Alarc{\'o}n, Schroetter, Kannavou, Habart, Bern{\'e}, Boersma, Cami, Dartois, Goicoechea, Gordon, \& Onaka}]{Chown2025}
Chown, R., Okada, Y., Peeters, E., {et~al.} 2025, Astronomy \& Astrophysics, 698, A86, \dodoi{10.1051/0004-6361/202452940}

\bibitem[{{Diaz-Santos} {et~al.}(2025){Diaz-Santos}, {Lai}, {Finnerty}, {Privon}, {Bonfini}, {Larson}, {Marshall}, {Armus}, \& {Charmandaris}}]{2025ascl.soft01001D}
{Diaz-Santos}, T., {Lai}, T. S.~Y., {Finnerty}, L., {et~al.} 2025, {CAFE: Continuum And Feature Extraction tool}, Astrophysics Source Code Library, record ascl:2501.001

\bibitem[{Doney {et~al.}(2016)Doney, Candian, Mori, Onaka, \& Tielens}]{Doney2016}
Doney, K.~D., Candian, A., Mori, T., Onaka, T., \& Tielens, A. G. G.~M. 2016, Astronomy \& Astrophysics, 586, A65, \dodoi{10.1051/0004-6361/201526809}

\bibitem[{Donnan {et~al.}(2023)Donnan, {Garc{\'i}a-Bernete}, Rigopoulou, {Pereira-Santaella}, {Alonso-Herrero}, Roche, {Hern{\'a}n-Caballero}, \& Spoon}]{Donnan2023}
Donnan, F.~R., {Garc{\'i}a-Bernete}, I., Rigopoulou, D., {et~al.} 2023, Monthly Notices of the Royal Astronomical Society, 519, 3691, \dodoi{10.1093/mnras/stac3729}



\bibitem[{Draine {et~al.}(2025)}]{Draine2025}
Draine, B.~T., Sandstrom, K., Dale, D.~A., {et~al.} 2025, The Astrophysical Journal Letters, 984, L42, \dodoi{10.3847/2041-8213/adc991}

\bibitem[{Duley \& Williams(1981)}]{Duley1981}
Duley, W.~W., \& Williams, D.~A. 1981, Monthly Notices of the Royal Astronomical Society, 196, 269, \dodoi{10.1093/mnras/196.2.269}

\bibitem[{Esposito {et~al.}(2025)Esposito, Fortenberry, Boersma, \& Allamandola}]{Esposito2025}
Esposito, V.~J., Fortenberry, R.~C., Boersma, C., \& Allamandola, L.~J. 2025, The Journal of Physical Chemistry A, 129, 244, \dodoi{10.1021/acs.jpca.4c07416}


\bibitem[{Geballe {et~al.}(1985)Geballe, Lacy, Persson, McGregor, \& Soifer}]{Geballe1985}
Geballe, T.~R., Lacy, J.~H., Persson, S.~E., McGregor, P.~J., \& Soifer, B.~T. 1985, The Astrophysical Journal, 292, 500, \dodoi{10.1086/163182}

\bibitem[{Geballe {et~al.}(1992)Geballe, Tielens, Kwok, \& Hrivnak}]{Geballe1992}
Geballe, T.~R., Tielens, A. G. G.~M., Kwok, S., \& Hrivnak, B.~J. 1992, The Astrophysical Journal, 387, L89, \dodoi{10.1086/186312}

\bibitem[{Geballe {et~al.}(1994)}]{Geballe1994}
Geballe, T.~R., Joblin, C., D'Hendecourt, L.~B., {et~al.} 1994, ApJL, 434, L15, \dodoi{10.1086/187561}

\bibitem[{Habart {et~al.}(2024)Habart, Peeters, Bern{\'e}, Trahin, Canin, \& Chown}]{Habart2024}
Habart, E., Peeters, E., Bern{\'e}, O., Trahin, B., Canin, A., \& Chown, R., {et~al.} 2024, Astronomy \& Astrophysics, 685, A73, \dodoi{10.1051/0004-6361/202346747}


\bibitem[{Hora {et~al.}(2026)Hora, Noh, Melnick, {et~al.}}]{Hora2026}
Hora, J.~L., Noh, J.~K., Melnick, G.~J., {et~al.} 2026, The Astrophysical Journal, 1001, 165, \dodoi{10.3847/1538-4357/ae5180}




\bibitem[{Inami {et~al.}(2018)Inami, Armus, Matsuhara, Charmandaris, {D{\'i}az-Santos}, Surace, Stierwalt, Ohyama, Howell, Marshall, Evans, Linden, \& Mazzarella}]{Inami2018}
Inami, H., Armus, L., Matsuhara, H., {et~al.} 2018, Astronomy \& Astrophysics, 617, A130, \dodoi{10.1051/0004-6361/201833053}

\bibitem[{Joblin {et~al.}(1996)Joblin, Tielens, Allamandola, \& Geballe}]{Joblin1996}
Joblin, C., Tielens, A. G. G.~M., Allamandola, L.~J., \& Geballe, T.~R. 1996, The Astrophysical Journal, 458, 610, \dodoi{10.1086/176843}

\bibitem[{Jones {et~al.}(1990)Jones, Duley, \& Williams}]{Jones1990}
Jones, A. P., Duley, W. W., \& Williams, D.A.\ 1990, QJRAS, 31, 567

\bibitem[{Knacke(1977)}]{Knacke1977}
Knacke, R.~F. 1977, Nature, 269, 132, \dodoi{10.1038/269132a0}

\bibitem[{Kondo {et~al.}(2012)Kondo, Kaneda, Oyabu, Ishihara, Mori, Yamagishi, Onaka, Sakon, \& Suzuki}]{Kondo2012}
Kondo, T., Kaneda, H., Oyabu, S., {et~al.} 2012, The Astrophysical Journal, 751, L18, \dodoi{10.1088/2041-8205/751/1/L18}

\bibitem[{Kondo {et~al.}(2024)}]{Kondo2024}
Kondo, T., Kondo, A., Murata, K.~L., {et~al.} 2024, Publications of the Astronomical Society of Japan, 76, 1041, \dodoi{10.1093/pasj/psae069}

\bibitem[Kwok(2022)]{Kwok2022}
Kwok, S.\ 2022, \apss, 367, 2, 16, \dodoi{10.1007/s10509-022-04045-6}

\bibitem[{Kwok \& Zhang(2011)}]{Kwok2011}
Kwok, S., \& Zhang, Y.\ 2011, Nature, 479, 80, \dodoi{10.1038/nature10542}

\bibitem[{Kwok \& Zhang(2013)}]{Kwok2013}
Kwok, S., \& Zhang, Y.\ 2013, ApJ, 771, 5, \dodoi{10.1088/0004-637X/771/1/5}

\bibitem[{Lai {et~al.}(2020)Lai, Smith, Baba, Spoon, \& Imanishi}]{Lai2020}
Lai, T. S.-Y., Smith, J. D.~T., Baba, S., Spoon, H. W.~W., \& Imanishi, M. 2020, The Astrophysical Journal, 905, 55, \dodoi{10.3847/1538-4357/abc002}

\bibitem[{Leger \& Puget(1984)}]{Leger1984}
Leger, A., \& Puget, J.~L. 1984, Astronomy and Astrophysics, 137, L5

\bibitem[{Li(2020)}]{Li2020}
Li, A. 2020, Nature Astronomy, 4, 339, \dodoi{10.1038/s41550-020-1051-1}

\bibitem[{Li \& Draine(2012)}]{Li2012}
Li, A., \& Draine, B.~T. 2012, The Astrophysical Journal, 760, L35, \dodoi{10.1088/2041-8205/760/2/L35}

\bibitem[{Lyu {et~al.}(2025)Lyu, Yang, Li, Sun, Rieke, Alberts, \& Shivaei}]{Lyu2025}
Lyu, J., Yang, X., Li, A., {et~al.} 2025, The Astrophysical Journal, 986, 156, \dodoi{10.3847/1538-4357/add538}



\bibitem[{McCarthy \& McGuire(2026)}]{McCarthy2026}
McCarthy, M.~C., \& McGuire, B.~A. 2026, Annual Review of Physical Chemistry, 77, 345, \dodoi{10.1146/annurev-physchem-082324-010544}

\bibitem[{McKinney {et~al.}(2026)McKinney, Eleazer, Pope, Sajina, Alberts, Stone, Sajkov, Vanicek, Kirkpatrick, Lai, Casey, Armus, Santos, Korkus, Cooper, House, Akins, Lambrides, Long, \& Yan}]{Mckinney2026}
McKinney, J., Eleazer, M., Pope, A., {et~al.} 2026, The Astrophysical Journal, 1001, 93, \dodoi{10.3847/1538-4357/ae4d46}

\bibitem[{{Mentzer} {et~al.}(2026){Mentzer}, {Li}, \& {Yang}}]{Mentzer2026}
{Mentzer}, C.~E., {Li}, A., \& {Yang}, X.~J. 2026, Astrophysical Journal Letters, 1005, L38, \dodoi{10.3847/2041-8213/ae7e86}

\bibitem[{Misselt {et~al.}(2025)Misselt, Witt, Gordon, Van De~Putte, Trahin, Abergel, {Noriega-Crespo}, Guillard, Zannese, Dell'ova, Baes, Klaassen, \& Ysard}]{Misselt2025}
Misselt, K., Witt, A.~N., Gordon, K.~D., {et~al.} 2025, Astronomy \& Astrophysics, 700, A158, \dodoi{10.1051/0004-6361/202554851}

\bibitem[{Mori {et~al.}(2012)Mori, Sakon, Onaka, Kaneda, Umehata, \& Ohsawa}]{Mori2012}
Mori, T.~I., Sakon, I., Onaka, T., Kaneda, H., Umehata, H., \& Ohsawa, R. 2012, The Astrophysical Journal, 744, 68, \dodoi{10.1088/0004-637X/744/1/68}

\bibitem[{Nagata {et~al.}(1988)Nagata, Tokunaga, Sellgren, Smith, Onaka, Nakada, \& Sakata}]{Nagata1988}
Nagata, T., Tokunaga, A.~T., Sellgren, K., {et~al.} 1988, The Astrophysical Journal, 326, 157, \dodoi{10.1086/166076}



\bibitem[{Onaka {et~al.}(2022)Onaka, Sakon, \& Shimonishi}]{Onaka2022}
Onaka, T., Sakon, I., \& Shimonishi, T. 2022, The Astrophysical Journal, 941, 190, \dodoi{10.3847/1538-4357/ac9b15}

\bibitem[{Papoular {et~al.}(1989)Papoular, Conrad, Giuliano, Kister, \& Mille}]{Papoular1989}
Papoular, R., Conrad, J., Giuliano, M., Kister, J., \& Mille, G.\ 1989, A\&A, 217, 204

\bibitem[{Peeters {et~al.}(2004)Peeters, Allamandola, Bauschlicher, Hudgins, Sandford, \& Tielens}]{Peeters2004}
Peeters, E., Allamandola, L.~J., Bauschlicher, Jr., C.~W., {et~al.} 2004, The Astrophysical Journal, 604, 252, \dodoi{10.1086/381866}

\bibitem[{Peeters {et~al.}(2024)}]{Peeters2024}
Peeters, E., Habart, E., Bern{\'e}, O., {et~al.} 2024, Astronomy \& Astrophysics, 685, A74, \dodoi{10.1051/0004-6361/202348244}

\bibitem[{{Pereira-Santaella} {et~al.}(2024){Pereira-Santaella}, {Gonz{\'a}lez-Alfonso}, {Garc{\'i}a-Bernete}, {Garc{\'i}a-Burillo}, \& Rigopoulou}]{Pereira-santaella2024}
{Pereira-Santaella}, M., {Gonz{\'a}lez-Alfonso}, E., {Garc{\'i}a-Bernete}, I., {Garc{\'i}a-Burillo}, S., \& Rigopoulou, D. 2024, Astronomy \& Astrophysics, 681, A117, \dodoi{10.1051/0004-6361/202347942}

\bibitem[{Riechers {et~al.}(2014)Riechers, Pope, Daddi, {et~al.}}]{Riechers2014}
Riechers, D.~A., Pope, A., Daddi, E., et al.\ 2014, ApJ, 796, 84, \dodoi{10.1088/0004-637X/796/2/84}

\bibitem[{Sabbi {et~al.}(2013)Sabbi, Anderson, Lennon, Van Der~Marel, Aloisi, Boyer, Cignoni, De~Marchi, De~Mink, Evans, Gallagher, Gordon, Gouliermis, Grebel, Koekemoer, Larsen, Panagia, Ryon, Smith, Tosi, \& Zaritsky}]{Sabbi2013}
Sabbi, E., Anderson, J., Lennon, D.~J., {et~al.} 2013, The Astronomical Journal, 146, 53, \dodoi{10.1088/0004-6256/146/3/53}

\bibitem[{Sabbi {et~al.}(2016)Sabbi, Lennon, Anderson, Cignoni, Marel, Zaritsky, Marchi, Panagia, Gouliermis, Grebel, Iii, Smith, Sana, Aloisi, Tosi, Evans, Arab, Boyer, Mink, Gordon, Koekemoer, Larsen, Ryon, \& Zeidler}]{Sabbi2016}
Sabbi, E., Lennon, D.~J., Anderson, J., {et~al.} 2016, The Astrophysical Journal Supplement Series, 222, 11, \dodoi{10.3847/0067-0049/222/1/11}

\bibitem[{Sakata {et~al.}(1990)Sakata, Wada, Onaka, \& Tokunaga}]{Sakata1990}
Sakata, A., Wada, S., Onaka, T., \& Tokunaga, A. T.\ 1990, ApJS, 353, 543, \dodoi{10.1086/168642}


\bibitem[{Schroetter {et~al.}(2024)Schroetter, Bern{\'e}, Joblin, Canin, Chown, Sidhu, Habart, Peeters, Lai, Candian, Chakraborty, Petrignani, Trahin, Van De~Putte, \& Alarc{\'o}n}]{Schroetter2024}
Schroetter, I., Bern{\'e}, O., Joblin, C., {et~al.} 2024, Astronomy \& Astrophysics, 685, A78, \dodoi{10.1051/0004-6361/202348974}

\bibitem[{Sellgren {et~al.}(1990)Sellgren, Tokunaga, \& Nakada}]{Sellgren1990}
Sellgren, K., Tokunaga, A.~T., \& Nakada, Y. 1990, The Astrophysical Journal, 349, 120, \dodoi{10.1086/168299}

\bibitem[{Sloan {et~al.}(1997)Sloan, Bregman, Geballe, Allamandola, \& Woodward}]{Sloan1997}
Sloan, G.~C., Bregman, J.~D., Geballe, T.~R., Allamandola, L.~J., \& Woodward, E. 1997, The Astrophysical Journal, 474, 735, \dodoi{10.1086/303484}

\bibitem[{Sloan {et~al.}(2014)Sloan, Lagadec, Zijlstra, Kraemer, Weis, Matsuura, Volk, Peeters, Duley, Cami, {Bernard-Salas}, Kemper, \& Sahai}]{Sloan2014}
Sloan, G.~C., Lagadec, E., Zijlstra, A.~A., {et~al.} 2014, The Astrophysical Journal, 791, 28, \dodoi{10.1088/0004-637X/791/1/28}

\bibitem[{Smith {et~al.}(2007)Smith, Draine, Dale, Moustakas, Kennicutt, Helou, Armus, Roussel, Sheth, Bendo, Buckalew, Calzetti, Engelbracht, Gordon, Hollenbach, Li, Malhotra, Murphy, \& Walter}]{Smith2007}
  Smith, J. D.~T., Draine, B.~T., Dale, D.~A., {et~al.} 2007, The Astrophysical Journal, 656, 770, \dodoi{10.1086/510549}

\bibitem[{Spilker {et~al.}(2023)Spilker, Phadke, Aravena, {et~al.}}]{Spilker2023}
Spilker, J.~S., Phadke, K.~A., Aravena, M., et al.\ 2023, Nature, 618, 708, \dodoi{10.1038/s41586-023-05998-6}


\bibitem[{Tokunaga(1997)}]{Tokunaga1997}
Tokunaga, A.~T. 1997, Diffuse Infrared Radiation and the IRTS, 124, 149

\bibitem[Tokunaga \& Bernstein(2021)]{Tokunaga2021}
Tokunaga, A.~T. \& Bernstein, L.~S.\ 2021, \apj, 916, 1, 52, \dodoi{10.3847/1538-4357/ac004b}

\bibitem[Tokunaga et al.(2025)]{Tokunaga2025}
Tokunaga, A.~T., Bernstein, L.~S., \& Onaka, T.\ 2025, MNRAS, 544, 4, 3280, \dodoi{10.1093/mnras/staf1874}

\bibitem[{Verstraete {et~al.}(1996)Verstraete, Puget, Falgarone, Drapatz, Wright, \& Timmermann}]{Verstraete1996}
Verstraete, L., Puget, J.~L., Falgarone, E., {et~al.} 1996, Astronomy and Astrophysics, 315, L337

\bibitem[{Walborn {et~al.}(2013)Walborn, Barb{\'a}, \& Sewi{\l}o}]{Walborn2013}
Walborn, N.~R., Barb{\'a}, R.~H., \& Sewi{\l}o, M.~M. 2013, The Astronomical Journal, 145, 98, \dodoi{10.1088/0004-6256/145/4/98}

\bibitem[{Witt \& Lai(2020)}]{Witt2020}
Witt, A.~N., \& Lai, T.~S.-Y. 2020, Astrophysics and Space Science, 365, 58, \dodoi{10.1007/s10509-020-03766-w}

\bibitem[{Yamagishi {et~al.}(2012)Yamagishi, Kaneda, Ishihara, Kondo, Onaka, Suzuki, \& Minh}]{Yamagishi2012}
Yamagishi, M., Kaneda, H., Ishihara, D., {et~al.} 2012, Astronomy \& Astrophysics, 541, A10, \dodoi{10.1051/0004-6361/201218904}

\bibitem[{Yang \& Li(2023)}]{Yang2023}
Yang, X.~J., \& Li, A. 2023, The Astrophysical Journal Supplement Series, 268, 12, \dodoi{10.3847/1538-4365/ace4c6}


\bibitem[{Yang \& Li(2025)}]{Yang2025}
---. 2025, The Astrophysical Journal, 983, 136, \dodoi{10.3847/1538-4357/adbd11}

\bibitem[{Yang {et~al.}(2016)Yang, Glaser, Li, \& Zhong}]{Yang2016}
Yang, X.~J., Glaser, R., Li, A., \& Zhong, J.~X.\ 2016, Monthly Notices of the Royal Astronomical Society, 462, 1551, \dodoi{10.1093/mnras/stw1740}

\bibitem[{Yang {et~al.}(2017)Yang, Glaser, Li, \& Zhong}]{Yang2017}
Yang, X.~J., Glaser, R., Li, A., \& Zhong, J.~X.\ 2017, New Astronomy Reviews, 77, 1, \dodoi{10.1016/j.newar.2017.01.001}




\bibitem[{Zagury(2021)}]{Zagury2021}
Zagury, F. 2021, Astronomy \& Astrophysics, 652, L5, \dodoi{10.1051/0004-6361/202141541}

\bibitem[{Zagury(2023)}]{Zagury2023}
Zagury, F. 2023, The Astrophysical Journal, 952, 116, \dodoi{10.3847/1538-4357/acdad0}

\bibitem[{Zagury(2025)}]{Zagury2025}
Zagury, F. 2025, The Astrophysical Journal, 981, 36, \dodoi{10.3847/1538-4357/adb0cc}

\bibitem[{Zhang {et~al.}(2025)Zhang, Hales, Peeters, Cami, Sidhu, \& Zhen}]{Zhang2025}
Zhang, C., Hales, J., Peeters, E., {et~al.} 2025, The Astrophysical Journal Supplement Series, 280, 4, \dodoi{10.3847/1538-4365/adea6b}

\end{thebibliography}


\begin{landscape}
\begin{table*}
\begin{center}
\caption{Spectral fitting results\label{tab:fitresult}}
{\small
\begin{tabular}{ccccccccccccc}
\hline
\textbf{Region} & \textbf{R.A.} & \textbf{Decl.} & \textbf{$\mathrm{I_{3.3}}$} & \textbf{$\mathrm{I_{3.4}}$} & \textbf{$\mathrm{I_{4.65}}$} & \textbf{$\mathrm{I_{6.2}}$} & \textbf{$\mathrm{I_{6.85}}$} & \textbf{$\mathrm{I_{7.7}}$} & \textbf{$\mathrm{I_{11.3}}$}\\
(1) & (2) & (3) & (4) & (5) & (6) & (7) & (8) & (9) & (10)\\
\hline
R1 & 5:38:46.7980 & -69:05:06.13 & 45.19 & 15.63 & 0.62 & 169.55 & 1.53 & 443.21 & 144.36 \\
R2 & 5:38:46.9580 & -69:05:06.33 & 29.74 & 10.44 & 0.52 & 114.92 & 1.49 & 321.57 & 100.21 \\
R3 & 5:38:47.0050 & -69:05:05.65 & 80.02 & 29.20 & 1.13 & 141.30 & 0.75 & 457.41 & 133.53 \\
R4 & 5:38:46.9630 & -69:05:04.91 & 30.64 & 11.71 & 0.68 & 131.88 & 1.59 & 359.36 & 117.54 \\
R5 & 5:38:46.8350 & -69:05:04.94 & 20.59 & 8.12 & 0.53 & 92.91 & 1.68 & 260.55 & 85.59 \\
R6 & 5:38:46.7450 & -69:05:05.43 & 42.24 & 14.72 & 0.73 & 112.48 & 1.42 & 307.73 & 100.11 \\
R7 & 5:38:47.0900 & -69:05:05.91 & 51.47 & 18.56 & 0.89 & 127.90 & 1.24 & 350.37 & 109.89 \\
R8 & 5:38:47.0800 & -69:05:05.25 & 18.74 & 7.52 & 0.36 & 92.95 & 1.34 & 260.09 & 92.29 \\
R9 & 5:38:46.7300 & -69:05:04.85 & 18.97 & 6.39 & 0.44 & 97.19 & 1.38 & 274.59 & 82.84 \\
R10 & 5:38:46.9140 & -69:05:04.55 & 24.16 & 9.69 & 0.46 & 94.61 & 1.46 & 249.42 & 85.17 \\
R11 & 5:38:46.8050 & -69:05:04.30 & 27.68 & 10.62 & 0.41 & 123.02 & 1.32 & 328.08 & 103.13 \\
R12 & 5:38:46.0950 & -69:05:04.69 & 11.48 & 5.17 & 0.37 & 60.91 & 1.35 & 182.04 & 65.97 \\
R13 & 5:38:46.8650 & -69:05:03.62 & 96.87 & 3.65 & 0.21 & 69.37 & 1.13 & 172.22 & 58.16 \\
\hline
\end{tabular}
}
\smallskip
\begin{minipage}{\textwidth}
\footnotesize
Columns: (1) Selected regions; (2)--(3) R.A. and Decl. (J2000) of the
aperture center; (4)--(10) Intensities of the UIE band in units of
$\mathrm{10^{-18}\,W\,m^{-2}}$. With the exception of the 3.3
and 4.65$\mum$ bands, each listed intensity represents the sum
of several subcomponents.
\end{minipage}
\end{center}
\end{table*}
\end{landscape}

\begin{table}[h!]
\begin{center}
\small
\caption{Spectral Modifications to PAHFIT\label{tab:modifications}}
\begin{tabular}{cccc}
\hline
$\lambda$ & $\lambda$ ranges & FWHM & FWHM ranges\\
($\mu$m) & ($\mu$m) & ($\mu$m) & ($\mu$m)\\
\hline
3.29 & [3.27, 3.31] & 0.0418 & [0.03762, 0.0836] \\
3.40 & [3.38, 3.42] & 0.0299 & [0.01, 0.06] \\
3.47 & [3.43, 3.50] & 0.04 & [0.01, 0.1] \\
3.51 & [3.49, 3.53] & 0.0299 & [0.01, 0.06] \\
3.56 & [3.54, 3.58] & 0.0299 & [0.01, 0.06] \\
5.88 & [5.85, 5.90] & 0.0528 & [0.03, 0.12] \\
6.02 & [6.0, 6.1] & 0.0569 & [0.02, 0.12] \\
6.212 & [6.15, 6.25] & 0.0932 & [0.05, 0.15] \\
6.32 & [6.2, 6.5] & 0.2571 & [0.15, 0.35] \\
6.828 & [6.805, 6.85] & 0.013 & [0.005, 0.02] \\
6.858 & [6.848, 8.868] & 0.015 & [0.005, 0.025] \\
6.878 & [6.868, 6.9] & 0.013 & [0.005, 0.02] \\
11.02 & [10.98, 11.1] & 0.0551 & [0.02, 0.1] \\
11.191 & [11.13, 11.23] & 0.023 & [0.01, 0.06] \\
11.211 & [11.18, 11.25] & 0.028 & [0.01, 0.06] \\
11.244 & [11.21, 11.28] & 0.043 & [0.01, 0.1] \\
11.291 & [11.25, 11.33] & 0.077 & [0.01, 0.15] \\
11.380 & [11.35, 11.42] & 0.362 & [0.1, 0.399] \\
3.05 ($\mathrm{H_2O\ ice}$ absorption) & [3.0, 3.1] & 0.406 & [0.4056, 0.528] \\
4.27 ($\mathrm{CO_2\ ice}$ absorption) & fixed & 0.033 & [0.0264, 0.396] \\
\hline
\end{tabular}
\smallskip
\end{center}
\end{table}

\small
\begin{longtable}{l cccc cccc cccc}
\caption{Central Wavelengths ($\lambda$), Widths ($\gamma$),
and Intensities for the Aliphatic C--H and C--D Stretches}
    \label{tab:fit_parameters} \\
    \hline
    \hline
    \multirow{2}{*}{Region} &
    \multicolumn{4}{c}{Aliphatic C--H stretch} &
    \multicolumn{4}{c}{Aliphatic C--H deformation} &
    \multicolumn{4}{c}{Aliphatic C--D stretch} \\
    \cmidrule(lr){2-5} \cmidrule(lr){6-9} \cmidrule(lr){10-13}
    & $\lambda_j$ & $\gamma_j$ & $P_j^\dag$ & $f_j^{\ddag}$ &
      $\lambda_j$ & $\gamma_j$ & $P_j^\dag$ & $f_j^{\ddag}$ &
      $\lambda_j$ & $\gamma_j$ & $P_j^\dag$ & $f_j^{\ddag}$\\
    & ($\mu$m) & ($\mu$m) & & &
      ($\mu$m) & ($\mu$m) & & &
      ($\mu$m) & ($\mu$m) & & \\
    \hline
    \endfirsthead

\caption{Central Wavelengths ($\lambda$), Widths ($\gamma$),
and Intensities for the Aliphatic C--H and C--D Stretches (Continued)}\\
    \hline
    \hline
    \multirow{2}{*}{Region} &
    \multicolumn{4}{c}{Aliphatic C--H stretch} &
    \multicolumn{4}{c}{Aliphatic C--H deformation} &
    \multicolumn{4}{c}{Aliphatic C--D stretch} \\
    \cmidrule(lr){2-5} \cmidrule(lr){6-9} \cmidrule(lr){10-13}
    & $\lambda_j$ & $\gamma_j$ & $P_j^\dag$ & $f_j^{\ddag}$ &
      $\lambda_j$ & $\gamma_j$ & $P_j^\dag$ & $f_j^{\ddag}$ &
      $\lambda_j$ & $\gamma_j$ & $P_j^\dag$ & $f_j^{\ddag}$ \\
    & ($\mu$m) & ($\mu$m) & & &
      ($\mu$m) & ($\mu$m) & & &
      ($\mu$m) & ($\mu$m) & & \\
    \hline
    \endhead

    \hline
    \endfoot

\hline
    \multicolumn{13}{l}{
        \begin{minipage}{\textwidth}
            \vspace{2pt}
            \footnotesize
            $^\dag$ $P_j \equiv \int_{\lambda_{j}}^{} \Delta
            F_{\lambda}\,d\lambda$, in units of
            $\mathrm{10^{-18}\,W\,m^{-2}}$, is the power emitted in
            the $j$-th subfeature.\\
            $^{\ddag}$ $f_j= P_j/\sum P_j$ is the fractional power
            emitted in the $j$-th subfeature.
        \end{minipage}
    } \\
    \endlastfoot

        \multirow{4}{*}{$R_{1}$} & 3.404 & 0.032 & 7.21 & 0.46 & 6.827 & 0.007 & 0.33 & 0.21 & 4.650 & 0.033 & 0.62 & -- \\*
        & 3.461 & 0.065 & 6.27 & 0.40 & 6.861 & 0.025 & 0.83 & 0.54 & & & & \\*
        & 3.517 & 0.026 & 1.77 & 0.11 & 6.891 & 0.020 & 0.37 & 0.24 & & & & \\*
        & 3.567 & 0.022 & 0.37 & 0.02 & & & & & & & & \\
        \hline
        \multirow{4}{*}{$R_{2}$} & 3.404 & 0.032 & 4.88 & 0.47 & 6.829 & 0.012 & 0.39 & 0.26 & 4.650 & 0.033 & 0.52 & -- \\*
        & 3.462 & 0.066 & 4.31 & 0.41 & 6.861 & 0.022 & 0.73 & 0.49 & & & & \\*
        & 3.517 & 0.023 & 1.04 & 0.10 & 6.892 & 0.020 & 0.37 & 0.25 & & & & \\*
        & 3.563 & 0.020 & 0.22 & 0.02 & & & & & & & & \\
        \hline
        \multirow{4}{*}{$R_{3}$} & 3.404 & 0.036 & 15.51 & 0.53 & -- & -- & 0.00 & 0.00 & 4.650 & 0.028 & 1.13 & -- \\*
        & 3.463 & 0.060 & 10.22 & 0.35 & 6.856 & 0.014 & 0.40 & 0.54 & & & & \\*
        & 3.517 & 0.025 & 3.06 & 0.10 & 6.879 & 0.019 & 0.35 & 0.46 & & & & \\*
        & 3.562 & 0.017 & 0.43 & 0.01 & & & & & & & & \\
        \hline
        \multirow{4}{*}{$R_{4}$} & 3.404 & 0.031 & 5.87 & 0.50 & 6.840 & 0.020 & 0.37 & 0.23 & 4.650 & 0.027 & 0.68 & -- \\*
        & 3.461 & 0.061 & 4.32 & 0.37 & 6.860 & 0.025 & 0.79 & 0.49 & & & & \\*
        & 3.516 & 0.029 & 1.53 & 0.13 & 6.877 & 0.020 & 0.43 & 0.27 & & & & \\*
        & -- & -- & 0.00 & 0.00 & & & & & & & & \\
        \hline
        \multirow{4}{*}{$R_{5}$} & 3.404 & 0.033 & 4.07 & 0.50 & 6.831 & 0.020 & 0.58 & 0.35 & 4.650 & 0.060 & 0.53 & -- \\*
        & 3.464 & 0.063 & 3.09 & 0.38 & 6.860 & 0.019 & 0.87 & 0.52 & & & & \\*
        & 3.518 & 0.023 & 0.87 & 0.11 & 6.883 & 0.013 & 0.23 & 0.14 & & & & \\*
        & 3.574 & 0.010 & 0.10 & 0.01 & & & & & & & & \\
        \hline
        \multirow{4}{*}{$R_{6}$} & 3.404 & 0.032 & 7.04 & 0.48 & 6.826 & 0.005 & 0.10 & 0.07 & 4.650 & 0.035 & 0.73 & -- \\*
        & 3.462 & 0.063 & 5.64 & 0.38 & 6.854 & 0.025 & 0.78 & 0.55 & & & & \\*
        & 3.517 & 0.024 & 1.71 & 0.12 & 6.878 & 0.020 & 0.54 & 0.38 & & & & \\*
        & 3.564 & 0.022 & 0.33 & 0.02 & & & & & & & & \\
        \hline
        \multirow{4}{*}{$R_{7}$} & 3.404 & 0.034 & 9.32 & 0.50 & -- & -- & 0.00 & 0.00 & 4.650 & 0.037 & 0.89 & -- \\*
        & 3.462 & 0.062 & 6.79 & 0.37 & 6.850 & 0.025 & 0.64 & 0.51 & & & & \\*
        & 3.517 & 0.027 & 2.09 & 0.11 & 6.876 & 0.020 & 0.60 & 0.49 & & & & \\*
        & 3.560 & 0.020 & 0.35 & 0.02 & & & & & & & & \\
        \hline
        \multirow{4}{*}{$R_{8}$} & 3.404 & 0.033 & 3.24 & 0.43 & 6.831 & 0.016 & 0.40 & 0.30 & 4.650 & 0.023 & 0.36 & -- \\*
        & 3.464 & 0.084 & 3.51 & 0.47 & 6.861 & 0.022 & 0.85 & 0.64 & & & & \\*
        & 3.518 & 0.021 & 0.52 & 0.07 & 6.887 & 0.005 & 0.08 & 0.06 & & & & \\*
        & 3.564 & 0.042 & 0.25 & 0.03 & & & & & & & & \\
        \hline
        \multirow{4}{*}{$R_{9}$} & 3.404 & 0.031 & 2.88 & 0.45 & 6.829 & 0.013 & 0.39 & 0.28 & 4.650 & 0.050 & 0.44 & -- \\*
        & 3.461 & 0.066 & 2.66 & 0.42 & 6.859 & 0.023 & 0.94 & 0.68 & & & & \\*
        & 3.518 & 0.024 & 0.69 & 0.11 & 6.883 & 0.006 & 0.05 & 0.04 & & & & \\*
        & 3.563 & 0.023 & 0.15 & 0.02 & & & & & & & & \\
        \hline
        \multirow{4}{*}{$R_{10}$} & 3.404 & 0.034 & 4.85 & 0.50 & 6.826 & 0.005 & 0.07 & 0.05 & 4.650 & 0.028 & 0.46 & -- \\*
        & 3.463 & 0.063 & 3.61 & 0.37 & 6.857 & 0.019 & 0.78 & 0.54 & & & & \\*
        & 3.517 & 0.026 & 1.07 & 0.11 & 6.881 & 0.017 & 0.60 & 0.41 & & & & \\*
        & 3.566 & 0.016 & 0.16 & 0.02 & & & & & & & & \\
        \hline
        \multirow{4}{*}{$R_{11}$} & 3.405 & 0.034 & 5.38 & 0.51 & 6.827 & 0.005 & 0.09 & 0.06 & 4.650 & 0.027 & 0.41 & -- \\*
        & 3.463 & 0.063 & 3.96 & 0.37 & 6.858 & 0.013 & 0.61 & 0.46 & & & & \\*
        & 3.517 & 0.024 & 1.08 & 0.10 & 6.881 & 0.019 & 0.63 & 0.48 & & & & \\*
        & 3.564 & 0.021 & 0.19 & 0.02 & & & & & & & & \\
        \hline
        \multirow{4}{*}{$R_{12}$} & 3.402 & 0.030 & 2.23 & 0.43 & 6.825 & 0.020 & 0.09 & 0.07 & 4.650 & 0.030 & 0.37 & -- \\*
        & 3.457 & 0.076 & 2.03 & 0.39 & 6.857 & 0.017 & 0.64 & 0.47 & & & & \\*
        & 3.510 & 0.042 & 0.82 & 0.16 & 6.881 & 0.020 & 0.62 & 0.46 & & & & \\*
        & 3.546 & 0.010 & 0.10 & 0.02 & & & & & & & & \\
        \hline
        \multirow{4}{*}{$R_{13}$} & 3.406 & 0.037 & 2.07 & 0.57 & 6.825 & 0.020 & 0.26 & 0.23 & 4.650 & 0.026 & 0.21 & -- \\*
        & 3.467 & 0.055 & 1.16 & 0.32 & 6.857 & 0.016 & 0.46 & 0.41 & & & & \\*
        & 3.518 & 0.023 & 0.42 & 0.11 & 6.878 & 0.019 & 0.41 & 0.36 & & & & \\*
        & -- & -- & 0.00 & 0.00 & & & & & & & & \\
        \hline
        \cline{2-13}
        \multirow{4}{*}{Average} & 3.404 & 0.033 & 5.73 & 0.49 & 6.829 & 0.013 & 0.28 & 0.19 & 4.650 & 0.034 & 0.57 & -- \\*
        & 3.462 & 0.065 & 4.43 & 0.38 & 6.858 & 0.020 & 0.72 & 0.53 & & & & \\*
        & 3.517 & 0.026 & 1.28 & 0.11 & 6.882 & 0.017 & 0.41 & 0.31 & & & & \\*
        & 3.563 & 0.020 & 0.24 & 0.02 & & & & & & & & \\
    \hline
\end{longtable}


\end{document}